\documentclass[11pt]{article}

\usepackage{geometry}
\usepackage{graphicx}
\usepackage{multirow}  
\usepackage{amsmath}
\usepackage{natbib}
\usepackage{amsfonts} 
\usepackage{booktabs}
\usepackage{pifont}
\usepackage[shortlabels]{enumitem}
\usepackage{titlesec}

\DeclareMathOperator*{\argmin}{argmin}
\newcommand{\cmark}{\ding{51}}
\newcommand{\xmark}{\ding{55}}
\titleformat*{\paragraph}{\itshape\mdseries}

\begin{document}

\title{High-dimensional regression in practice: an empirical study of finite-sample prediction, variable selection and ranking}

\author{Fan Wang$^1$ \and Sach Mukherjee$^2$ \and Sylvia Richardson$^1$ \and Steven M.\ Hill$^1$ \vspace{0.1cm} \and
$^{1}$MRC Biostatistics Unit, University of Cambridge, Cambridge, UK\\
$^{2}$German Centre for Neurodegenerative Diseases (DZNE), Bonn, Germany
}
\date{\today}

\maketitle

\begin{abstract}
Penalized likelihood approaches are widely used for high-dimensional regression. 
Although many methods have been proposed and the associated theory is now well-developed, the relative efficacy of different approaches in finite-sample settings, as encountered in practice, remains incompletely understood. 
There is therefore a need for empirical investigations in this area that can offer practical insight and guidance to users. In this paper we present a large-scale comparison of penalized regression methods. 
We distinguish between three related goals: prediction, variable selection and variable ranking. 
Our results span more than 2,300 data-generating scenarios, including both synthetic and semi-synthetic data (real covariates and simulated responses), allowing us to systematically consider the influence of various factors (sample size, dimensionality, sparsity, signal strength and multicollinearity). 
We consider several widely-used approaches (Lasso, Adaptive Lasso, Elastic Net, Ridge Regression, SCAD, the Dantzig Selector and Stability Selection). 
We find considerable variation in performance between methods. 
Our results support a `no panacea' view, with no unambiguous winner across all scenarios or goals, even in this restricted setting where all data align well with the assumptions underlying the methods. 
The study allows us to make some recommendations as to which approaches may be most (or least) suitable given the goal and some data characteristics.
Our empirical results complement existing theory and provide a resource to compare methods across a range of scenarios and metrics.

\vspace{0.5cm}
\end{abstract}

\section{Introduction}
In a wide range of applications it is now routine to encounter regression problems where the number of features or covariates $p$ exceeds the sample size  $n$, often greatly. 
Even in the simple case of linear models with independent Gaussian noise, estimation is nontrivial and requires specific assumptions. A common and often appropriate assumption is that of sparsity, where only a subset of the variables (the active set) have non-zero coefficients, with the number $s_0$ of such active variables usually assumed much smaller than $p$. 

Penalized  methods augment the regression log-likelihood with a penalty term that encodes a structural assumption such as sparsity. Recent years have seen much progress in theory and methodology for penalized regression (see \citealp{Buhlmann2011}, for a lucid account).
However, while the theoretical developments have been remarkable and insightful, they cannot go as far as telling the user which method to use in a given finite-sample setting. 
Meanwhile, rapid methodological progress has meant a wide range of plausible approaches to choose amongst. 

The present study performs a systematic empirical comparison of a number of penalized regression methods, which could provide some guidance for users when selecting methods for specific applications.
We consider seven popular approaches (Lasso, Adaptive Lasso, Elastic Net, Ridge Regression, SCAD, the Dantzig Selector and Stability Selection) and a  range of data-generating scenarios. 
It is obvious that large departures from modeling assumptions can produce poor results. 
Here our intention is not so much to look at robustness to such departures, but rather to  look at variation in performance even in the favorable case where assumptions broadly hold (i.e. for data generated from sparse linear models). 
             
In the simulations, we vary  a number of factors in a relatively fine-grained manner within an essentially full factorial design (i.e.  all combinations of factors).
Furthermore, in addition to synthetic data (covariates and responses are simulated), we also consider semi-synthetic data (real covariates but simulated responses, using gene expression data from cancer samples) which allows us to study method performance under a more realistic covariate correlation structure.
We distinguish between three goals: prediction, variable selection and variable ranking.
 We consider  variable ranking in addition to selection due to the fact that in many applications, users are interested in guidance for follow-up studies or data acquisition. 
Then, highlighting  variables in a suitable rank order is particularly important. 

We find that for many scenarios there is substantial variation in performance between methods (i.e. choice of method is influential).
However, there is no unambiguous winner across scenarios (i.e details of the data-generating set-up matter), and this is despite the fact that we focus on a relatively narrow class of scenarios broadly favorable to the approaches employed.
Relative performance also depends on the specific goal.

Our study allows some broad recommendations to be made based on the goal and on characteristics of the data that are known, or can be determined, by the user (e.g. correlation structure).
We find that Lasso and Adaptive Lasso are usually competitive for ranking when there is no or very weak correlation between variables, and Ridge Regression is often a good choice in more highly correlated scenarios.
For prediction, Lasso is competitive in most scenarios (correlated or uncorrelated).
Choice of method for selection depends on whether the user would rather keep false positives low or maximize the number of active variables discovered.
For the former, our results suggest Stability Selection is the best option, and for the latter, Adaptive Lasso performs well when variables have no or very weak correlation and Elastic Net when variables are more highly correlated.
Lasso typically offers a reasonable compromise between controlling false positives and discovering true positives.

We also find evidence of an interesting ``phase transition''-like behavior for SCAD, where it goes from being the best performing approach to the worst as scenario difficulty increases.
SCAD is therefore highly variable and so carries more risk as a choice of approach.
Ridge Regression and Adaptive Lasso can also perform particularly poorly relative to other approaches in some scenarios for prediction.
Furthermore, our results and associated simulation and plotting code (see the ``Code and data availability'' section below) provide a resource, allowing users to check in detail how the methods considered here fare against each other across many scenarios and also to extend the study with other (existing or novel) approaches.

In addition to the main simulation study, we extend some data-generating scenarios in specific directions to further explore properties of the methods. Specifically, we investigate how performance changes under a different covariate correlation structure to that explored in the main study, we explore sensitivity of Stability Selection to its tuning parameters and we examine the impact of heterogeneous regression coefficients on selection performance.

A number of previous papers have examined the empirical performance of penalized regression methods.
 \citet{Meinshausen2010} consider large $p$ problems from a selection perspective. 
\citet{Buhlmann2014} is a more comprehensive study using  semi-synthetic data and evaluating screening or ranking properties in high dimensional settings.
\citet{Hastie2017} consider both low- and high-dimensional settings with a focus on prediction.
In contrast to previous work, our design is considerably more comprehensive and systematic. 
We use finer grids on factors including $n,p,s_0$ and signal-to-noise ratio (SNR) so that our results cover a wider range of designs, allowing us to more fully investigate the trends in relative performance. 
We also consider several types of multicollinearity, so we can better understand this practically important factor. Furthermore, we  evaluate all three of prediction, selection and ranking, using specific performance metrics for each.  To limit scope we do not consider Bayesian approaches here but note that there have been some interesting empirical comparisons of frequentist and Bayesian methods \citep[including][]{celeux2012,Bondell2012,Perrakis2019}.

The remainder of the paper is organized as follows. 
In Section \ref{sec:methods}, we outline the methods compared and describe our simulation strategy, including the data-generating factors considered. We also give details of how the methods are implemented and the performance metrics used. 
Section \ref{sec:systematic_comparison} presents the results from our main simulation study.
For each goal we present some key observations and provide a summary with some recommendations.
Results from additional simulations appear in Section \ref{sec:additional_simulations}.
We conclude with a discussion in Section~\ref{sec:discussion}.

\section{Methods} \label{sec:methods}

\subsection{Model setting and notation}
We focus on the best studied high-dimensional regression setting, namely the sparse linear model with independent Gaussian noise.  That is, we consider models of the form 
\begin{equation} \label{eq:linear_model}
\mathbf{y}=\mathbf{X} \boldsymbol{\beta}+\boldsymbol{\epsilon},
\end{equation}
\noindent
where $\mathbf{y}$=$(y_\textnormal{1},y_\textnormal{2}, \dots ,y_\textnormal{n})^T$ is a vector of responses, $\mathbf{X}=[\mathbf{x}_1, \dots, \mathbf{x}_p]$ a $n\times p$ design matrix, $\boldsymbol{\beta}=(\beta_1, \dots ,\beta_p)^T$  a vector of  (true)  coefficients and $\boldsymbol{\epsilon}=(\epsilon_1, \epsilon_2, \dots ,\epsilon_n)^T$ are the errors. We use $S= \{ j : \beta_j\neq 0 \}$ to denote the active set with 
$s_0=|S|$ the number of active variables (below, we also refer to active variables as ``signals'').
We focus on the case where $p>n$ and where 
$s_0$ is small  (i.e. a sparse setting). Unless otherwise noted, 
$\boldsymbol{\epsilon}\sim N_n(\mathbf{0},\sigma^\textnormal{2} \it{I_n}), \, \sigma>0$, where 
$N_n$ is the $n$-dimensional Gaussian and
$\it{I_n}$ the $n {\times} n$ identity matrix.

\subsection{The methods considered} \label{sec:algorithms}

A general penalized estimate for linear regression takes the following form:
\begin{equation}\label{eq:penEst}
\hat{\boldsymbol{\beta}}_{\lambda}=\argmin_{\beta} \frac{1}{2n}\|\mathbf{y}-\mathbf{X}\boldsymbol{\beta}\|^2+\sum_{j=1}^p P_{\lambda}(\beta_j)
\end{equation}
where $P_{\lambda}(\beta_j)$ is a penalty function applied to each component of $\boldsymbol{\beta}$ and $\lambda \geq 0$ is a tuning parameter that controls the amount of penalization. 
We consider several specific methods, outlined below.

\smallskip

\noindent
\textbf{Lasso}. The Lasso estimator \citep{Tibshirani1996} takes the form given in (\ref{eq:penEst}) with an $L_1$-norm penalty:
$P_{\lambda}(\beta_j)=\lambda|\beta_j|$.
This shrinks coefficients towards zero, with some  set to exactly zero, and $\lambda$ controls the amount of shrinkage and degree of sparsity.

The theoretical properties of the Lasso have been well-studied and an extensive treatment can be found in \citet{Buhlmann2011}. 
We provide a very brief summary of the conditions for consistent selection and prediction. 
Allowing $p\! \gg\! n$, under a sparsity assumption on $\boldsymbol{\beta}$, Lasso is consistent for prediction for values of $\lambda$ in a suitable range of the order $\sqrt{\log(p)/n}$.
Additional assumptions can be made on the design matrix $\mathbf{X}$ to improve the rate of convergence for prediction error and to obtain consistency for estimation. 
For consistent variable selection, further non-trivial assumptions need to be made. 
One is a ``beta-min'' assumption that requires coefficients for active variables to be sufficiently large.
If we then further assume a restrictive assumption on the design matrix $\mathbf{X}$, called the irrepresentable condition \citep{Zhao2006} (or equivalently the neighborhood stability assumption; \citealt{Meinshausen2006}), which places restrictions on correlation between variables, then Lasso is consistent for variable selection for $\lambda \gg \sqrt{\log(p)/n}$.

We highlight three important points arising from the above. 
First, that the conditions required for consistent selection are much stronger than those for consistent prediction.
Second, that $\lambda$ should be larger for consistent variable selection than for consistent prediction. 
Third, that the prediction-optimal $\lambda$ (estimated using e.g. cross-validation) can lead to inclusion of many false positives \citep{Meinshausen2006}.

\smallskip

\noindent
\textbf{Ridge Regression.}
Ridge Regression \citep{Hoerl1970} uses an $L_2$-norm penalty in (\ref{eq:penEst}):
$P_{\lambda}(\beta_j)=\lambda\beta_j^2$. This  shrinks coefficients towards zero, but results in non-sparse solutions because it is not singular at the origin. It also has a grouping effect where correlated variables have similar estimates.
Note that Ridge Regression is the only method considered here that does not perform variable selection {\it per se}.

\smallskip 

\noindent
\textbf{Elastic Net.} The Elastic Net estimator \citep{Zou2005} is (\ref{eq:penEst}) with a penalty
\begin{equation}
P_{\lambda}(\beta_j)=\lambda \left(\alpha |\beta_j|+(1-\alpha)\beta_j^2\right).
\end{equation}
That is, $L_1$- and $L_2$-norm penalties combined with an additional  parameter $\alpha \in [0,1]$ ($\alpha{=}1$ and $\alpha{=}0$ correspond to Lasso and Ridge respectively). 
 This combines some of the benefits of Ridge while giving sparse solutions. 
In the $p>n$ setting,  Lasso can select at most $n$ variables, but Elastic Net has no such limitation.

\smallskip

\noindent
\textbf{SCAD.} SCAD \citep{Fan2001} uses the following penalty in (\ref{eq:penEst}): 
             \begin{equation}
             P_\lambda(\beta_j) = \left\{\begin{array}{lll}
             \lambda|\beta_j|\,, & if & |\beta_j| \leq \lambda\\
            -\frac{|\beta_j|^2-2a\lambda|\beta_j|+\lambda^2}{2(a-1)}\,, & if & |\beta_j|\in(\lambda,a\lambda]\\
             \frac{(a+1)\lambda^2}{2}\,, & if & |\beta_j|>a\lambda
             \end{array}\right.
             \end{equation} 
where $a>2$ and $\lambda>0$.  This is a non-convex, quadratic spline function  by which small coefficients are shrunk towards zero with a Lasso penalty, while large coefficients are not penalized. The resulting estimator is, unlike Lasso, nearly unbiased for large coefficients. \citet{Fan2001} and \citet{Fan2004} also show that SCAD enjoys an oracle property (assuming some regularity conditions) -- it is simultaneously consistent for variable selection and estimation, where the latter is as efficient (asymptotically) as the ideal case when the true model is known in advance. For further details on the properties of SCAD, see \citet{Fan2010a} and references therein.

\smallskip

\noindent
\textbf{Adaptive Lasso.}
Adaptive Lasso \citep{Zou2006} uses a Lasso penalty with weights in (\ref{eq:penEst}):
$P_{\lambda}(\beta_j)=\lambda\omega_j|\beta_j|$.
Similar in spirit to SCAD, Adaptive Lasso aims to eliminate the bias in the Lasso by shrinking larger coefficients less than smaller ones.
This coefficient-specific regularization is achieved using the weights $\omega_j$, which are taken to have the form $\omega_j=1/|\tilde{\beta}_j|^{\gamma}$, where $\tilde{\beta}_j$ is an initial estimate for $\beta_j$ and $\gamma{>}0$. 
Larger initial estimates give rise to smaller weights and so receive less shrinkage.
The ordinary least squares estimate or Ridge Regression estimate are suggested as initial estimates by \citet{Zou2006}.
Adaptive Lasso also enjoys the oracle property (for suitable choices of $\lambda$).

\smallskip

\noindent
\textbf{Dantzig Selector.}
The Dantzig Selector estimator \citep{Candes2005} takes a different form to that in (\ref{eq:penEst}), namely:
\begin{equation}   
	\hat{\boldsymbol{\beta}}_{\lambda}=\argmin_{\beta}\left\{\|\boldsymbol{\beta}\|_1 : \|\mathbf{X}^T\left(\mathbf{Y}-\mathbf{X}\boldsymbol{\beta}\right)\|_\infty \leq \lambda \right\}.
\end{equation} 
The Dantzig Selector and the Lasso are closely connected as discussed in \citet{Bickel2009} and under certain conditions on the design matrix, Lasso and Dantzig provide the same solution \citep{Meinshausen2007dantzig,James2009a}.

\smallskip

\noindent
\textbf{Stability Selection.}
This is a general approach by which to combine variable selection with data subsampling to obtain more stable selection and  control the number of false positives. 
Specifically, $M$ random data subsamples of size $\tilde{n}<n$ are generated by sampling without replacement. Applying a variable selection procedure, with regularization parameter $\lambda$, to these datasets gives 
a score $\hat{\Pi}_{\lambda,j}$ indicating the frequency with which variable $j$ is selected among the $M$  iterations. 
Let $\Lambda$ denote the set of considered values for the regularization parameter. Then, a set of ``stable variables'' is obtained by choosing those variables that have selection probabilities larger than a cutoff value $\pi_{\mathrm{thr}}\in(0,1)$ for any $\lambda\in\Lambda$.

 In contrast to the methods described above, Stability Selection does not require setting of the parameter $\lambda$, but instead requires the cutoff  $\pi_{\mathrm{thr}}$ to be chosen. \citet{Meinshausen2010} provide theoretical results showing how $\pi_{\mathrm{thr}}$ can be chosen to achieve a user-specified upper bound $\tilde{V}$ on the expected number of false positives $\mathbb{E}[V]$, assuming a fixed set of regularization parameters $\Lambda$. Alternatively, the user can fix $\pi_{\mathrm{thr}}$ and then the theory shows how $\Lambda$ should be chosen to achieve the desired upper bound on $\mathbb{E}[V]$.
In our study we use the Lasso as the variable selection procedure with Stability Selection.

\subsection{Simulation set-up} \label{subsec:designs}

We generate values for the response vector using model 
(\ref{eq:linear_model}). 
We set $\boldsymbol{\beta}$ to have $s_0$ non-zero entries (all set to 3 except in Section \ref{subsec:heterogeneous_betas} where we consider  heterogeneous coefficients) 
with  $\sigma$ then set to obtain a desired SNR, defined here as
$\mathrm{SNR} = \sqrt{\boldsymbol{\beta}^\textnormal{T} \textbf{X}^\textnormal{T}\mathbf{X}\boldsymbol{\beta}/(n \sigma^2)}$.

We consider synthetic data, where both covariates and responses are simulated, and semi-synthetic data, where covariates are real and responses are simulated.

\subsubsection{Synthetic data}
We consider the following two designs with synthetic covariates:
\begin{itemize}
\item  \textbf{Independence design.} All $p$ covariates are i.i.d. standard normal.

\item \textbf{Pairwise correlation design.} The $p$ covariates are partitioned into $B$ blocks, each of size $p^B=p/B$. All covariates are standard normal but with correlation between any pair of covariates within the same block set to $\rho$. Covariates in different blocks are independent of each other. The number of active variables within a block is $s_0^B$ for the first $s_0/s_0^B$ blocks, with the remaining blocks containing no active variables.

\end{itemize}

\subsubsection{Semi-synthetic data} \label{sec:semi-synth_setup}
We consider semi-synthetic data using real covariates from The Cancer Genome Atlas (TCGA) study.  
We use gene expression data from TCGA ovarian cancer samples \citep{TCGA}
\footnote{Specifically, we use the dataset provided in the Supplementary Appendix of \citet{Tucker2014}; the dataset is available at \url{http://bioinformatics.mdanderson.org/Supplements/ResidualDisease}.}. 
The dataset contains 594 samples and expression levels for 22,277 genes. 
The samples are a mixture of primary tumor (569), recurrent tumor (17) and normal tissue (8). 
We randomly subsample the samples and genes to obtain a $n\times p$ design matrix $\mathbf{X}=[\mathbf{x}_1, \dots, \mathbf{x}_p]$.
Those samples not included in $\mathbf{X}$ are used as test data. 

Signals are allocated among the $p$ predictors to give either ``low'' or ``high'' correlation designs, using an approach similar to \citet{Buhlmann2014}:

\begin{itemize}
\item  \textbf{``Low'' correlation design.}
We allocate $s_0$ signals at random among $\mathbf{x}_1, \dots, \mathbf{x}_p$.

\item \textbf{``High'' correlation design.}
We use the following procedure to form correlated blocks: 
\begin{enumerate}[(i)]
\item Form a block of $p^B{=}10$ predictors consisting of the two predictors $\tilde{\mathbf{x}}_1$ and $\tilde{\mathbf{x}}_2$ that are most correlated and the eight other predictors that are most correlated with $\tilde{\mathbf{x}}_1$

\item Allocate $s_0^B$ signals to this block by designating $\tilde{\mathbf{x}}_1$ and the $s_0^B-1$ predictors that are most correlated to it as signals

\item Repeat steps (i) and (ii), but remove from consideration any predictors already allocated to a block, and continue repeating until $s_0$ signals have been allocated.
\end{enumerate}
\end{itemize}

The $p$ variables in each of our simulation scenarios are selected completely at random from the original dataset, so the correlation structure among the $p$ variables is representative of the original data.
In the ``low'' correlation design, the correlation between a given signal and any other variable is, on average, the same as the average correlation between all $p$ variables. 
In the ``high'' correlation design, the average correlation between all $p$ variables will follow the same distribution as in the ``low'' correlation design. However, by identifying correlated blocks and allocating signals within these blocks as described above, a given signal is now more likely to have higher correlation with some non-signals and, for $s_0^B{>}1$, with some other signals.

\subsubsection{Systematic exploration of data-generating factors}
We consider the effects of the various data-generating factors in a systematic way via 2,394 simulation scenarios, each corresponding to a different configuration. 
The values considered for each factor are shown in Table~\ref{tab:setting_parameters} and we cover the majority of combinations of the factors.
One exception is for correlation designs we exclude some combinations of $s_0^B$ and $B=p/p^B$ 
which violate the necessary constraint $s_0^B\geq s_0/B$ (see Table \ref{tab:correlation_design_summary}).
Also, SNR=0.5 is not considered for the synthetic correlation design.

\begin{table*}[t]
\renewcommand{\arraystretch}{1.5}
\begin{center}
\caption{Factors varied in the simulation study and values considered. Note that for the correlation designs, the $s_0^B$ signals per block applies to the first $s_0/s_0^B$ blocks only.} 
 \label{tab:setting_parameters}
\begin{tabular}{c|c|c|}
\cline{2-3}
&  \multicolumn{1}{ c| }{Factors}  & \multicolumn{1}{ c| }{Values considered} \\ \cline{1-3}
\multicolumn{1}{ |c|  }{\multirow{4}{5cm}{\centering All designs} } 
																			& Sample size, $n$  													 & 100, 200, 300 \\ \cline{2-3}
\multicolumn{1}{ |c|  }{}             & Dimensionality, $p$ 												 & 500, 1000, 2000, 4000 \\ \cline{2-3}
\multicolumn{1}{ |c|  }{}             & Sparsity, $s_0$ 														 & 10, 20, 40  \\ \cline{2-3}
\multicolumn{1}{ |c|  }{}             & Signal-to-noise ratio, SNR & $0.5^{\ast}$, 1, 2, 4  \\ \cline{1-3}
\multicolumn{1}{ |c|  }{\multirow{3}{5cm}{\centering Synthetic (pairwise) correlation design only} } 
																			& Block size, $p^B$                            & 10, 100 \\ \cline{2-3}
\multicolumn{1}{ |c|  }{}             & Pairwise correlation within a block, $\rho$  & 0.5, 0.7, 0.9  \\  \cline{2-3}
\multicolumn{1}{ |c|  }{}             & Number of signals per block, $s_0^B$  & 1, 2, 5 \\ \cline{1-3}
\multicolumn{1}{ |c|  }{\multirow{2}{5cm}{\centering Semi-synthetic (``low''/``high'') correlation designs only}}
                                      & Block size, $p^B$                            & 10 \\ \cline{2-3}
\multicolumn{1}{ |c|  }{}             & Number of signals per block, $s_0^B$  & 1, 2, 5 \\ \cline{1-3}							
\multicolumn{2}{@{}l}{ \scriptsize{\textsuperscript{$\ast$} All designs except synthetic pairwise correlation design} }
\end{tabular}
\end{center}
\end{table*}

\subsection{Method implementation} \label{sec:implement}
Tuning parameters are set to reflect the way methods would typically be used by users.
For Lasso, Elastic Net, Ridge Regression, SCAD, Adaptive Lasso and Dantzig Selector, $\lambda$ is set via 10-fold cross-validation (CV). 
Following \citet{Buhlmann2014}, we implement two versions of Elastic Net with $\alpha=0.3$ and $\alpha=0.6$, referred to as heavy Elastic Net (HENet) and light Elastic Net (LENet) respectively. For SCAD, we set $a=3.7$, as recommended by \citet{Fan2001}. For Adaptive Lasso (AdaLasso), we use the Ridge Regression estimate as the initial estimate to calculate the weights and set $\gamma{=}1$. For Stability Selection, we set the number of iterations to $M{=}100$ with subsample size $\tilde{n}=\lfloor 0.632 n \rfloor$ and selection probability cutoff $\pi_{thr}=0.6$ (the R package defaults; see below). 
We do not place any explicit control on the expected number of false positives $\mathbb{E}[V]$ (i.e. we consider the full range of regularization parameters $\Lambda$).
An exception to this is for selection in the semi-synthetic data analysis, where we set $\tilde{V}$, the upper bound on $\mathbb{E}[V]$, to 10.
However, we assess sensitivity to these tuning parameters in Section \ref{sec:additional_simulations}. 

We use available R packages to implement the methods: \texttt{glmnet} for Lasso, Elastic Net, Ridge Regression and Adaptive Lasso \citep{Friedman2010a}; \texttt{ncvreg} for SCAD \citep{Breheny2011}; \texttt{flare} for Dantzig Selector \citep{Li2015}; and \texttt{c060} for Stability Selection \citep{Sill2014}. Covariates are standardized  and the response vector is centered. We run all methods on all simulation scenarios with the exception of Dantzig and AdaLasso: 
Dantzig is run only for the synthetic independence design, and synthetic correlated design with $p=500$ and $p=1000$, due to its computational demands under multicollinearity for large $p$; 
Adaptive Lasso is not run for the synthetic correlated design.
For each simulation scenario, we show results averaged across 64 simulated datasets.

\subsection{Performance metrics} \label{sec:metrics}
We distinguish between prediction, variable selection and ranking and use  the following metrics.

\smallskip

\noindent
{\bf Prediction.} To assess predictive performance we use the root mean squared error (RMSE). 
For each simulation scenario, we generate training data  with sample size $n$ and  test data with sample size $n_{\mathrm{test}}{=}500$. Models are fitted on training data to obtain coefficient estimates $\hat{\boldsymbol{\beta}}_{cv}$ and prediction error, calculated as 
$\mathrm{RMSE}=
\|\mathbf{y}_{\mathrm{test}}-\mathbf{X}_{\mathrm{test}}\hat{\boldsymbol{\beta}}_{cv}||_\textnormal{2} / \sqrt{n_{\mathrm{test}}}$,
where $\mathbf{y}_{\mathrm{test}}$ and $\mathbf{X}_{\mathrm{test}}$ are the test responses  and design matrix respectively.
Stability Selection focuses on variable selection and we therefore do not include it in assessment of predictive performance.

\bigskip

\noindent
{\bf Variable selection.}  For assessment of variable selection, we use true positive rate (TPR) and positive predictive value (PPV): 
\begin{gather}
\mathrm{TPR}=\frac{\mathrm{TP}}{\mathrm{TP}+\mathrm{FN}}\in [0,1]; \ \ \ \mathrm{PPV}=\frac{\mathrm{TP}}{\mathrm{TP}+\mathrm{FP}}\in [0,1],
\label{eq:TPR_PPV_definition}
\end{gather}
where TP, FP and FN are the number of true positives, false positives and false negatives respectively. Ridge Regression does not perform variable selection {\it per se} and is therefore excluded from this evaluation.

\bigskip

\noindent
{\bf Variable ranking.} For ranking, we assess performance using the partial area under the receiver operating characteristic curve (pAUC). 
This is the area under the curve obtained 
when restricting to a maximum of 50 false positives ($\mathrm{FPR}=\frac{50}{p-s_0}$).
The pAUC calculation requires a score under which to rank variables $j$.
For Ridge Regression, we rank by $s_j=|(\hat{\beta}_{cv})_j|$ and for Stability Selection by $s_j=\max_{\lambda\in\Lambda}\hat{\Pi}_{\lambda,j}$. For the other methods (Lasso, Elastic Net, SCAD and Dantzig Selector), we could use $|(\hat{\beta}_{cv})_j|$ as for Ridge, but due to sparsity this would involve ranking many covariates with $(\hat{\beta}_{cv})_j=0$. We instead consider the set of estimated active sets $\{\hat{S}_{\lambda}:\lambda\in\Lambda\}$ where $\Lambda$ is the set of candidate regularization parameters. We consider a covariate to be more important the longer it remains in $\hat{S}_{\lambda}$ as $\lambda$ increases and more sparsity is induced. This motivates defining ranking scores as: $s_j=\max\{\tilde{\lambda}\in\Lambda:j\in\hat{S}_{\lambda}\mathrm{\ for\ all\ } \lambda\leq\tilde{\lambda}, \lambda\in\Lambda\}$ or $s_j=0$ if $j\notin\hat{S}_{\lambda_{\min}}$, where $\lambda_{\min}=\min\{\lambda\in\Lambda\}$.

\begin{table}[t]
\begin{center}
\caption{Combinations of $p, p^B, s_0$ and $s_0^B$ explored in the (synthetic and semi-synthetic) correlation designs. \cmark  \hspace{0.02cm} indicates that the combination is included and \xmark \hspace{0.02cm} indicates that the combination is not included. 
For $p^B{=}10$, * denotes all combinations of $p$ and $s_0$. } \label{tab:correlation_design_summary}
\begin{tabular}{ccccccccc}
  \toprule
                         &  \multirow{2}{*}{$p$} & \multirow{2}{*}{$p^B$} & \multirow{2}{*}{$B{=}\frac{p}{p^B}$} & \multirow{2}{*}{$s_0$} & &\multicolumn{3}{c}{$s_0^B$}\\
  					     &  	                 & 						  &									   &						& & 1      & 2      & 5       \\
  \midrule 
  \multirow{9}{2.5cm}{\centering Synthetic (pairwise) correlation design only}
  						 &                       &                        &                                    & 10                     & & \xmark & \cmark & \cmark  \\
                         &  500                  & 100                    & 5                                  & 20                     & & \xmark & \xmark & \cmark \\
                         &                       &                        &                                    & 40                     & & \xmark & \xmark & \xmark\\
  \cline{2-9}
                         &                       &                        &                                    & 10                     & & \cmark & \cmark & \cmark \\
                         &  1000                 & 100                    & 10                                 & 20                     & & \xmark &\cmark &\cmark \\
                         &                       &                        &                                    & 40                     & & \xmark & \xmark & \cmark  \\
  \cline{2-9}
                         &                       &                        &                                    & 10                     & & \cmark & \cmark & \cmark \\
                         &  2000                 & 100                    & 20                                 & 20                     & & \cmark & \cmark &\cmark \\
                         &                       &                        &                                    & 40                     & & \xmark & \cmark & \cmark \\
  \cline{2-9} 
                         &                       &                        &                                    & 10                     & & \cmark & \cmark &\cmark    \\
                         & 4000                  & 100                    & 40                                 & 20                     & & \cmark & \cmark & \cmark \\
                         &                       &                        &                                    & 40                     & & \cmark & \cmark & \cmark \\
  \midrule
  \multirow{3}{2.5cm}{\centering All correlation designs} 
  						 &  	                 & 						  &									   &						& &        &        &         \\
                         &  *                    & 10                     & *                                  &*                       & &\cmark& \cmark & \cmark   \\
 						 &  	                 & 						  &									   &						& &        &        &         \\
  \bottomrule
\end{tabular}
\end{center}
\end{table}

%%%%%%%%%%%%%%%%%%%%%%%%%%%%%%%%%%%%%%%%%%%%%%%%%%%%%%%%%%%%%%%%%%%%%%%%%%%%%%%%%%%%%%%%%%%%%%%%%%%%%%
%%%%%%%%%%%%%%%%%%%%%%%%%%%%%%%%%%%%%%%%%%%%%%%%%%%%%%%%%%%%%%%%%%%%%%%%%%%%%%%%%%%%%%%%%%%%%%%%%%%%%%

\begin{figure*}[t]
\centering
\includegraphics[height=5in]{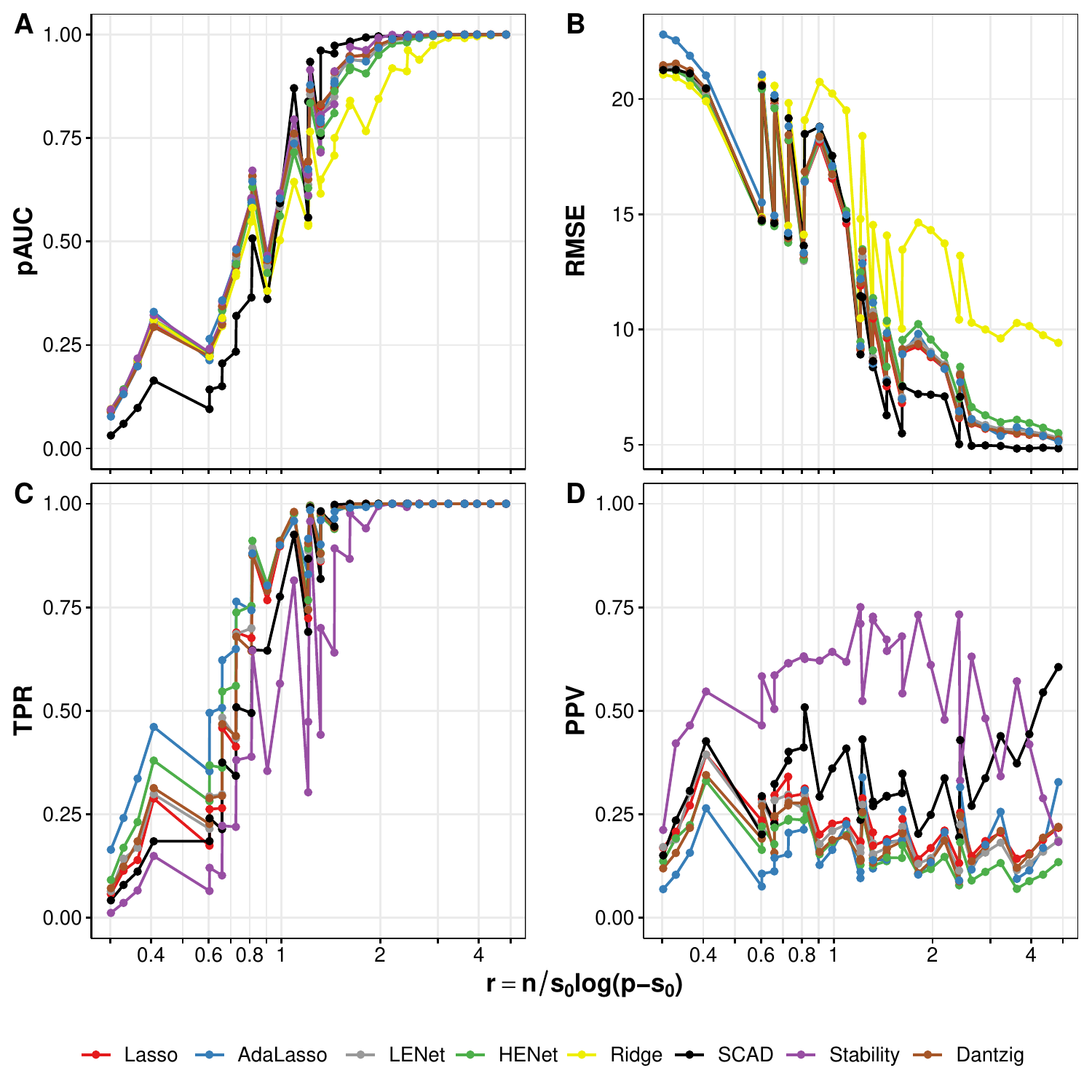}
\caption{Ranking (A), prediction (B) and selection (C,D) performance versus the rescaled sample size $r=n/(s_0log(p-s_0))$  for synthetic independence design scenarios with SNR=2. 
Line color indicates method.
Note that Stability Selection and Ridge Regression are not included in the assessment of prediction and selection performance respectively.
See Section~\ref{sec:metrics} for details of metrics; pAUC = partial area under the receiver operating characteristic curve, RMSE = root mean squared error, TPR = true positive rate, PPV = positive predictive value.} 
\label{fig:r_ind_snr2}
\end{figure*}

\section{Main results} \label{sec:systematic_comparison}

Due to the large number of simulation regimes, we focus below on the key  patterns. 
All performance data and plotting code are made available on GitHub, allowing specific scenarios to be investigated further (see the ``Code and data availablity'' section). 
Figures~S1-S21, referred to below, can be found in Supplementary Material.

We first present summary observations that hold across all the simulation scenarios. 
We then present results for each metric in turn: ranking in Section~\ref{sec:rank}, prediction in Section~\ref{sec:pred} and selection in Section~\ref{sec:sel}.
In each of these three sections, we first present key observations for the synthetic independence design and then key observations for the correlation designs (both semi-synthetic and synthetic designs).
We then end each section by providing a summary with a recommendation regarding choice of method.

\subsection{Observations from across all simulation scenarios} \label{sec:general_obs}

\paragraph{An approximate guide to simulation scenario difficulty.}
Figure \ref{fig:r_ind_snr2} shows the performance metrics versus rescaled sample size $r$, for the synthetic independence design with SNR=2. 
The quantity $r$ equals $n/(s_0 \log(p-s_0))$ \citep[see][]{Wainwright2009} and is motivated by scaling results for consistent Lasso variable selection. 
Large (small) values of $r$ can be interpreted as large (small) sample size relative to dimensionality and sparsity.
We observe a clear overall trend of better pAUC (Fig. \ref{fig:r_ind_snr2}A) and TPR (Fig. \ref{fig:r_ind_snr2}C) for all methods as $r$ increases, with performance leveling off for larger values of $r$. 
The trend is similar for RMSE as $r$ increases (Fig. \ref{fig:r_ind_snr2}B).  
The behavior of PPV is method-dependent and the overall trend is non-monotonic as $r$ increases (Fig. \ref{fig:r_ind_snr2}D).
Performance with varying $r$ was qualitatively similar for other SNR values and also for correlation designs (see Fig.~S1 for independence design with SNR=0.5 and Fig.~S2 for a semi-synthetic correlation design with SNR=2). 
Therefore, although the motivation for $r$ lies in asymptotic theory for variable selection, we found that $r$ and SNR together serve as a useful approximate guide to the difficulty of each simulation scenario for all three tasks (selection, ranking and prediction). We make use of this characterization below.

\paragraph{LENet is between Lasso and HENet.}
The performance of LENet is invariably between that of Lasso and HENet for all metrics.
For example, ranking performance of LENet lies between Lasso and HENet for 98\% of synthetic data scenarios where there is a ``salient'' difference in pAUC between Lasso and HENet (for our purposes here, we take a difference in pAUC of larger than 0.01 to be ``salient'').
We therefore exclude LENet below to aid presentation.

\paragraph{Dantzig Selector is similar to Lasso.}
The Dantzig Selector mostly performed similarly to Lasso (see red and brown lines in Fig.~\ref{fig:r_ind_snr2} and see also Fig.~S3), in line with theory \citep[e.g.][]{Meinshausen2007dantzig, Efron2007}. 
However, Dantzig is more computationally expensive than Lasso \citep{Meinshausen2007dantzig}. For example, when $(n,p,s_0){=}(100,500,10)$ and SNR=1 in the synthetic independence design, Dantzig takes around 1,500 seconds to compute the whole solution path, while Lasso takes less than one second. 
In the interest of brevity, we also exclude Dantzig in the presentation of results below.

\paragraph{No overall winner; large differences.}
For all metrics, there is no one method that consistently performs best across all or the majority of the scenarios. 
Moreover, relative differences in performance can be large in some scenarios. 
Even in the textbook context of synthetic independence design scenarios shown in Figure~\ref{fig:r_ind_snr2}D, the median percentage relative decrease in PPV between the methods with the highest and lowest scores is 77\%. 
Across all 2,394 scenarios considered, the median percentage relative decrease is 46\% for pAUC, 14\% for RMSE, 61\% for TPR and 68\% for PPV.

\begin{figure*}[t]
\centering
\includegraphics[height=4.66in]{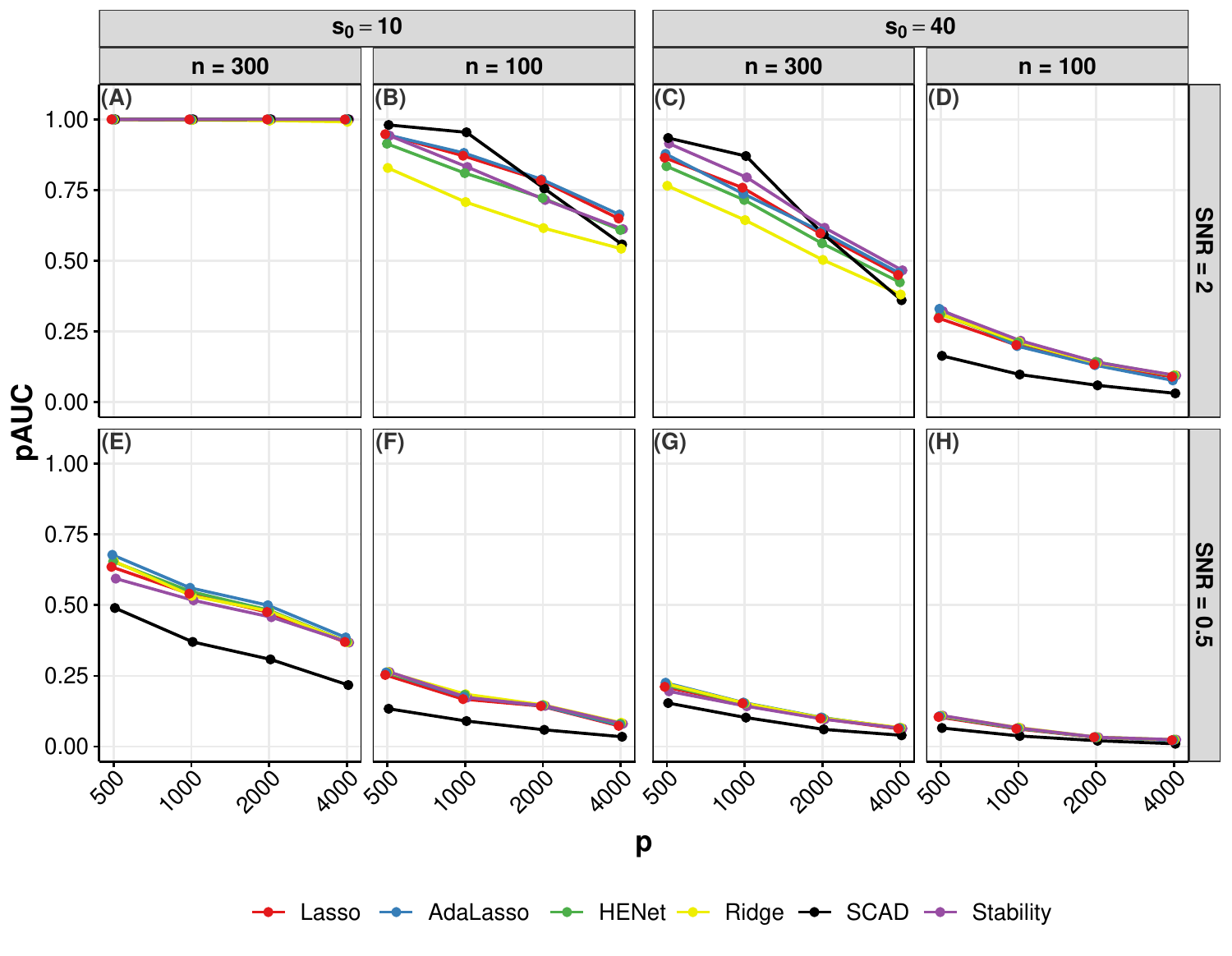}
\caption{Ranking performance (pAUC) versus $p$ for a subset of synthetic independence design scenarios.
Each panel represents a different combination of $n$, $s_0$ and SNR.
Line color indicates method and $x$-axis is on a log-scale. 
See also Figure~S4.
}
\label{fig:pauc_ind}
\end{figure*}

\begin{figure*}[t]
\centering
\includegraphics[height=5in]{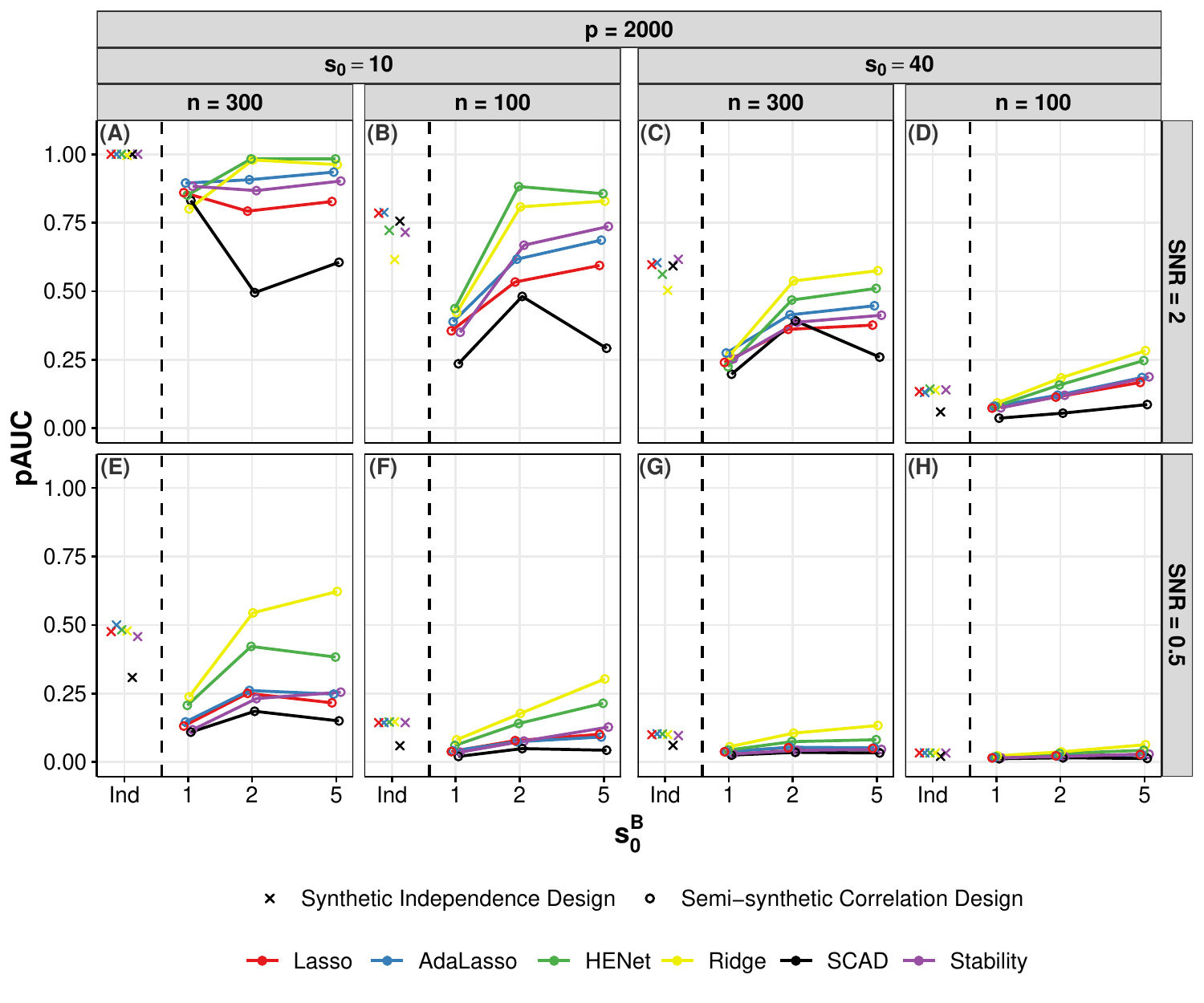}
\caption{Ranking performance (pAUC) versus $s_0^B$ (number of signals per block) for a subset of semi-synthetic ``high'' correlation designs.
Each panel represents a different combination of $n$, $s_0$ and SNR.
All results shown are for $p{=}2000$ (see Figure~S7 for results with $p{=}500$).
For comparison, results for the corresponding independence design scenarios are also shown in each panel (``Ind'').
Line color indicates method and $x$-axis is on a log-scale. 
}
\label{fig:pauc_cor_real}
\end{figure*}

%%%%%%%%%%%%%%%%%%%%%%%%%%%%%%%%%%%%%%%%%%%%%%%%%%%%%%%%%%%%%%%%%%%%%%%

\subsection{Ranking} \label{sec:rank}

%%%%%%%%%%

\subsubsection{Independence design - synthetic data} \label{sec:rank_ind}

Figure~\ref{fig:pauc_ind} shows ranking performance for a subset of independence design scenarios (see also Figure~S4 where performance of pairs of methods are plotted against each other for all independence design scenarios).

\paragraph{SCAD transition in performance.} \label{para:ind_rank_SCADtrans}
The performance of SCAD relative to other approaches varies substantially across scenarios. SCAD can offer the best performance in ``easier'' scenarios (e.g. Fig. \ref{fig:pauc_ind}B, black line), but does not retain this advantage as scenario difficulty increases. In particular, SCAD undergoes a transition from best to worst performing method with an unfavorable change in $n$, $p$, $s_0$ or SNR (see Fig.~\ref{fig:pauc_ind}C for such a transition with increasing $p$).

\paragraph{An $L_2$ penalty, AdaLasso and Stability Selection provide no substantive benefit over Lasso.} \label{para:ind_rank_L2}
Apart from SCAD in ``easy'' settings, none of the approaches perform notably better than Lasso (see red lines in Fig.~\ref{fig:pauc_ind} and Fig.~S4).
Moreover, Stability Selection, HENet and Ridge sometimes perform worse than Lasso (e.g. Fig.~\ref{fig:pauc_ind}B).
AdaLasso performs essentially the same as Lasso (Fig.~S4), but can give small gains in pAUC over Lasso when SNR is small (see blue line in Fig.~\ref{fig:pauc_ind}E for $p{=}500$).

%%%%%%%%%%%%%

\begin{figure*}[t]
\centering
\includegraphics[height=4.8in]{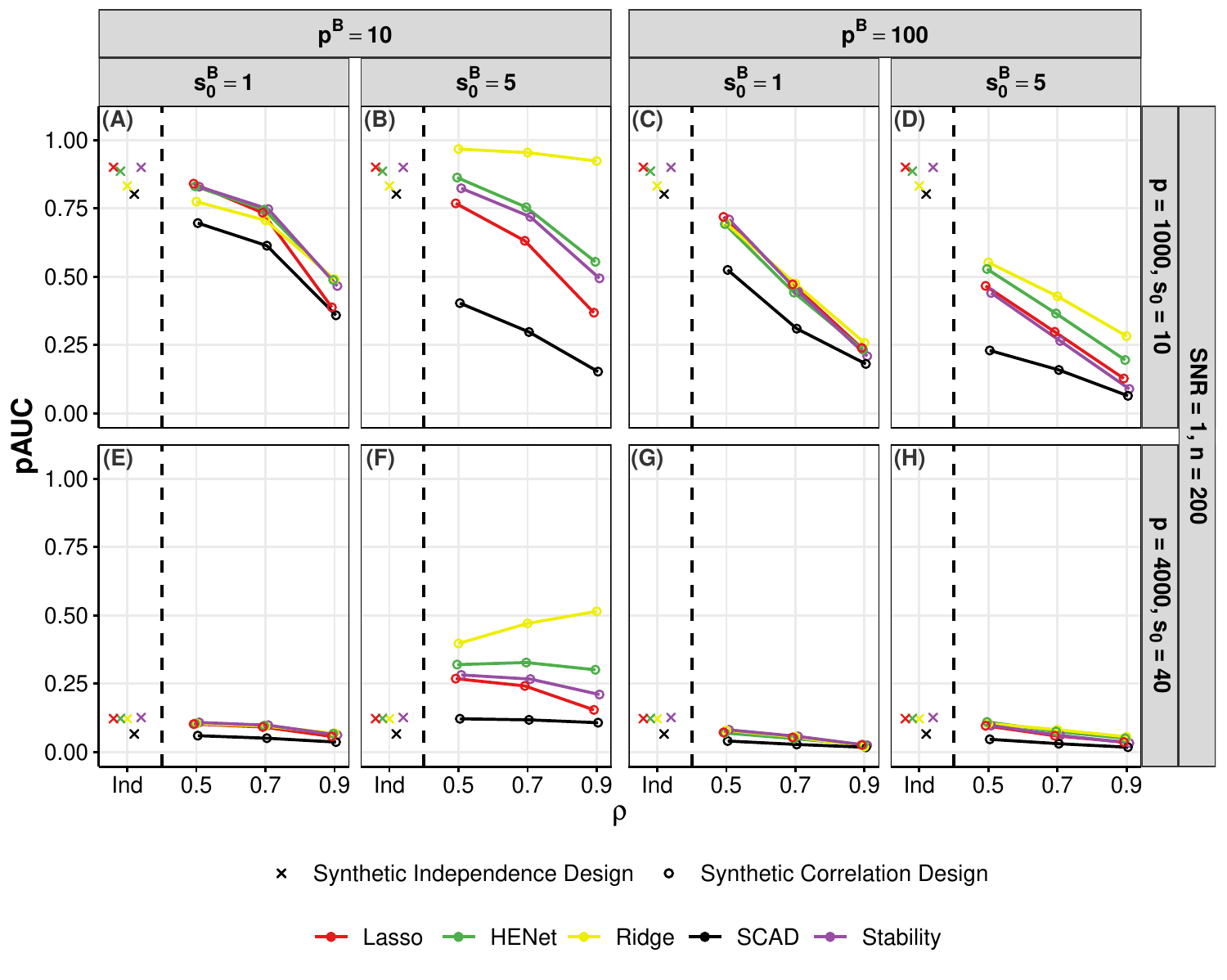}
\caption{Ranking performance (pAUC) versus $\rho$ (correlation strength) for a subset of synthetic pairwise correlation designs.
Each row represents a different combination of $p$ and $s_0$, while each column represents a different combination of  $p^B$ (block size) and $s_0^B$ (number of signals per block).
All results shown are for SNR=1 and $n{=}200$ (see Figure~S8 for SNR=2 and 4).
The top row has $(n,p,s_0)=(200,1000,10)$, giving $r{=}2.9$, and the bottom row has $(n,p,s_0)=(200,4000,40)$, giving $r{=}0.6$.
For comparison, results for the corresponding independence design scenarios are also shown (``Ind''; these data points are identical across the panels in each row).
Line color indicates method.
}
\label{fig:pauc_cor_syn}
\end{figure*}

\subsubsection{Correlation designs} \label{sec:rank_cor}

For the semi-synthetic data, we focus on the ``high'' correlation design (see Section~\ref{sec:semi-synth_setup}) because results for the ``low'' correlation design are in good agreement with those from the synthetic independence design (see Figs.~S5 and Fig.~S6). 
This is because the covariates are very weakly correlated on average (mean absolute correlation coefficient between covariate pairs is 0.08).
Performance tends to be a bit worse for the ``low'' correlation design than for the independence design (Fig.~S5).
We also note that, for ranking, AdaLasso typically performs slightly better than Lasso in the ``low'' correlation design, whereas they mostly had equal performance in the independence design (Fig.~S6).

The semi-synthetic and synthetic data results are broadly similar, so we focus on the semi-synthetic data results and mainly use the synthetic data to investigate the impact of varying correlation strength $\rho$ and block size $p^B$; these parameters were either fixed ($p^B{=}10$) or not directly controlled (in the case of $\rho$) for the semi-synthetic data.

Figure~\ref{fig:pauc_cor_real} shows ranking performance, as a function of number of signals per block $s_0^B$, for a subset of the ``high'' correlation semi-synthetic design scenarios with $p{=}2000$ (analogous results for $p{=}500$ are shown in Figure~S7).
Results for the synthetic independence design are also shown in each figure panel for reference (denoted by ``Ind'').
Figure~\ref{fig:pauc_cor_syn} shows ranking performance, as a function of correlation strength $\rho$, for a subset of pairwise correlation synthetic design scenarios.
To aid presentation of results, we fix $(n,p,s_0){=}(200,4000,40)$ or $(200,1000,10)$ which give $r{=}0.6$ (``hard'') or $r{=}2.9$ (``easy'') respectively, and also fix SNR=1 (analogous results for SNR=2 and 4 are shown in Figure~S8).

\paragraph{Improved performance relative to the independence design for some scenarios.} \label{para:cor_rank_cor_vs_ind}
Correlated covariates have a negative effect on ranking performance relative to the synthetic independence design when there is one signal per block (compare crosses with corresponding $s_0^B{=}1$ circles in Figs.~\ref{fig:pauc_cor_real} and \ref{fig:pauc_cor_syn}).
Performance then often improves as $s_0^B$ increases, particularly for HENet and Ridge regression (e.g. yellow and green lines in Fig.~\ref{fig:pauc_cor_real}B for semi-synthetic data; contrast also the first and second columns in Fig.~\ref{fig:pauc_cor_syn} for synthetic data).
This can lead to an improvement in performance relative to the independence design when $s_0^B{>}1$, with the largest improvements typically for HENet and Ridge in ``harder'' settings with small $r$ or SNR.
For example, in Figure~\ref{fig:pauc_cor_real}B where $r{=}1.3$ and SNR=2, HENet and Ridge have an increase in pAUC of 0.13 and 0.21 respectively relative to the independence design when $s_0^B{=}5$. 

For the synthetic data we also find that an increase in block size $p^B$ has an opposite effect to $s_0^B$, with a decrease in pAUC (contrast first and third columns in Fig.~\ref{fig:pauc_cor_syn}).
Increasing correlation strength $\rho$ typically has a detrimental effect.
Only in the case of ``harder'' scenarios (small $r$ or SNR) with small block size and several signals per block, performance can be enhanced by increasing $\rho$, most notably for Ridge Regression (see e.g. yellow line in Fig.~\ref{fig:pauc_cor_syn}F where $r{=}0.60$, $p^B{=}10$ and $s_0^B{=}5$). 

Taken together, the above means that it is in ``hard'' scenarios when block size $p^B$ is small and blocks consist of highly correlated variables of which several are active (i.e. large $\rho$ and $s_0^B$) that we see the largest gains from correlation relative to the independence design, for HENet and Ridge (contrast yellow and green crosses and circles in Fig.~\ref{fig:pauc_cor_syn}F).

We also find that SCAD tends to be the most negatively affected by correlation (see e.g. black in Fig.~\ref{fig:pauc_cor_real}A).

\paragraph{HENet and Ridge Regression outperform other methods.} \label{para:cor_rank_L2_benefit}
The positive influence of correlation on the ranking performance of HENet and Ridge Regression means that they now have the best pAUC scores in most scenarios with small block sizes and $s_0^B{>}1$, with Ridge outperforming HENet.
For example, for the semi-synthetic data scenario in Figure~\ref{fig:pauc_cor_real}E where SNR=0.5, Ridge substantially outperforms all other approaches when $s_0^B{=}5$, with an improvement in pAUC of 0.24 over the second best method, HENet. 
HENet itself also improves over Stability Selection with a difference in pAUC of 0.13. 
There was no such benefit from an $L_2$ penalty in the corresponding independence design scenario (crosses in Fig.~\ref{fig:pauc_cor_real}E).

We also observe in the small block size ($p^B{=}10$) synthetic data results that the gains in pAUC from an $L_2$ penalty over Lasso become larger as correlation strength $\rho$ increases (contrast yellow and red lines for $\rho{=}0.5$ and $\rho{=}0.9$ in Fig.~\ref{fig:pauc_cor_syn}B).
These advantages from an $L_2$ penalty are either smaller or not present at all in the corresponding larger block size scenarios with $p^B{=}100$ and $s_0^B{>}1$ (fourth column in Fig.~\ref{fig:pauc_cor_syn}), suggesting that the \emph{proportion} of covariates in a block that are signals is important. 
We investigated this by increasing $s_0^B$ to 40 in the $p^B{=}100$ scenarios shown in Figure~\ref{fig:pauc_cor_syn}H (where $r{=}0.60$, SNR=1 and $s_0^B{=}5$), and indeed found that salient improvements over Lasso are then obtained with an $L_2$ penalty: pAUC=0.42, 0.13 and 0.07 for Ridge, HENet and Lasso respectively when $\rho=0.9$.

The largest benefits from an $L_2$ penalty are therefore for scenarios with small, highly correlated blocks with many signals per block.
In general, benefits from an $L_2$ penalty appear to be more prevalent for the semi-synthetic data than the synthetic data. This is likely due to the covariate correlation structure being less rigid for the semi-synthetic data, with covariates being weakly correlated across blocks as opposed to independent.

\paragraph{SCAD transition in performance.} \label{para:cor_rank_SCADtrans}
SCAD again displays its characteristic transition behavior with decreasing $r$ or SNR in the correlation design (see e.g. Fig.~S7), but due to it typically being the most negatively affected by correlation, the number of ``easy'' scenarios where SCAD performs best is reduced. 

SCAD's sensitivity to correlation means there can also be a transition with increasing $s_0^B$. 
In ``easy'' settings with large $r$ or SNR, SCAD can perform best when there is only one signal per block (and also in the corresponding independence design), but perform worst when there are many signals per block (see e.g. Fig.~S7G).
We also have a transition with increasing $\rho$ for the synthetic data (e.g. Fig.~S8D).

\paragraph{Stability selection and AdaLasso mostly outperform Lasso.} \label{para:realCor_rank_SS}
Stability Selection and AdaLasso remain competitive relative to Lasso, as in the independence design (Figs.~\ref{fig:pauc_cor_real} and \ref{fig:pauc_cor_syn}).
Moreover, they now offer notable improvements over Lasso for some scenarios with sufficiently large $s_0^B$, and $r$ or SNR (see e.g. purple and blue lines vs. red line in Fig.~\ref{fig:pauc_cor_real}B and purple vs. red line in Fig.~\ref{fig:pauc_cor_syn}B ).
However, they are usually outperformed by HENet and Ridge Regression, except in a few ``easy'' scenarios (e.g. $s_0^B{=}1$ in Fig.~\ref{fig:pauc_cor_real}A; Fig.~S7C).

%%%%%%%%%%%%%

\subsubsection{Summary and recommendations} \label{sec:summary_rank}

For settings with uncorrelated or very weakly correlated covariates\footnote{Here we are assuming that uncorrelated variables are also independent, so that the independence design simulations apply. For very weak correlation, the semi-synthetic ``low'' correlation design applies.}, Lasso or AdaLasso are usually competitive for ranking and so can be considered as good choices. 
When one is confident of being in an ``easy'' scenario with sufficiently large $r$ and SNR; SCAD could be considered here as it may perform notably better than Lasso and AdaLasso, but using SCAD carries more risk due to the high variability arising from its transition behavior. 

For settings with more highly correlated covariates, we confirm that Ridge Regression is a good option since it outperforms or is competitive with the other approaches in most scenarios. 
Since SCAD rarely outperformed other methods and is very sensitive to changes in scenario properties, we would suggest it is not a good option for correlated settings.

%%%%%%%%%%%%%%%%%%%%%%%%%%%%%%%%%%%%%%%%%%%%%%%%%%%%%%%%%%%%%%%%%%%%%%%%%%%%

\subsection{Prediction} \label{sec:pred}

\subsubsection{Independence design - synthetic data} \label{sec:pred_ind}

\begin{figure*}[t]
\centering
\includegraphics[height=4.66in]{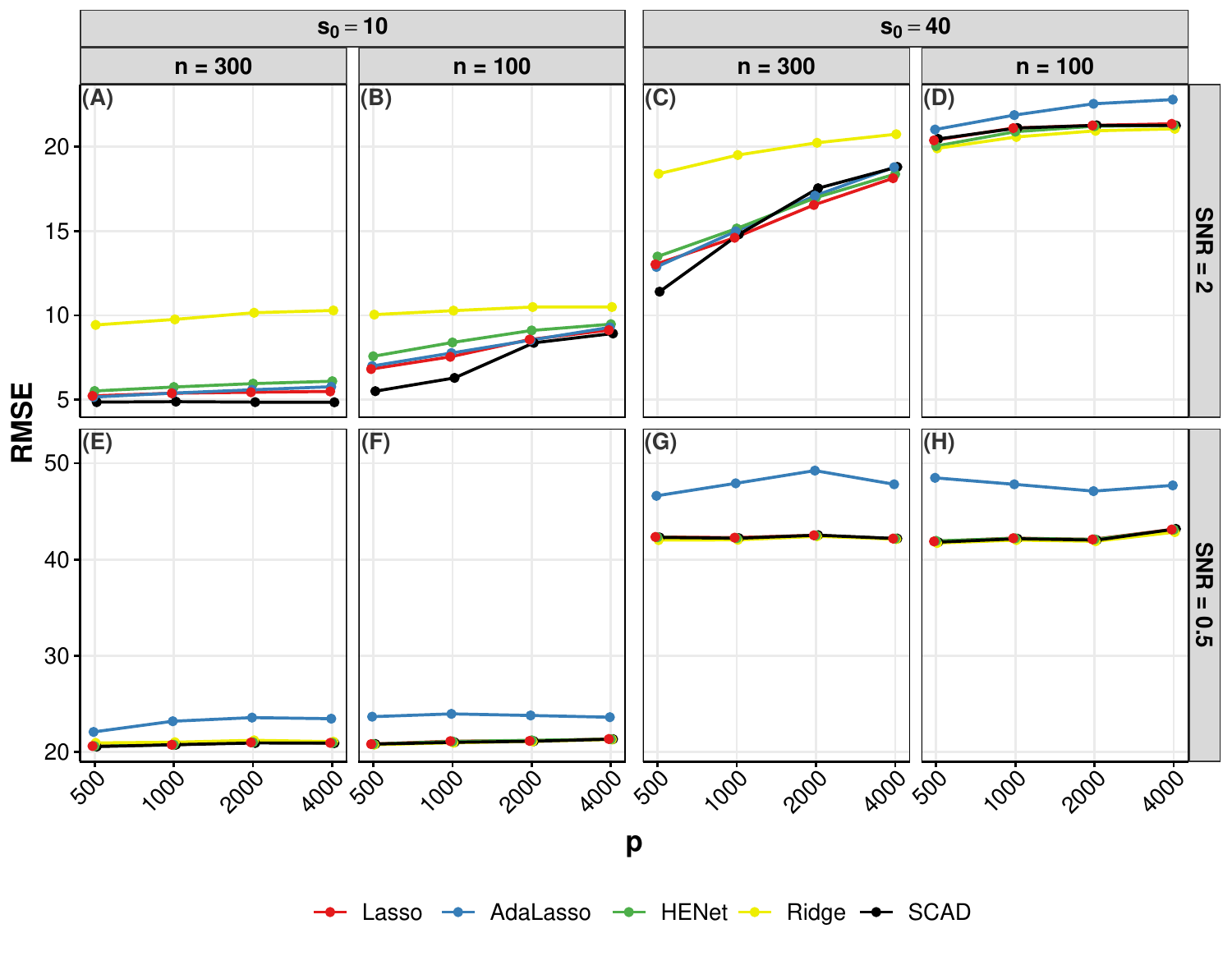}
\caption{Prediction performance (RMSE) versus $p$ for a subset of synthetic independence design scenarios.
Each panel represents a different combination of $n$, $s_0$ and SNR.
Line color indicates method. 
Note that $y$-axis scales vary across rows.
See also Figure~S9.
}
\label{fig:pred_ind}
\end{figure*}

Figure~\ref{fig:pred_ind} shows predictive performance for a subset of independence design scenarios (see also Figure~S9 where performance of pairs of methods are plotted against each other for all independence design scenarios).

\paragraph{An $L_2$ penalty and AdaLasso provide no substantive benefit over Lasso.} \label{para:ind_pred_L2}
An $L_2$ penalty offers very little benefit for prediction, with Ridge performing substantially worse than all the other methods in many scenarios of moderate-to-large SNR (see e.g. Fig.~\ref{fig:pred_ind}A).
When SNR is small, HENet and Ridge perform similarly to Lasso (see e.g. Fig.~\ref{fig:pred_ind}E-H).
The exception is for small $r$ scenarios, where small improvements in prediction error can be seen for HENet and Ridge relative to Lasso (see e.g. $p{=}500$ in Fig.~\ref{fig:pred_ind}D). 
AdaLasso performs similar to or worse than Lasso and performs particularly badly for smaller SNR, where it has the highest prediction error (see e.g. blue line in Fig.~\ref{fig:pred_ind}E-H).

\paragraph{SCAD transition in performance.} \label{para:ind_pred_SCADtrans}
SCAD has a similar transition property for prediction as for ranking (see above), but with the difference that SCAD does not become the worst performing method as scenario difficulty increases; Ridge or AdaLasso still performs worse (black line in Fig.~\ref{fig:pred_ind}C).

%%%%%%%%%%%%

\subsubsection{Correlation designs} \label{sec:pred_cor}
For prediction performance in the ``low'' correlation semi-synthetic design, see Figure~S10, where performance of pairs of methods are plotted against each other. Relative performance of methods agrees well with the synthetic independence design (Fig.~S9).

\begin{figure*}[t]
\centering
\includegraphics[height=5in]{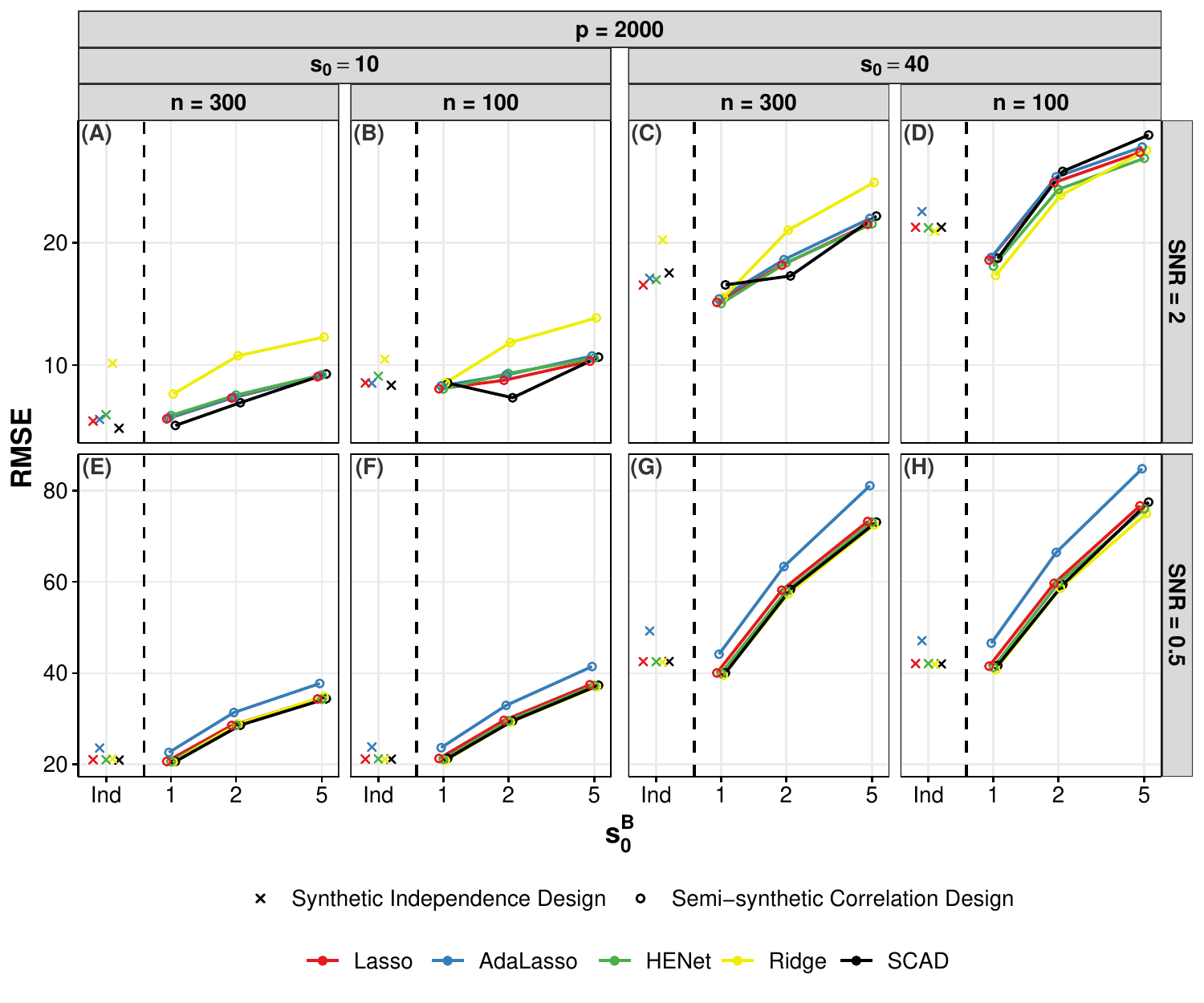}
\caption{Prediction performance (RMSE) versus $s_0^B$ for a subset of semi-synthetic ``high'' correlation designs.
Each panel represents a different combination of $n$, $s_0$ and SNR.
All results shown are for $p{=}2000$ (see Figure~S11 for results with $p{=}500$).
For comparison, results for the corresponding independence design scenarios are also shown in each panel (``Ind'').
Line color indicates method, $x$-axis is on a log-scale and $y$-axis scales vary across rows. 
}
\label{fig:pred_cor_real}
\end{figure*}

\begin{figure*}[t]
\centering
\includegraphics[height=4.8in]{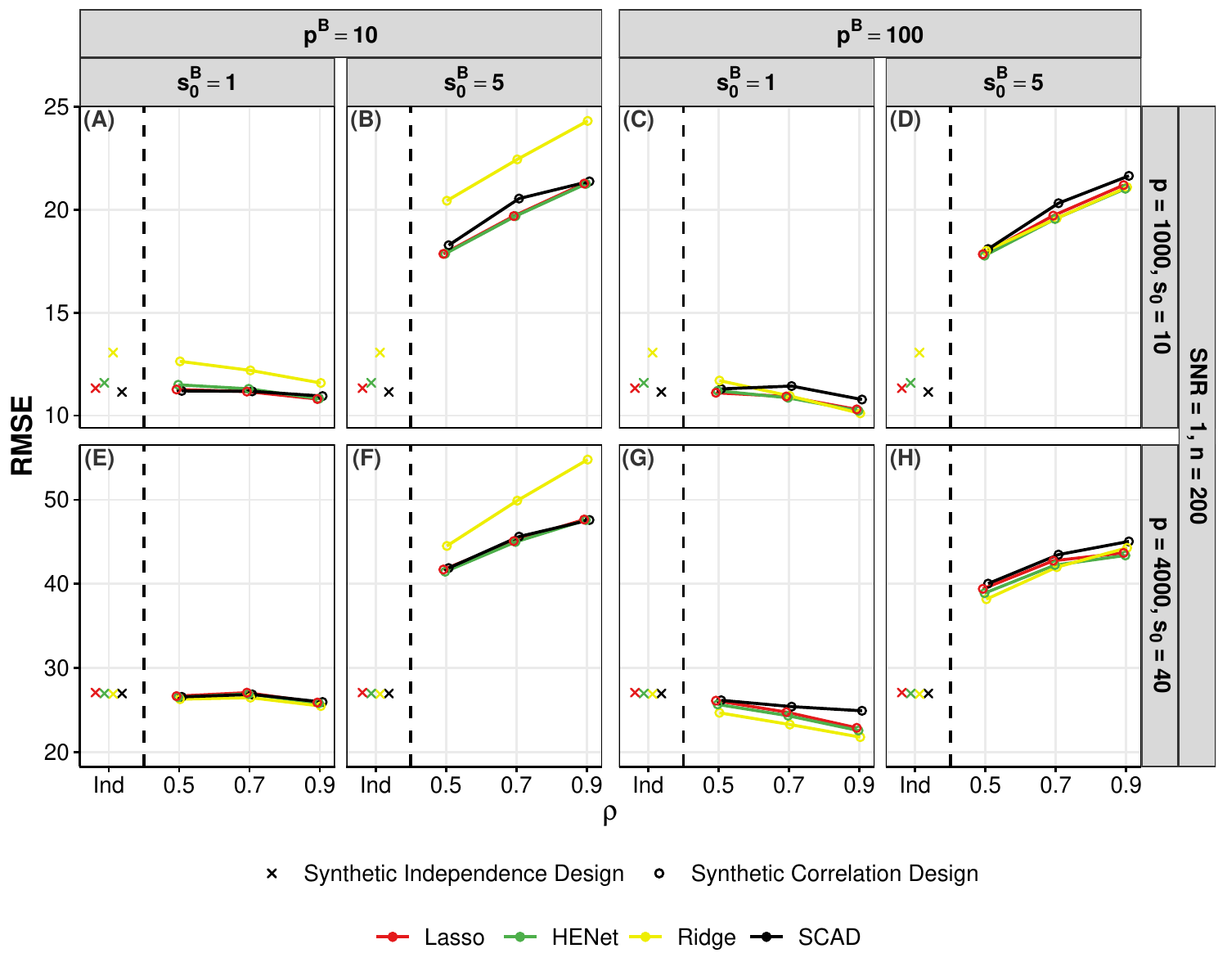}
\caption{Prediction performance (RMSE) versus $\rho$ for a subset of synthetic pairwise correlation designs.
Each row represents a different combination of $p$ and $s_0$, while each column represents a different combination of  $p^B$ and $s_0^B$.
All results shown are for SNR=1 and $n{=}200$ (see Figure~S12 for SNR=2 and 4).
The top row has $(n,p,s_0)=(200,1000,10)$, giving $r{=}2.9$, and the bottom row has $(n,p,s_0)=(200,4000,40)$, giving $r{=}0.6$.
For comparison, results for the corresponding independence design scenarios are also shown (``Ind''; these data points are identical across the panels in each row).
Line color indicates method and $y$-axis scales vary across rows
}
\label{fig:pred_cor_syn}
\end{figure*}

Figure~\ref{fig:pred_cor_real} shows predictive performance for a subset of the ``high'' correlation semi-synthetic design scenarios with $p{=}2000$ (analogous results for $p{=}500$ are shown in Figure~S11) and Figure~\ref{fig:pred_cor_syn} shows predictive performance for a subset of pairwise correlation synthetic design scenarios with SNR=1 (analogous results for SNR=2 and 4 are shown in Figure~S12).

\paragraph{Performance improvements relative to the independence design when $s_0^B{=}1$.} \label{para:realCor_pred_cor_vs_ind}
Predictive performance worsens with increasing number of signals per block $s_0^B$ (see Fig.~\ref{fig:pred_cor_real}) and this is primarily due to an increase in the variance of the response $\mathbf{y}$ as a result of the correlation between signals.
For the same reason, in the synthetic data design, increasing correlation strength $\rho$ leads to higher predictive error when blocks contain more than one signal (see second and fourth columns of Fig.~\ref{fig:pred_cor_syn}).

When there is one signal per block ($s_0^B{=}1$), the signals are uncorrelated (or very weakly correlated for semi-synthetic data) and so there is no or little increase in the variance of $\mathbf{y}$ relative to the independence design.
The correlation between the signal and non-signals in each block can then result in a decrease in predictive error relative to the independence design, and we observe this for the semi-synthetic data (compare crosses and $s_0^B=1$ circles in Fig.~\ref{fig:pred_cor_real}D) and synthetic data (compare crosses and circles in Fig.~\ref{fig:pred_cor_syn}G).
For the latter, we also find that an increase in correlation strength $\rho$ and increase in block size $p^B$ leads to larger decreases in RMSE (compare Figs.~\ref{fig:pred_cor_syn}E and \ref{fig:pred_cor_syn}G).

The method that shows the largest improvements relative to the independence design is typically Ridge Regression.
For example, in Figure~\ref{fig:pred_cor_real}A for $s_0^B{=}1$, Ridge Regression (yellow circle) has a 25\% decrease in RMSE relative to the independence design (yellow cross), while all other methods show little change in RMSE.
Ridge regression may benefit the most because it has a non-sparse solution and, due to the correlation between signals and non-signals in each block, the correlated designs are also, in a sense, non-sparse.

\paragraph{An $L_2$ penalty and AdaLasso still provide no substantive gains over Lasso.} \label{para:realCor_pred_L2_benefit}
As for the independence design, Ridge and HENet do not substantively outperform the other approaches for prediction in any of the scenarios considered here, and this is the case even though Ridge often benefits the most from correlation (see above). 
In ``easier'' scenarios, Ridge still performs notably worse than other approaches (e.g. Fig.~\ref{fig:pred_cor_real}A),
but in ``hard'' scenarios with small $r$, Ridge can marginally outperform other methods.
For example, for the ``hard'' semi-synthetic data scenario in Fig.~\ref{fig:pred_cor_real}D where $r=0.3$ and SNR=2, Ridge has a 7\% decrease in RMSE relative to Lasso when $s_0^B{=}1$.
HENet also performs marginally better than Lasso in these ``scenarios'', but typically marginally worse than Ridge (HENet has a 3\% decrease in RMSE relative to Lasso in the above example).
Similar behavior is observed for ``hard'' synthetic data scenarios and this is particularly noticeable for large correlated blocks (yellow line in Fig.~\ref{fig:pred_cor_syn}G).

AdaLasso remains similar to or, for small SNR, worse than the other approaches (see blue lines in Fig.~\ref{fig:pred_cor_real}).

\paragraph{SCAD transition in performance.} \label{para:pairCor_pred_SCADtrans}
SCAD again shows transition behavior, offering modest gains over other methods when $r$ and SNR are large, and $s_0^B$ is small, but becoming worse than Lasso, HENet and sometimes Ridge as scenario difficulty, $s_0^B$ or $\rho$ increases. 
For example, SCAD performs best when $r{=}3.9$ and $s_0^B{=}1$ (Fig.~\ref{fig:pred_cor_real}A), but worst when $r{=}0.3$ and $s_0^B{=}5$ (Fig.~\ref{fig:pred_cor_real}D).

%%%%%%%%%%%%

\subsubsection{Summary and recommendations} \label{sec:summary_pred}

In settings with uncorrelated or very weakly correlated variables, predictive performance of methods relative to each other is mostly similar to that for ranking, so we make a similar recommendation: that is, use Lasso, or potentially SCAD if there is confidence that the scenario at hand is ``easy''. 
The key difference from ranking is that we would not recommend AdaLasso because it can perform much worse than Lasso. 

For more highly correlated settings, Lasso is mostly competitive and so can be considered a ``safe'' option. 
Ridge Regression may provide some small gains in ``harder'' scenarios, particularly for large correlated blocks, but can perform much worse than other approaches in ``easier'' settings. Therefore, HENet could be a good option here as it can still offer some gains over Lasso, but is not as sensitive to the scenario difficulty, remaining competitive where Ridge performs poorly.
SCAD and AdaLasso may not be good options since they do not result in substantive benefits over Lasso or HENet and can both perform much worse than other methods in some scenarios.

%%%%%%%%%%%%%%%%%%%%%%%%%%%%%%%%%%%%%%%%%%%%%%%%%%%%%%%%%%%%%%%%%%%%%%

\subsection{Selection} \label{sec:sel}

\subsubsection{Independence design - synthetic data} \label{sec:sel_ind}

\begin{figure*}[tp]
\centering
\includegraphics[height=7in]{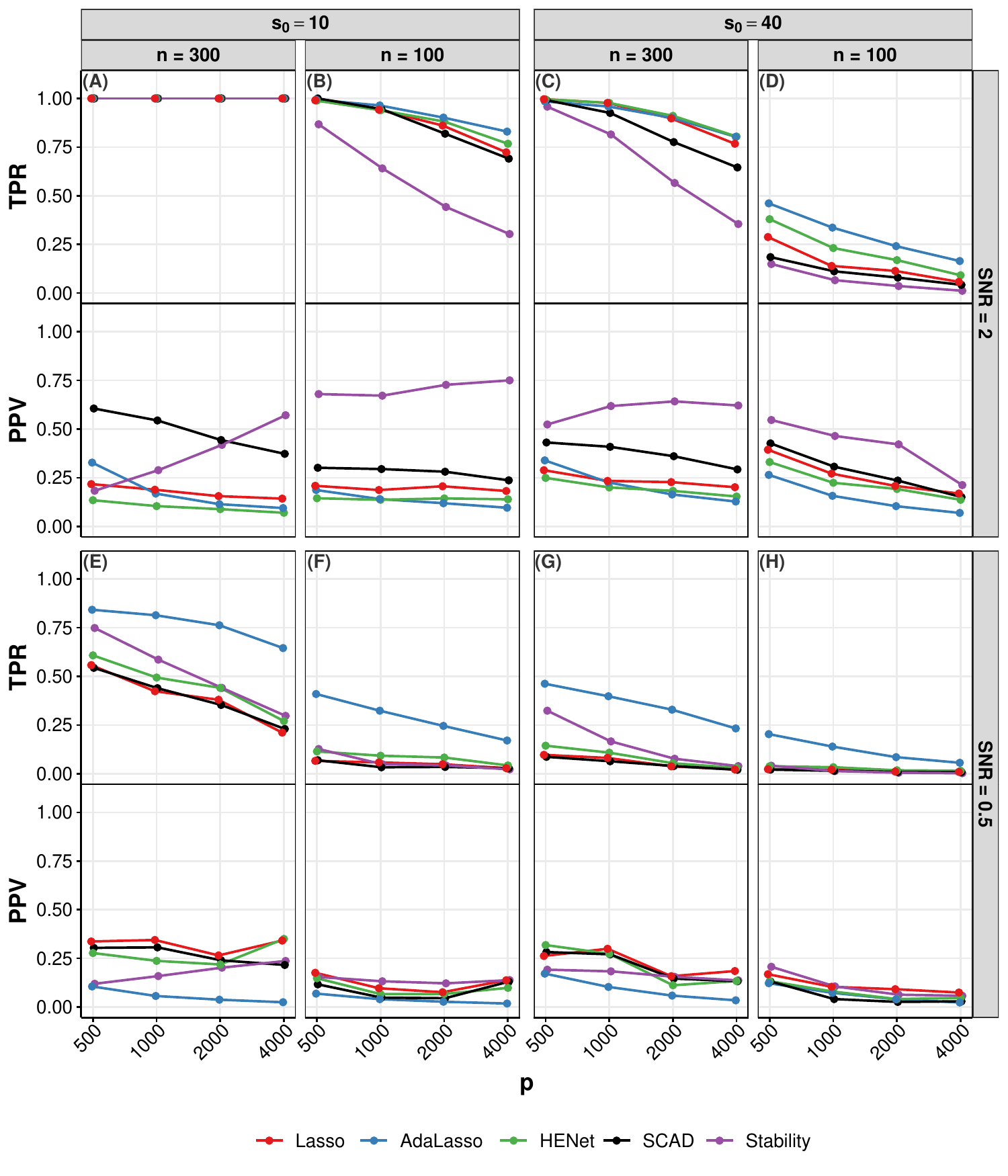}
\caption{Selection performance (TPR and PPV) versus $p$ for a subset of synthetic independence design scenarios.
Each panel shows TPR (top) and PPV (bottom) for a different combination of $n$, $s_0$ and SNR.
Line color indicates method. 
See also Figures~S13 and S14.
}
\label{fig:sel_ind}
\end{figure*}

Figure~\ref{fig:sel_ind} shows selection performance for a subset of independence design scenarios. See also Figures~S13 and S14 where performance of pairs of methods are plotted against each other for all independence design scenarios.

\paragraph{Stability Selection or SCAD often best for PPV; trade-off between PPV and TPR.} \label{para:ind_sel_SS}
All methods achieve optimal TPR{=}1 when $r$ and SNR are sufficiently large, but can at the same time have substantial differences in terms of PPV (see e.g. Fig.~\ref{fig:sel_ind}A; range of PPVs${\approx}0.1{-}0.6$). 
SCAD typically offers the best PPV in these ``easiest'' scenarios, followed by Stability Selection and AdaLasso\footnote{Note that this inferior performance of Stability Selection relative to SCAD could in part be due to the lack of false positive control in the implementation of Stability Selection used here.}.

In scenarios where TPR is sub-optimal (small-to-moderate values of $r$ or SNR), as could be expected, the relative performance of two methods typically follows the rule: if method $A$ has a higher TPR than method $B$, then method $A$ will have a lower PPV (see e.g. Fig.~\ref{fig:sel_ind}D). 
For the majority of these scenarios, Stability Selection has the highest PPV and lowest TPR.
SCAD performs similar to or better than Lasso, HENet and AdaLasso in terms of PPV, but similar or worse in terms of TPR (see e.g. Figs.~\ref{fig:sel_ind}B-D, S13 and S14). 
Lasso, HENet and AdaLasso fail to obtain a PPV higher than 0.55 across all scenarios, contrasting with a maximum PPV greater than 0.8 for SCAD or Stability Selection.
The range of PPVs across methods decreases as SNR decreases, and for SNR=0.5, Stability Selection no longer has an advantage over the other approaches (Fig.~\ref{fig:sel_ind}E-H).

\paragraph{HENet and  AdaLasso provide gains over Lasso for TPR.} \label{para:ind_sel_L2}
There is a benefit of using an $L_2$ penalty or AdaLasso for TPR, but it comes at the cost of poorer false positive control. Across the majority of scenarios, HENet has small gains in TPR (of at most 0.1) over Lasso, but the converse is true for PPV (see e.g. red and green lines in Fig.~\ref{fig:sel_ind}D). 
AdaLasso offers the highest TPR, particularly for small SNR where it provides large gains (of up to 0.35) over all the other approaches, but again its PPV suffers (see e.g. blue lines in Fig.~\ref{fig:sel_ind}E).

%%%%%%%%%%%

\subsubsection{Correlation designs} \label{sec:sel_cor}
For selection performance in the ``low'' correlation semi-synthetic design, see Figures~S15 and S16, where performance of pairs of methods are plotted against each other.
Relative performance of methods agrees well with the synthetic independence design (Figs.~S13 and S14).

\begin{figure*}[tp]
\centering
\includegraphics[height=7in]{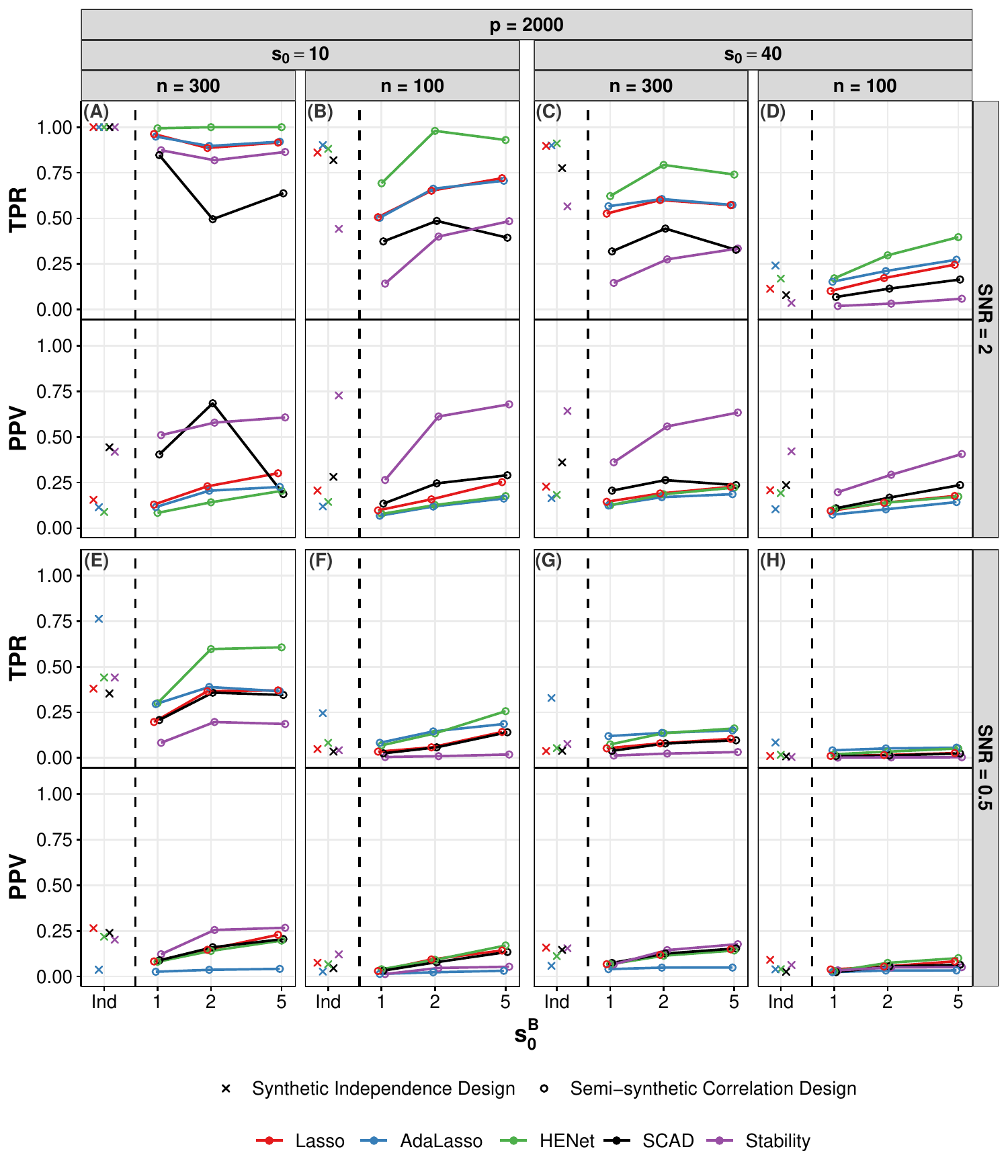}
\caption{Selection performance (TPR and PPV) versus $s_0^B$ for a subset of semi-synthetic ``high'' correlation design scenarios.
Each panel shows TPR (top) and PPV (bottom) for a different combination of $n$, $s_0$ and SNR.
All results shown are for $p{=}2000$ (see Figure~S17 for results with $p{=}500$).
For comparison, results for the corresponding independence design scenarios are also shown in each panel (``Ind'').
Line color indicates method and $x$-axis is on a log-scale.
}
\label{fig:sel_cor_real}
\end{figure*}

\begin{figure*}[tp]
\centering
\includegraphics[height=6.8in]{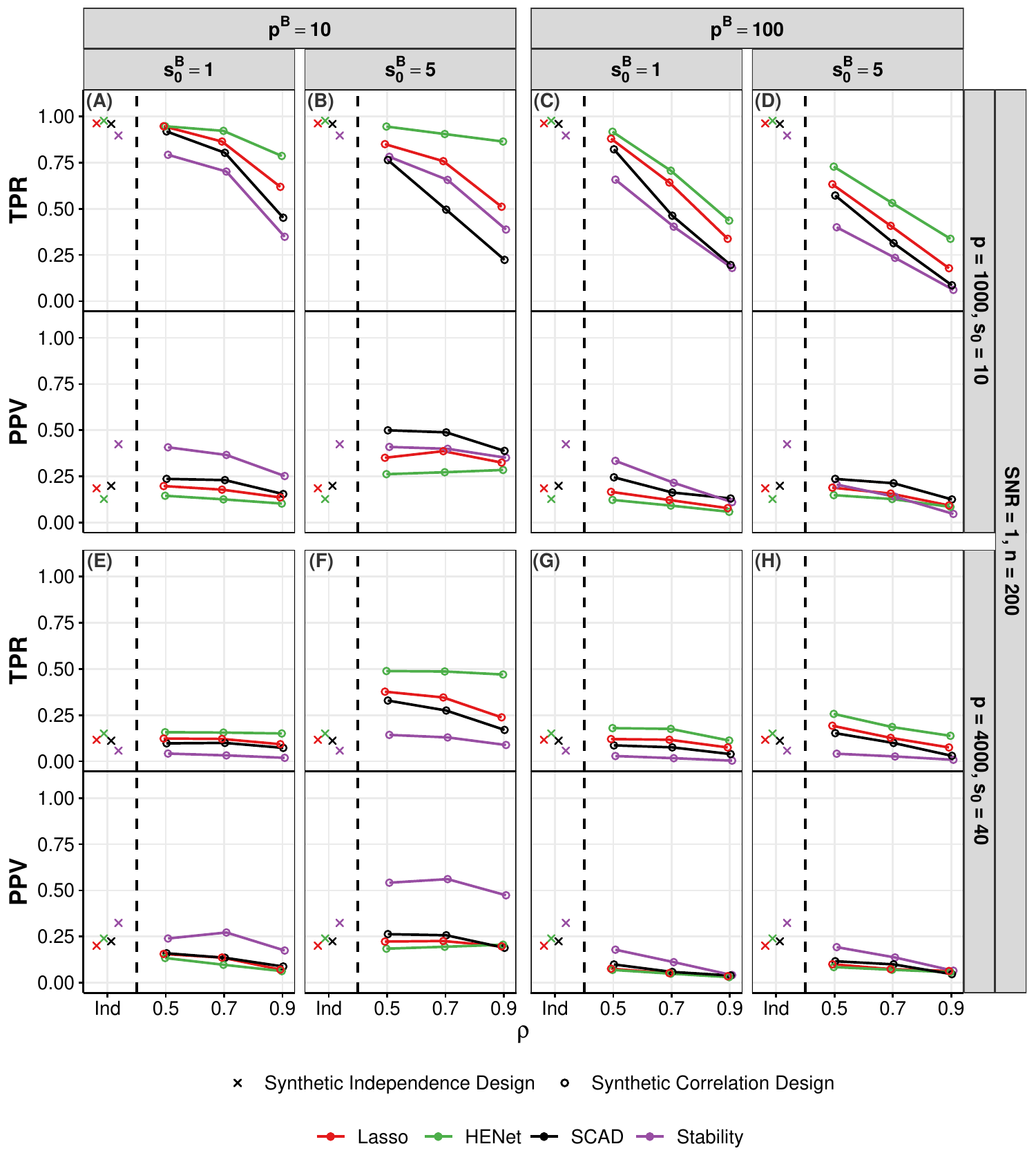}
\caption{Selection performance (TPR and PPV) versus $\rho$ for a subset of synthetic pairwise correlation design scenarios.
Each panel shows TPR (top) and PPV (bottom) for a different combination of $p$ and $s_0$ (rows), and $p^B$ and $s_0^B$ (columns).
All results shown are for SNR=1 and $n{=}200$ (see Figure~S18 for SNR=2 and 4).
The top row has $(n,p,s_0)=(200,1000,10)$, giving $r{=}2.9$, and the bottom row has $(n,p,s_0)=(200,4000,40)$, giving $r{=}0.6$.
For comparison, results for the corresponding independence design scenarios are also shown (``Ind''; these data points are identical across the panels in each row).
Line color indicates method.
}
\label{fig:sel_cor_syn}
\end{figure*}

Figure~\ref{fig:sel_cor_real} shows selection performance for a subset of the ``high'' correlation semi-synthetic design scenarios with $p{=}2000$ (analogous results for $p{=}500$ are shown in Figure~S17) and Figure~\ref{fig:sel_cor_syn} shows selection performance for a subset of pairwise correlation synthetic design scenarios with SNR=1 (analogous results for SNR=2 and 4 are shown in Figure~S18).

\paragraph{Improved performance relative to the independence design for some scenarios.} \label{para:cor_sel_cor_vs_ind}
The influence of correlation design parameters on selection performance is in line with that seen for ranking in Section~\ref{sec:rank_cor}. 
In particular, we find that the largest benefits from correlation relative to the independence design are again in ``hard'' scenarios (small SNR or $r$) with small blocks, strong correlation and several signals per block.
For example, in Figure~\ref{fig:sel_cor_real}D for semi-synthetic data with $r{=}0.3$, SNR=2 and $s_0^B{=5}$, all methods have increased TPR relative to the independence design, with the largest increase in TPR of 0.23 for HENet.
Similarly, in Figure~\ref{fig:sel_cor_syn}F for synthetic data with $r{=}0.6$, SNR=1, $p^B{=}10$, $s_0^B{=}5$ and $\rho{=}0.9$, the largest increase in TPR is 0.32, again for HENet.
These increases in TPR do not necessarily come with decreases in PPV of corresponding magnitude; HENet has a similar PPV to the independence design.

\paragraph{HENet provides increased gains over Lasso for TPR, while also being competitive for PPV.} \label{para:cor_sel_L2_benefit}
In the independence design, we found that	HENet has small gains in TPR over Lasso, but has worse PPV.
In the ``hard'', correlated scenarios described above where HENet benefits from correlation, HENet can give more substantial improvements in TPR over Lasso, while also remaining competitive in terms of PPV.
In the semi-synthetic data example from above (Fig.~\ref{fig:sel_cor_real}D), HENet has an increase in TPR of 0.15 relative to Lasso when $s_0^B{=}5$; the corresponding increase for the independence design was 0.05.
At the same time, PPV remains competitive at 0.17 for HENet and 0.18 for Lasso.
This behavior is in line with Elastic Net enjoying the grouping effect property for correlated variables.

\paragraph{Stability Selection can be best for PPV, but is sensitive to correlation.} \label{para:cor_sel_SS}
As for the independence design, Stability Selection typically performs best in terms of PPV, followed by SCAD, and they perform worse in terms of TPR (purple and black lines in Figs.~\ref{fig:sel_cor_real} and \ref{fig:sel_cor_syn}).
However, Stability Selection and SCAD are sensitive to correlation. 
For example, in the SNR=0.5 semi-synthetic data scenario with $s_0^B{=}1$ shown in Figure~\ref{fig:sel_cor_real}B, the substantial improvements in PPV provided by Stability Selection in the independence design are mostly lost. 
Also in line with the independence design, the advantage Stability Selection provides for PPV reduces as SNR decreases, with little to no advantage remaining for SNR=0.5; here, all approaches have a similar performance, with AdaLasso typically performing worst (Fig.~\ref{fig:sel_cor_real}E-H).

\paragraph{AdaLasso no longer competitive for TPR} \label{para:cor_sel_Ada}
AdaLasso offered the best performance for TPR in the independence design, but this is no longer the case as HENet has a similar or higher TPR, and AdaLasso is still not competitive for PPV (semi-synthetic data; green and blue lines in Fig.~\ref{fig:sel_cor_real}).

%%%%%%%%%%%

\subsubsection{Summary and recommendations} \label{sec:summary_sel}

Since there is a trade-off between PPV and TPR, the best method to use depends on the aim. If the aim is primarily to have a low false positive rate, then Stability Selection is a good choice for both correlated and uncorrelated covariates, since it is likely to provide the best PPV.
If the focus is more on maximizing the number of signals selected, then AdaLasso results in a TPR that dominates the other methods in most uncorrelated and very weakly correlated scenarios. However, it loses its advantage in more highly correlated designs, where HENet performs best.
Lasso could be used to obtain a compromise between the two aims.
If the scenario at hand is thought to be particularly ``easy'' with high $r$ or SNR and covariates are uncorrelated or very weakly correlated, SCAD may provide the best PPV while retaining a competitive TPR.

%%%%%%%%%%%%%%%%%%%%%%%%%%%%%%%%%%%%%%%%%%%%%%%%%%%%%%%%%%%%%%%%%%%%%%%%%%%%%%%%%%%%%%%%%%%%%%%%%
%%%%%%%%%%%%%%%%%%%%%%%%%%%%%%%%%%%%%%%%%%%%%%%%%%%%%%%%%%%%%%%%%%%%%%%%%%%%%%%%%%%%%%%%%%%%%%%%%

\section{Additional investigations} \label{sec:additional_simulations}
Below we extend the main simulations above in three directions. Section~\ref{sec:toeplitz} investigates a synthetic data Toeplitz correlation design, Section~\ref{subsec:stability} explores sensitivity of Stability Selection to its tuning parameters and Section~\ref{subsec:heterogeneous_betas} investigates the ability of methods to detect weak signals when coefficients are heterogeneous.

\subsection{Toeplitz correlation design} \label{sec:toeplitz}

We now consider method performance for synthetic data with a Toeplitz correlation design.
This is as for pairwise correlation, but with covariates $\mathbf{x}_{j_1}$ and $\mathbf{x}_{j_2}$ within the same block having correlation $0.95^{|j_1-j_2|}$. 
We only consider block sizes of $p^B{=}100$ that have two active variables per block, $s_0^B=2$, with their positions, $j_1'$ and $j_2'$, within a block chosen such that $|j_1'-j_2'|=7$, to give a correlation of $0.95^7\approx 0.7$.

\begin{figure*}[!t]
\centering
\includegraphics[height=5.9in]{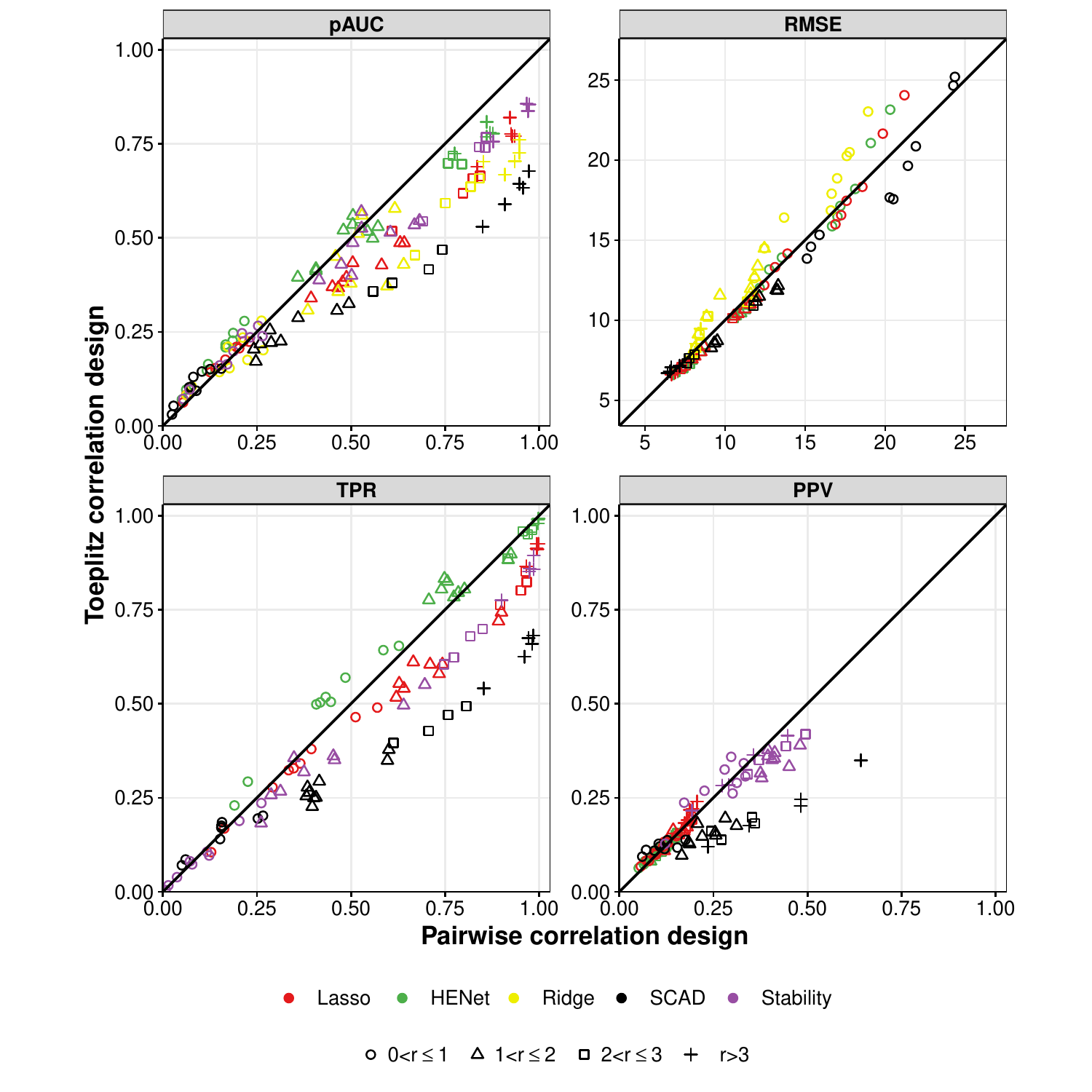}
\caption{Comparison between Toeplitz correlation and pairwise correlation designs for ranking, prediction and selection performance.
Performance in the Toeplitz correlation design is plotted against performance in the corresponding pairwise correlation design with $\rho=0.7, s_0^B=2$ and $p^B=100$.
Each point corresponds to a method (indicated by color) and a single $(n,p,s_0)$ triplet (the resulting value of the rescaled sample size $r$ is indicated by symbol). 
Results shown are for SNR=2 (see Figs.~S19 and S20 for SNR=1 and SNR=4) and are averages over 64 replicates.}
\label{fig:pairwise_toeplitz_all_metrics}
\end{figure*}

Figure \ref{fig:pairwise_toeplitz_all_metrics} compares performance in the Toeplitz design against that in the corresponding pairwise correlation design ($\rho=0.7$) for $\mathrm{SNR}=2$ and all possible combinations of $n$, $p$ and $s_0$ (see Figs.~S19 and S20 for SNR=1 and SNR=4 respectively).
Performance is typically similar for the two designs or worse in the Toeplitz design. For prediction, Ridge Regression is most negatively affected by Toeplitz correlation, while SCAD is most affected for the other metrics.

On the one hand, the pairwise correlation design could be considered more difficult than the Toeplitz design because the average correlation between signals and non-signals (within a block) is higher for pairwise than for Toeplitz (0.7 vs. 0.19). However, on the other hand, the Toeplitz design could be considered more difficult because there are several non-signals that are more strongly correlated with the signals than the signals are with each other; for the pairwise correlation design all signals and non-signals within a block are correlated with equal strength.
The generally poorer performance observed for the Toeplitz design therefore suggests that having strongly correlated signals and non-signals is more detrimental than a higher average correlation.

Relative performance of methods in the Toeplitz design is generally consistent with that seen for the corresponding pairwise correlation design.
For ranking, the impact of an $L_2$ penalty (relative to Lasso) is larger under the Toeplitz design than the pairwise design, with Ridge performing relatively well when SNR=1, but poorly when SNR=4.

\begin{figure*}[tp]
\centering
\includegraphics[height=6.5in]{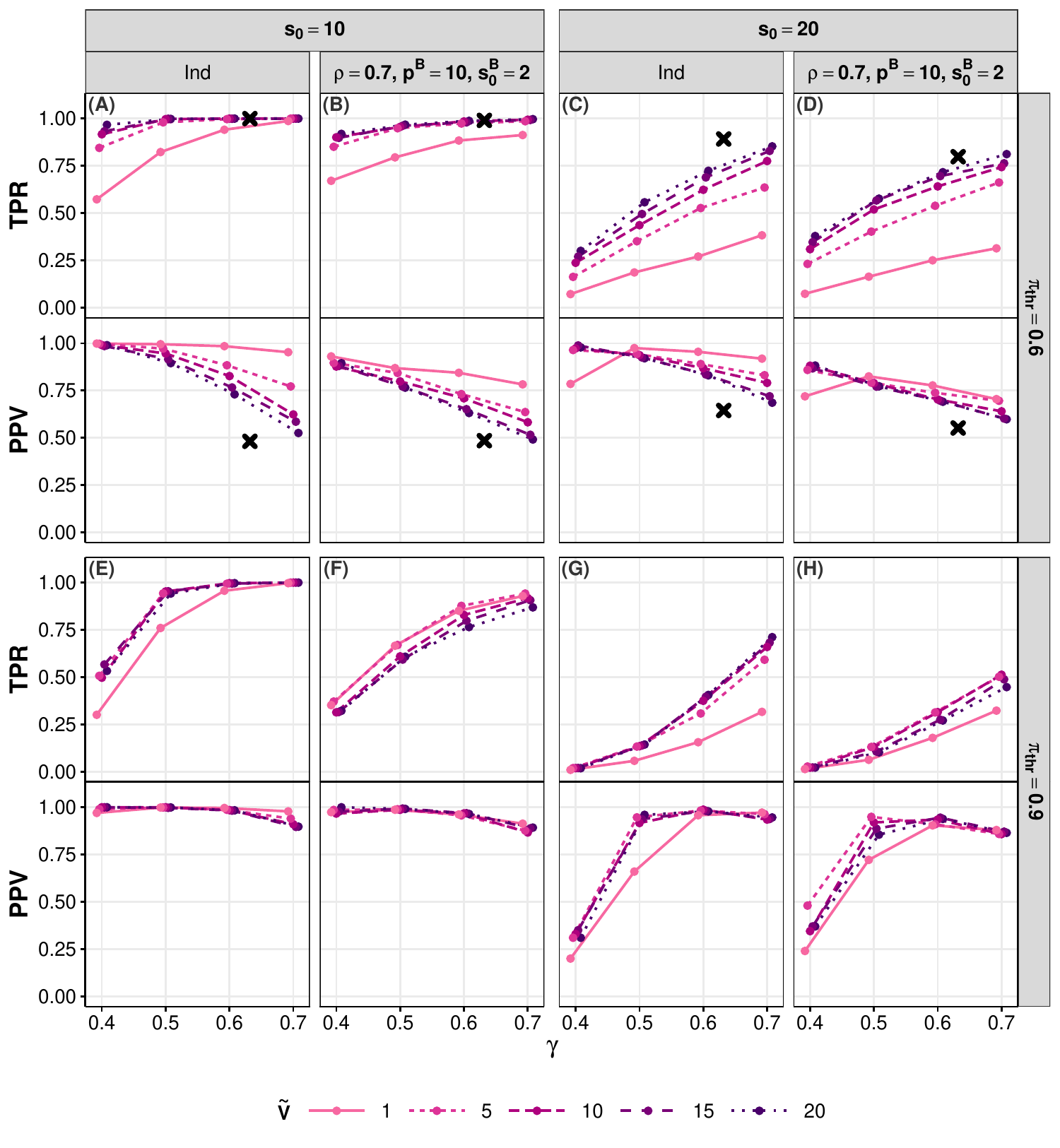}
\caption{Stability Selection tuning parameter sensitivity.
Each panel shows TPR (top) and PPV (bottom) versus subsample proportion $\gamma$.
Within each panel, line type indicates $\tilde{V}$, the upper bound for the expected number of false positives.
Top row of panels corresponds to threshold probability $\pi_{thr}{=}0.6$ and bottom row to $\pi_{thr}{=}0.9$.
Each column corresponds to a different simulation scenario: the synthetic independence design (``Ind'') with $n{=}200$, $p{=}1000$, SNR=2, and $s_0{=}10$ or 20, or the corresponding synthetic pairwise correlation design scenarios with $\rho{=}0.7$, $p^B{=}10$ and $s_0^B{=}2$.
Black crosses in the top row of panels show performance observed in the main simulations where $\pi_{thr}{=}0.6$, $\gamma{=}0.632$ and there was no explicit false positive control $\tilde{V}$.
Results are averages over 100 replicates.}
\label{fig:stability_selection}
\end{figure*}

\subsection{Stability Selection tuning parameters} \label{subsec:stability}

\begin{figure*}[t]
\centering
\includegraphics[height=4in]{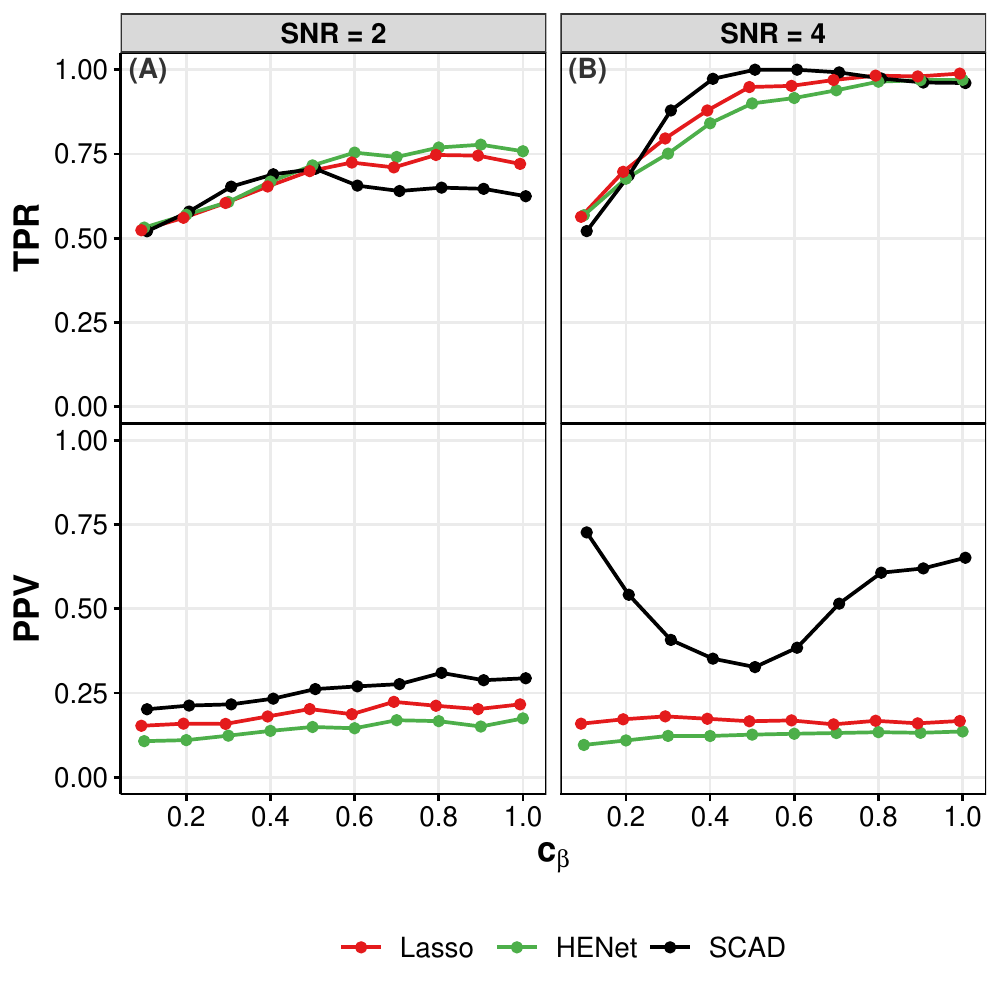}
\caption{Influence of heterogeneous regression coefficients on selection performance.
TPR (solid lines) and PPV (dotted lines) are plotted against the coefficient scaling factor $c_{\beta}$ for the independence design with $(n,p,s_0)=(300,4000,40)$ and SNR=2 (A) or SNR=4 (B).
In the data-generating linear model, half of the signals have coefficient $\beta'$ and the other half have coefficient $c_{\beta}\beta'$ (see text for details).
Note that $c_{\beta}=1$ gives the main simulation set-up with homogeneous coefficients.
Line color indicates method.
Results are averages over 50 replicates.}
\label{fig:heterogeneous_betas}
\end{figure*}

Stability Selection has several tuning parameters: the subsample size $\tilde{n}$, an upper bound $\tilde{V}$ for $\mathbb{E}[V]$ (the expected number of false positives), and either a threshold $\pi_{thr}$ on the selection probabilities or a set of regularization parameters to consider $\Lambda$ (see Section~\ref{sec:algorithms}).
Making appropriate choices for these parameters is non-trivial.
Here, we explore the effects of varying $\tilde{n}$, $\tilde{V}$ and $\pi_{thr}$ on selection performance.

We simulated data (as described in Section~\ref{subsec:designs}) with SNR=2, $n{=}200$, $p{=}1000$ and $s_0{=}10$ or 20 (giving $r{=}2.90$ or 1.45 respectively) for the independence design, and the pairwise correlation design with $p^B{=}10$, $s_0^B{=}2$ and $\rho{=}0.7$. 
We applied Stability Selection with all possible combinations of the following tuning parameter values: $\tilde{V} \! \in \! \left\{1, 5, 10, 15, 20\right\}$, $\pi_{thr}\! \in \! \left\{0.6, 0.9\right\}$ and $\tilde{n}{=}\lfloor n\gamma \rfloor$ where $\gamma \! \in \! \left\{0.4, 0.5, 0.6, 0.7\right\}$ is the subsample proportion. 

Figure~\ref{fig:stability_selection} shows that, in general, as $\tilde{V}$ or $\gamma$ increase, or $\pi_{thr}$ decreases, the number of selected variables increases, resulting in higher TPR, but lower PPV.
An exception is for $s_0{=}20$, where, for the most conservative choices of the parameters ($\gamma{=}0.4,\tilde{V}{=}1$ and $\pi_{thr}{=}0.9$), in addition to a very poor TPR, PPV is also low on average (see solid line, $\gamma{=}0.4$ in Fig.~\ref{fig:stability_selection}G and H).
Here, selection is too stringent and the majority of signals are missed.
When the underlying model size is smaller ($s_0{=}10$), the most conservative parameter choices are again sub-optimal in terms of performance (Fig.~\ref{fig:stability_selection}E and F), but the same is also true for the least conservative choices ($\gamma{=}0.7$,$\tilde{V}{=}20$ and $\pi_{thr}{=}0.6$; Fig.~\ref{fig:stability_selection}A and B).
However, in the scenarios considered here, being too stringent seems to have a more deleterious effect on performance than being too lenient.

Results from the main simulations, where we set $\tilde{n}{=}\lfloor 0.632n \rfloor$, $\pi_{thr}{=}0.6$ and had no explicit false positive control $\tilde{V}$ (i.e. the full range of regularization parameters $\Lambda$ was considered; see Section~\ref{sec:implement}), are indicated by crosses in Figure~\ref{fig:stability_selection}A-D. 
Performance in the main simulations is most similar to that of the largest $\tilde{V}$ considered here ($\tilde{V}{=}20$), but with better TPR and worse PPV (except for $s_0{=}10$ where TPR is already optimal and so there is only a decrease in PPV).

\subsection{Heterogeneous coefficients} \label{subsec:heterogeneous_betas}
In the main simulations, all non-zero coefficients were assigned the same value. Here, we consider detection of signals with heterogeneous coefficients for three methods: Lasso, HENet and SCAD. 
We simulated data (for the independence design) as described in Section~\ref{sec:methods}, except instead of $s_0$ active variables all having coefficient 3, half of them had coefficient $\beta'$ and the other half had coefficient $c_\beta \beta'$ where $c_\beta \! \in \! \left[0,1\right]$.
We chose $\beta'{=}\sqrt{18/(1+{c_\beta}^2)}$ such that with fixed SNR, $\mathbb{E}(\sigma^2)$  remains the same as in the homogeneous $\beta$'s case.
Note that $c_\beta{=}1$ gives the main simulation set-up with homogeneous coefficents. 
Informed by the main simulations, we set $n{=}300$, $s_0{=}40$, $p{=}4000$ and SNR=2 or 4, guaranteeing that when non-zero coefficients all take the same value, we are in a relatively ``easy'' scenario where the majority of the signals can be detected.

Figure \ref{fig:heterogeneous_betas} shows that as $c_\beta$ decreases, signals with smaller coefficients are less likely to be detected, resulting in a decrease in TPR.
All methods fail to detect the very weak signals when $c_\beta$=0.1 (i.e. only the stronger 50\% of the signals are detected giving TPR$\approx$0.5).
Consistent with the main simulations, SCAD has better false positive control (higher PPV) than Lasso and Elastic Net when SNR is large, and this is especially the case when $c_\beta$ is near 0.1 or 1 (contrast black line with red and green lines for PPV in Fig.~\ref{fig:heterogeneous_betas}B).
The ``U'' shape of the SCAD PPV curve here is likely due to the fact that bias is largest when $c_\beta$ is moderate, which leads to selection of more variables to compensate (SCAD is known to be nearly unbiased for strong signals; for large $c_{\beta}$ all signals are relatively strong, while for small $c_{\beta}$ the $s_0/2$ weaker signals have such a small influence that the underlying model is well-approximated by a model with $s_0/2$ strong signals and no weak signals).
In contrast, Lasso and Elastic Net are biased estimators, so their PPVs are not as affected.
SCAD also seems to have higher power to detect the weaker signals when SNR is large and $c_\beta$ is moderate (see TPR in Fig.~\ref{fig:heterogeneous_betas}B). 
However, as observed in the main simulations, SCAD is more sensitive to SNR and so is less competitive in ``harder'' scenarios (SNR=2; Figure~\ref{fig:heterogeneous_betas}A). 
Lasso has higher PPV than HENet and this is largely unaffected by changes in $c_{\beta}$.
Differences in TPR between HENet and Lasso decrease as $c_{\beta}$ decreases, until they both have a similar performance for $c_{\beta}{=}0.1$ (note that which method performs best depends on SNR).

\section{Discussion} \label{sec:discussion}

Our results complement theory by shedding light on the finite-sample relative performance of methods. 
Many of our results do align with available theory.  For instance, SCAD is known to have nearly unbiased estimates for coefficients that are large (relative to noise), explaining why it tends to have better selection performance in  ``easy'' scenarios. 
However, some conditions of theoretical results (asymptotic or finite-sample) can be hard to verify in practice, and the results do not directly provide insight into the performance of a method relative to others, making it difficult to pick a suitable approach in any given finite-sample setting. 
Our results suggest that there is no one method which clearly dominates others in all scenarios, even in the relatively narrow set of possibilities considered here (e.g. we did not consider heavy tailed noise, non-sparsity, non-block-type covariance etc.). 
Relative performance depends on many factors, and also on the specific metric(s) of interest. 

A challenge of translating results of our empirical study into practice is that not all of the factors will be known to a user in a given setting, specifically those that are related to the unknown signals (e.g. $s_0$).
However, domain knowledge may provide some indication as to, for example, whether SNR is likely to be high or low, or as to the likely number of signals, which could then give an idea of the ``difficulty'' of the problem.
Nevertheless, with the above caveats, we have been able to make some general observations that in turn have allowed some broad recommendations to be made (see Sections~\ref{sec:summary_rank}, \ref{sec:summary_pred} and \ref{sec:summary_sel}).
These recommendations are primarily based on covariate correlation and focus on which approach is most likely to perform well across a broad range of scenarios.
The synthetic independence design and semi-synthetic ``low'' correlation design resulted in similar method performance, so we have made a single, joint recommendation for uncorrelated and very weakly correlated scenarios, for each metric.
For example, for ranking we have recommended Lasso or AdaLasso for uncorrelated or very weakly correlated covariates, and Ridge Regression when variables are more strongly correlated.
We have also highlighted when a method may be a risky choice.
For example, SCAD is double-edged, dominating in ``easier'' scenarios but deteriorating rapidly when conditions become difficult.
Therefore, its high variability means that it should only be used when one is sure that the scenario at hand is very ``easy''.
Six out of the seven approaches considered in our study have been recommended for at least one of the goals (further to the above, we recommended Stability Selection for PPV, and Elastic Net for TPR in correlated settings).
The Dantzig Selector is not recommended in any setting, since it is usually similar or worse than Lasso, and is more computationally expensive.

The overall average correlation between pairs of covariates is weak in all correlated designs (due to the block structure for synthetic data and reflecting the correlation in the real data set for the semi-synthetic data).
However, despite this weak average correlation, we have found that method performance in the synthetic pairwise correlation design and semi-synthetic ``high'' correlation design can still differ greatly to performance in the synthetic independence design (or semi-synthetic ``low'' correlation design).
This is because an important factor for method performance is the magnitude of correlation between signals, or between signals and non-signals.

For prediction, where we mostly recommended Lasso, Ridge does particularly badly in many ``easier'' scenarios, but it is worth pointing out that most scenarios considered here were unfriendly to Ridge in the sense of being highly sparse, and with low overall correlation (across all predictors). 
In many areas such as biomedicine, signals can be weak and so SNR may be at the low end of the values considered here, or possibly even smaller.
In such difficult settings, Ridge may be a good option and our results indeed suggest this, as the only scenarios where we saw any benefit of an $L_2$ penalty for prediction were those with small $r$ and SNR.

We focused on simulations from the sparse linear model to better understand the variability of performance in a broadly favorable setting. Extending this systematic empirical approach to (the huge range of) less favorable settings, spanning many kinds of model mis-specification, could be illuminating, but experimental design would be nontrivial. 
As one example, we revisited a ``low'' correlation scenario from the semi-synthetic data analysis, but with a non-Gaussian error distribution.
Figure~S21 shows method performance for all metrics and provides details of data generation. Method performance deteriorates as non-normality increases. SCAD is the most affected and mirrors its previous behavior, with a transition in performance from best to worst as non-normality increases for ranking and prediction. 

Our comparison focused on seven popular penalized linear regression methods, but there are of course many others that have been proposed, and some of these are also well-known.
For example, there are relatively well-known extensions of Lasso that have been proposed for data where covariates can be grouped \citep[Group Lasso;][]{Yuan2006} or ordered \citep[Fused Lasso;][]{Tibshirani2005}.
While, for reasons of tractability, our comparison was restricted to seven methods, we make our simulation code and method performance data available, allowing users to add in other approaches of interest into the comparison without the need to regenerate the results for the seven methods considered here.

Choices of tuning parameters can be crucial. In line with known results, we saw that standard cross-validation often yielded overly large models for Lasso and Elastic Net. An interesting alternative is proposed in \citet{Lim2016}, where cross-validation is based on an estimation stability metric. Compared to traditional cross-validation, this approach significantly reduces the false positive rate while slightly sacrificing the true positive rate, and achieves similar prediction but higher accuracy in parameter estimation. For Stability Selection, in \citet{Meinshausen2010discussion} the author points out that there is no established lower bound for the expected number of true positives, and the tuning parameters $\pi_{thr}$ and $\tilde{V}$ have significant influences on the true positive rate. They also found in their simulation study that the number of false positives is usually smaller than the specified $\tilde{V}$. This suggests that less stringent $\tilde{V}$ can help improve signal detection without sacrificing false positive control too much, thus providing a better balance between the two. This is reflected in our results in Section \ref{subsec:stability}. 

We explicitly defined the true model in terms of exact sparsity (i.e. some coefficients being precisely zero). Although this is the best studied case, in practice such a notion of sparsity may not be realistic and a more reasonable assumption may be that there are a few strong signals, several moderate signals and even more weak signals, but the majority of variables are irrelevant with small, but sometimes non-zero coefficients. In this case, since it may not be possible to find all relevant variables, a good method might be expected to detect all strong and moderate signals while removing the weaker ones. In this vein, \citet{Zhang2008} consider the problem where weak signals exist outside the ideal model, such that their total signal strength is below a certain level. The authors prove that the Lasso estimate has model size of the correct order, and the selection bias is controlled by the weak signal coefficients and a threshold bias. 

Due to the comprehensive nature of our simulation study, we focused on summarizing the predominant trends and relationships across the scenarios.
There will always be some scenarios which are exceptions to these summaries, but this in itself motivates the need for extensive simulation studies.
If a simulation study has limited scope then the derived conclusions may not generalize beyond the few scenarios considered. 
So while such studies may be useful in exploring and understanding the properties of a method, they may have limited practical implications for an end-user.
In contrast, a large-scale simulation study, such as the one presented here, can reveal which approaches perform well across a broad range of scenarios. 
These approaches may then translate into being a good or ``safe'' choice for the user's setting.
In addition, the study can offer some insight as to whether certain methods are best avoided, because they have high variability across scenarios in the study.

\section*{Code and data availability}
All analysis was performed in R \citep{R}. 
Scripts for generating the main simulation data sets, applying the regression methods, assessing performance and plotting results are available at \url{https://github.com/fw307/high_dimensional_regression_comparison}, together with performance metric data from the main simulation. \\

\subsection*{Acknowledgements}
This work was supported by the UK Medical Research Council (University Unit Programme numbers MC\_UU\_00002/2 and MC\_UU\_00002/10).

\bibliographystyle{abbrvnat}
\bibliography{Record}

\newpage

\renewcommand{\thefigure}{S\arabic{figure}}
\setcounter{figure}{0}

\begin{center}
{\LARGE {High-dimensional regression in practice:\\ an empirical study of finite-sample prediction, variable selection and ranking}}\\
\vspace{0.5cm}
{\LARGE {Supplementary Material}}\\
\vspace{0.5cm}
{\large Fan Wang$^1$, Sach Mukherjee$^2$, Sylvia Richardson$^1$ and Steven M.\ Hill$^1$}\\
\vspace{0.2cm}
			{\small 1. MRC Biostatistics Unit, University of Cambridge, Cambridge, UK\\
			 2. German Centre for Neurodegenerative Diseases (DZNE), Bonn, Germany}
\end{center}

\section*{Supplementary Figures}

\begin{figure*}[h]
\centering
\includegraphics[height=5in]{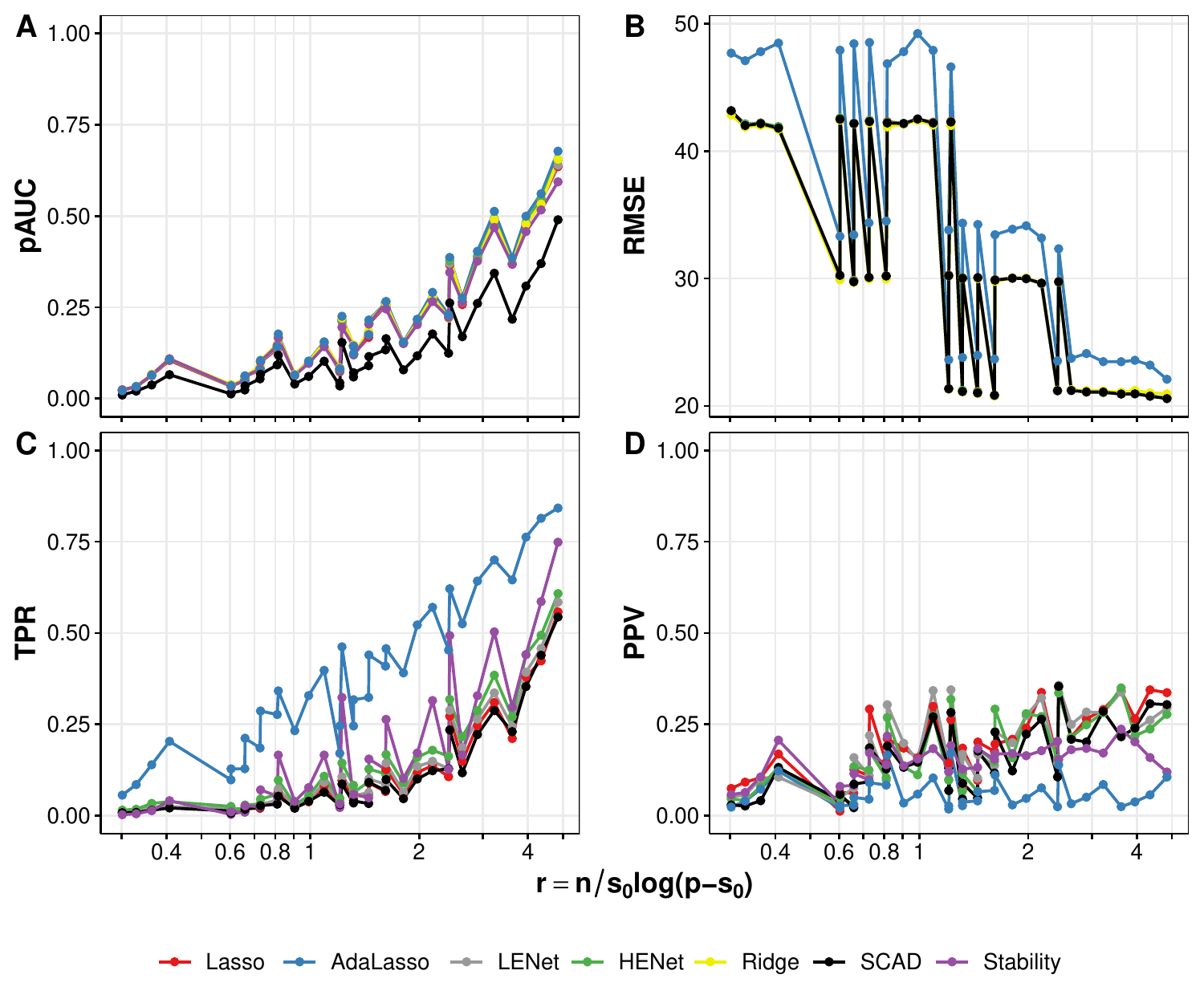}
\caption{Ranking (A), prediction (B) and selection (C,D) performance versus the rescaled sample size $r=n/(s_0log(p-s_0))$  for synthetic independence design scenarios.
As Figure~1 in Main Text, but with SNR=0.5 (instead of SNR=2).}
\label{fig:r_independence_snr1}
\end{figure*}

\begin{figure*}
\centering
\includegraphics[height=5in]{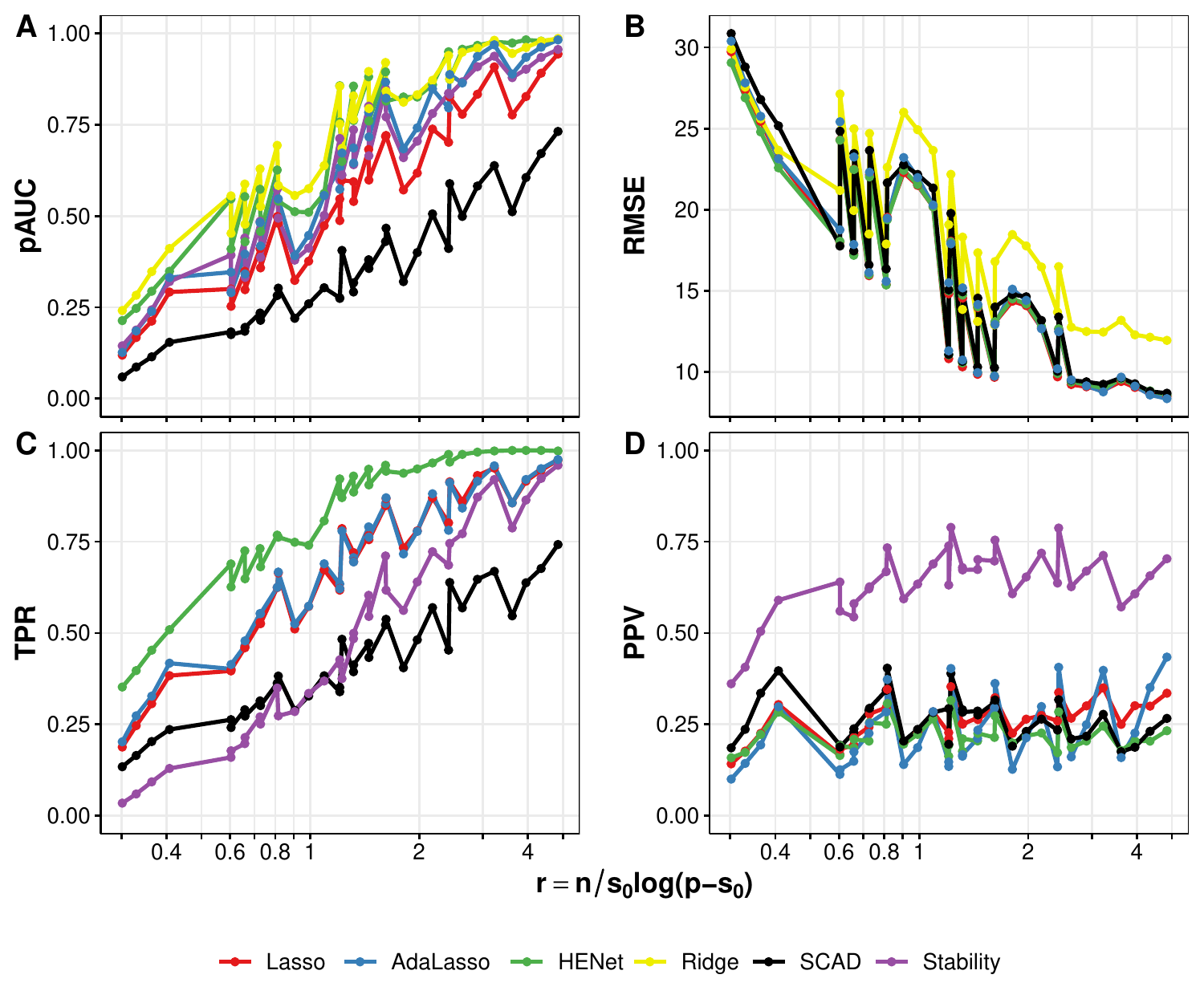}
\caption{Ranking (A), prediction (B) and selection (C,D) performance versus the rescaled sample size $r=n/(s_0log(p-s_0))$  for a semisynthetic ``high''-correlation design scenario.
As Figure~1 in Main Text, but for a semisynthetic ``high''-correlation design with SNR=2 and $s_0^B=5$.}
\label{fig:r_independence_snr4}
\end{figure*}

\begin{figure*}
\centering
\includegraphics[height=6in]{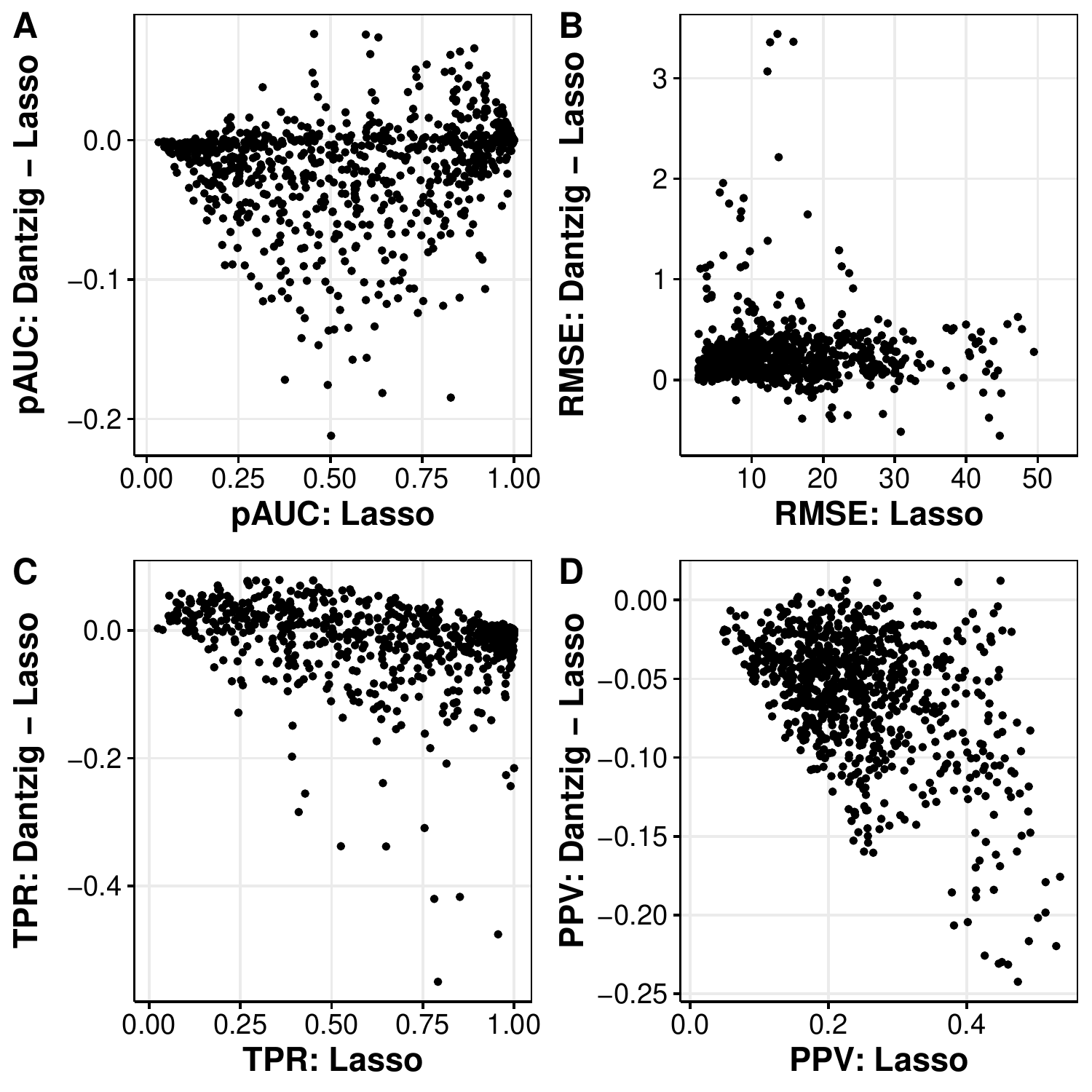}
\caption{Difference in performance between Dantzig and Lasso (Dantzig - Lasso) versus Lasso performance for ranking (A), prediction (B) and selection (C,D) in synthetic data scenarios.
Each point plotted represents a synthetic data scenario (both independence design and correlation design scenarios are plotted).
For A, C and D, negative values on the $y$-axis indicate that Lasso is outperforming Dantzig. For B, a positive value indicates the same.
}
\label{fig:dantzig_vs_lasso}
\end{figure*}

\begin{figure*}
\centering
\includegraphics[height=7in]{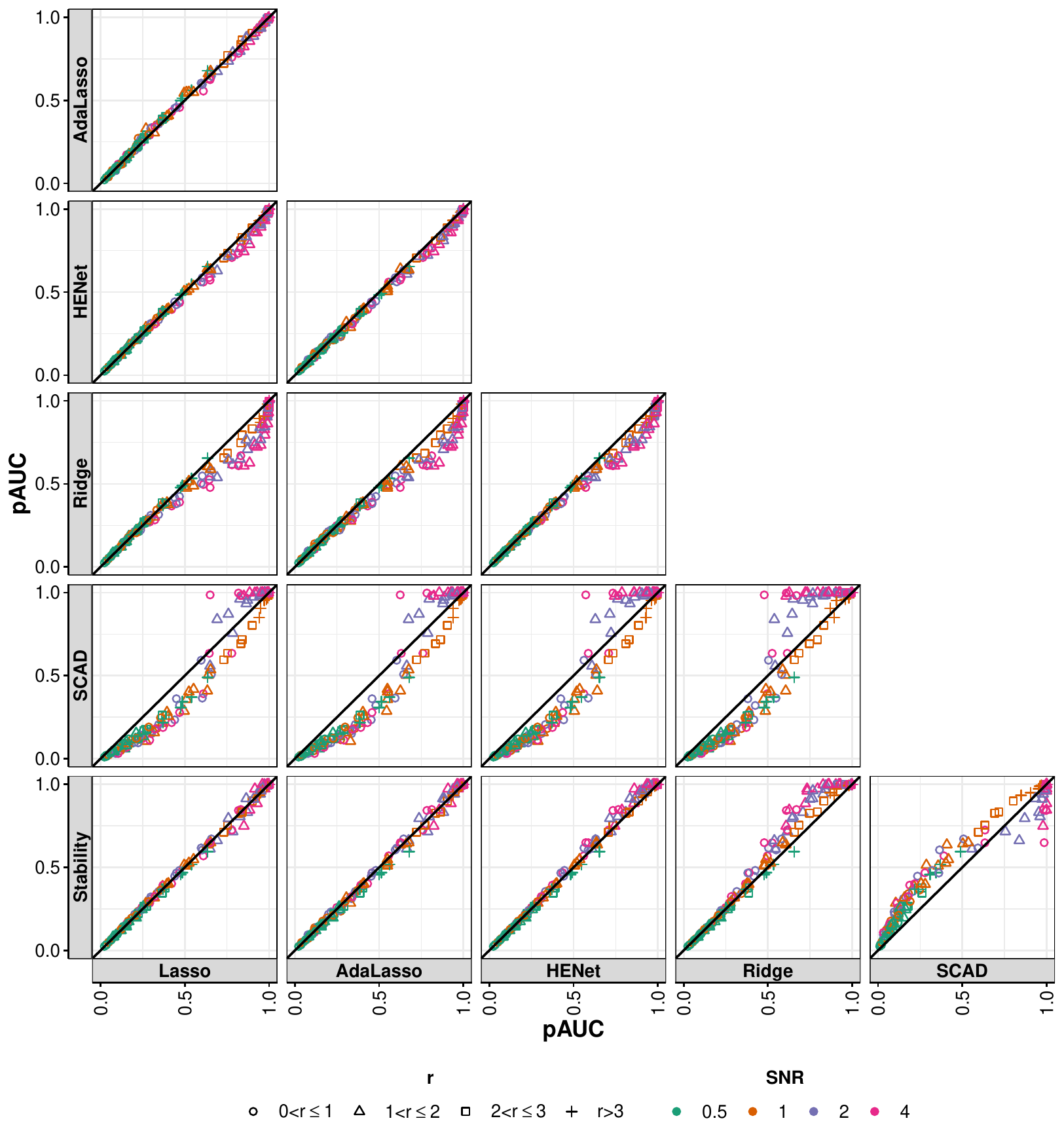}
\caption{A comparison of method performance in synthetic independence design scenarios: ranking.
Each panel plots the ranking performance of one method versus the ranking performance of another method. Each data point within a panel corresponds to an independence design scenario with color indicating SNR and symbol representing the value of the rescaled sample size $r$ (categorized).}
\label{fig:method_vs_method_independence_pauc}
\end{figure*}

\begin{figure*}
\centering
\includegraphics[height = 7in]{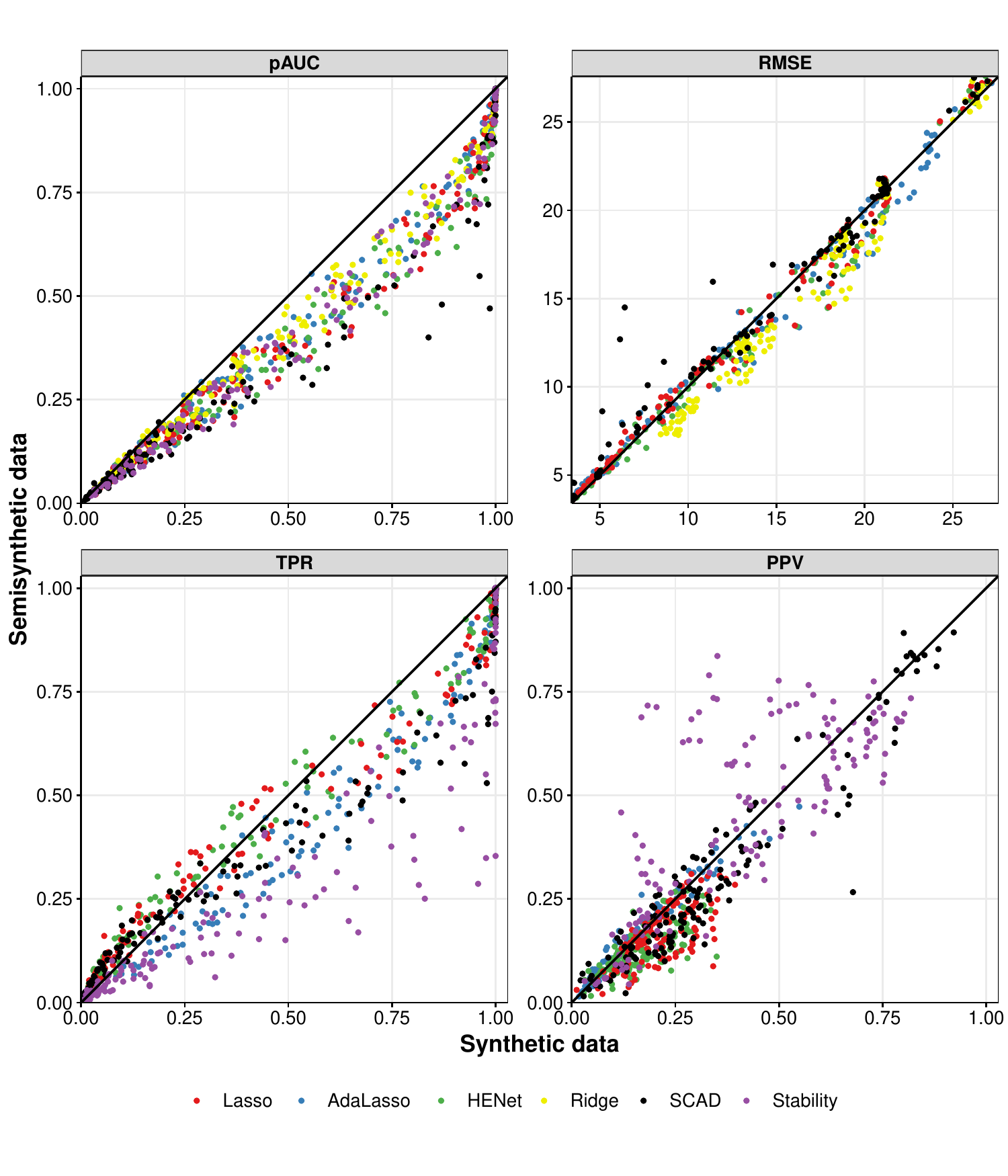}
\caption{A comparison of method performance in the synthetic independence design and semisynthetic ``low'' correlation design.
Each panel shows a different metric and each data point within a panel corresponds to a specific scenario (defined by $n$, $p$, $s_0$ and SNR), with color indicating method.}
\label{fig:ind_vs_semi_synth_low}
\end{figure*}

\begin{figure*}
\centering
\includegraphics[height=7in]{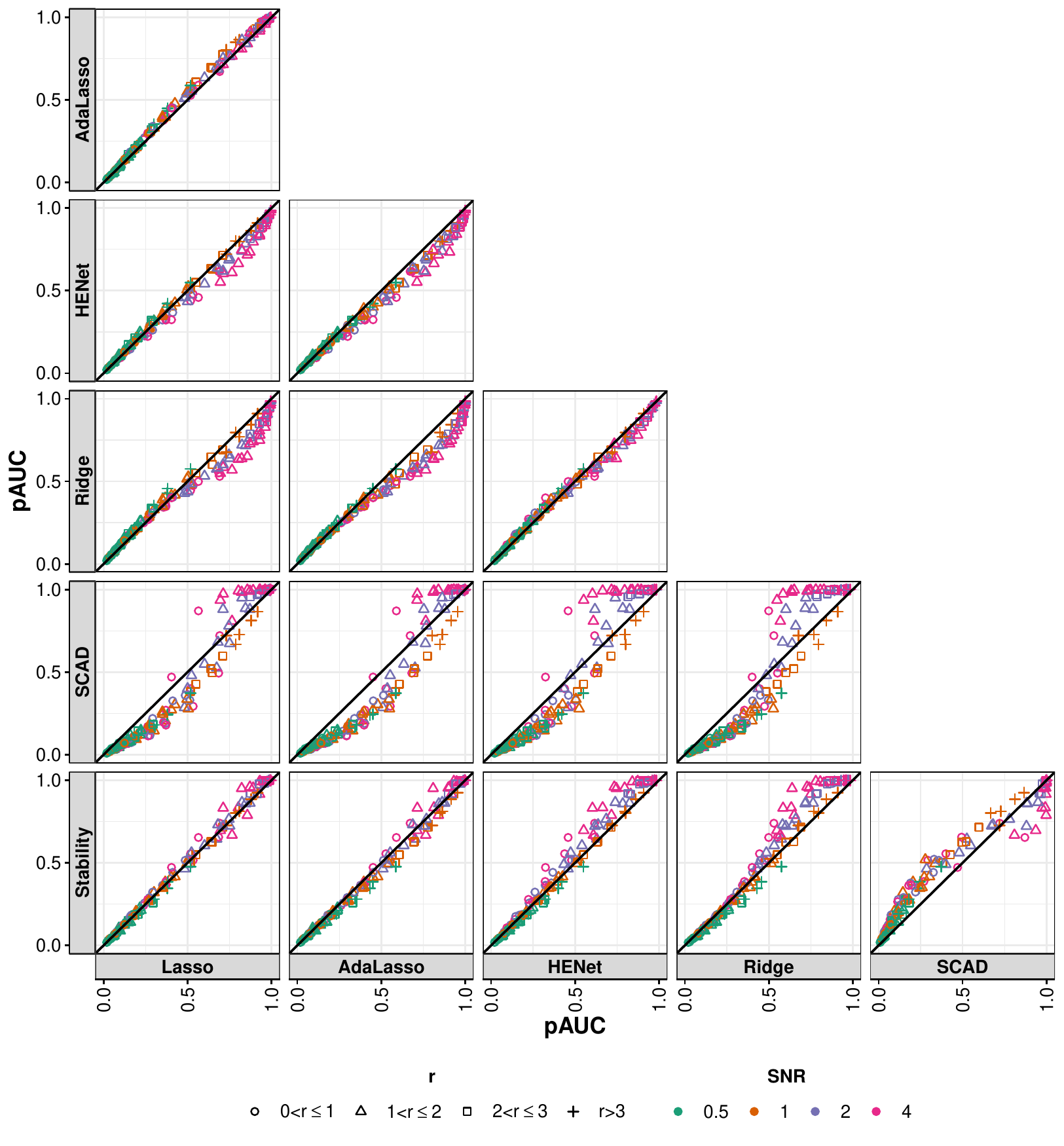}
\caption{A comparison of method performance in semisynthetic ``low''-correlation design scenarios: ranking.
Each panel plots the ranking performance of one method versus the ranking performance of another method. Each data point within a panel corresponds to a scenario with color indicating SNR and symbol representing the value of the rescaled sample size $r$ (categorized).}
\label{fig:method_vs_method_independence_pauc}
\end{figure*}

\begin{figure*}
\centering
\includegraphics[height=8in]{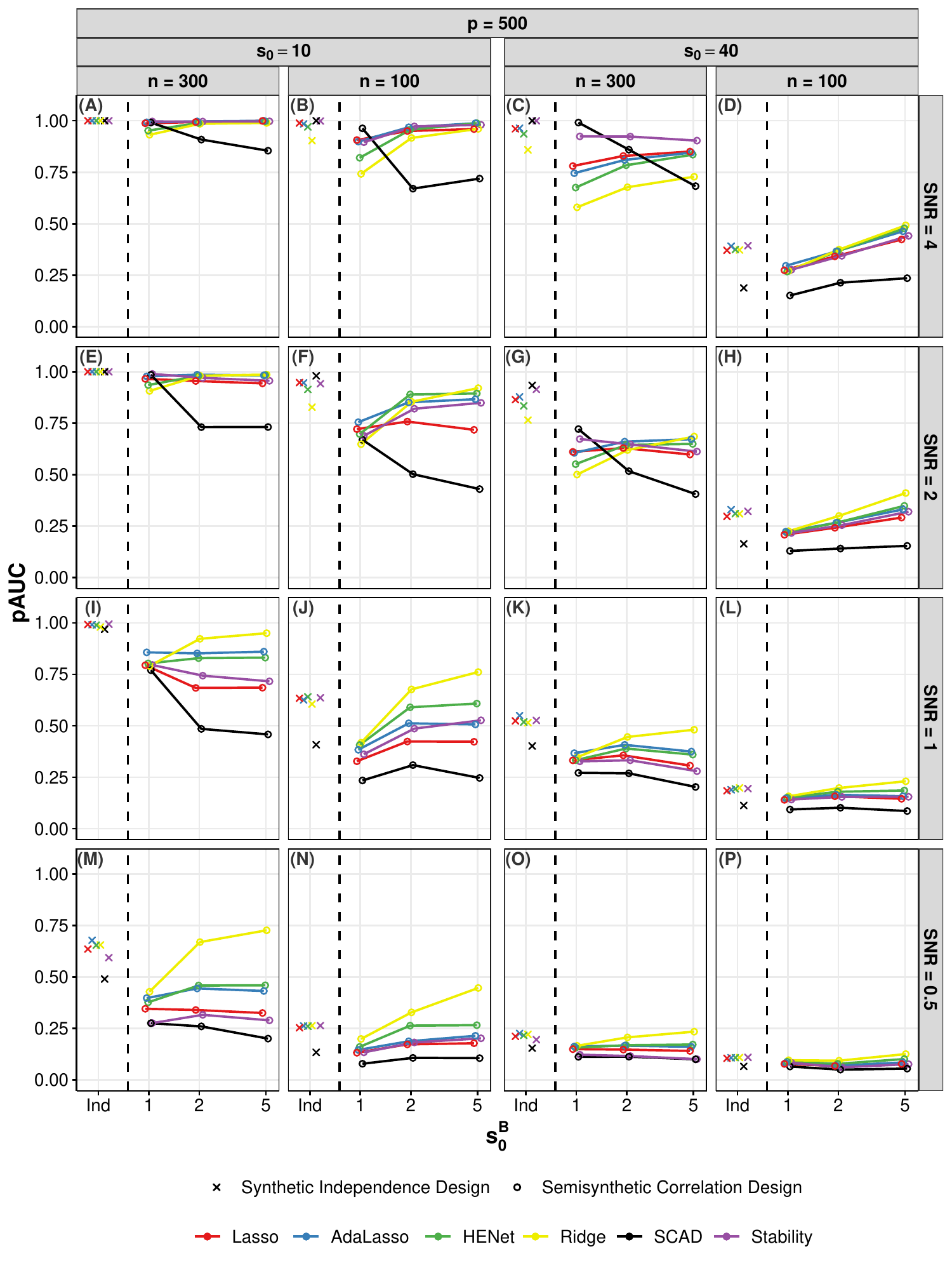}
\caption{Ranking performance (pAUC) versus $s_0^B$ (number of signals per block) for a subset of semisynthetic ``high''-correlation designs.
As Figure~3 in Main Text, but with $p{=}500$ (instead of $p{=}2000$) and all values of SNR are shown.
}
\end{figure*}

\begin{figure*}
\centering
\includegraphics[height=8in]{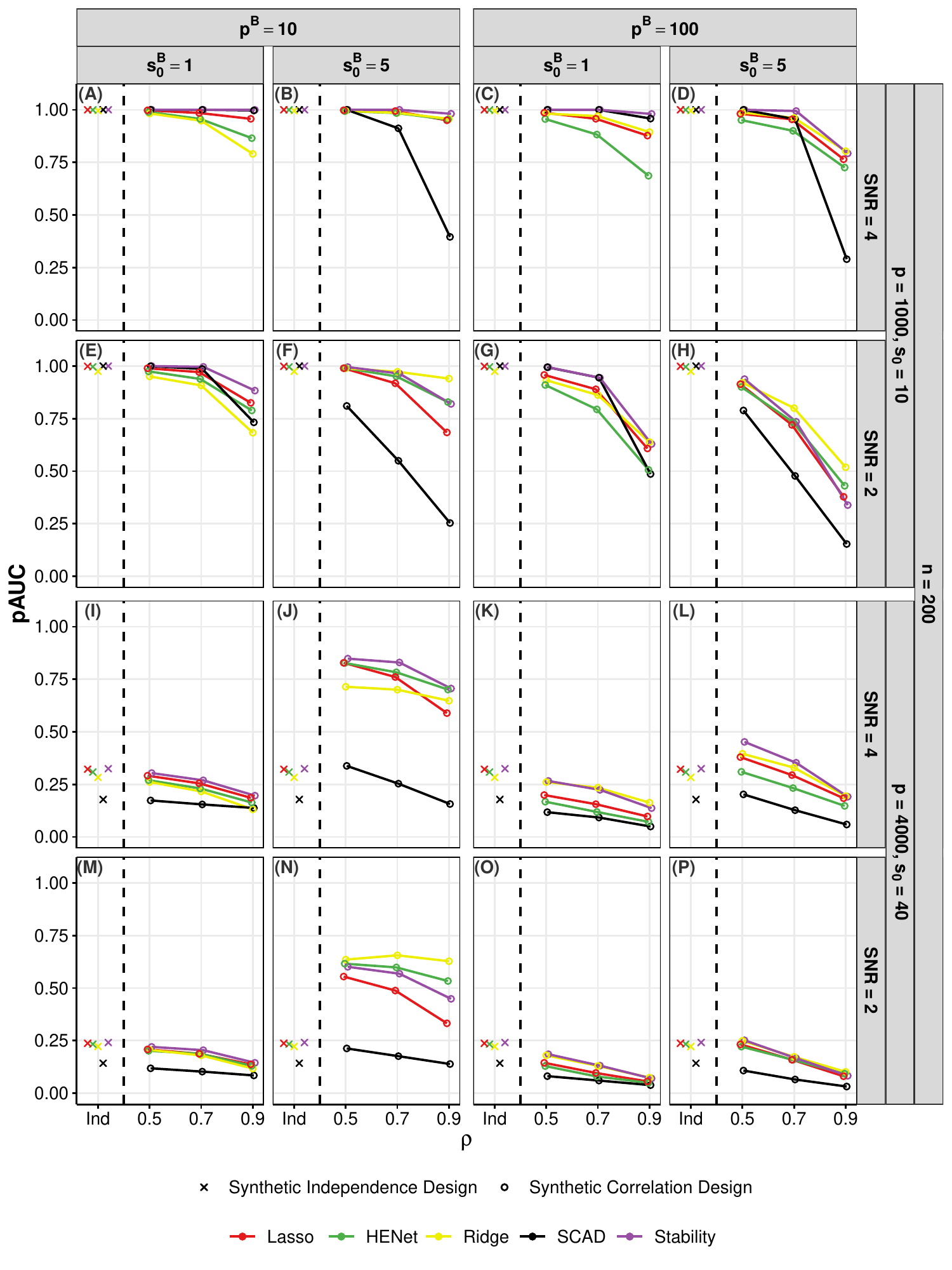}
\caption{Ranking performance (pAUC) versus $\rho$ (correlation strength) for a subset of synthetic pairwise correlation designs.
As Figure~4 in Main Text, but with SNR=2 and 4 (instead of SNR=1).
}
\end{figure*}

\begin{figure*}
\centering
\includegraphics[height=7in]{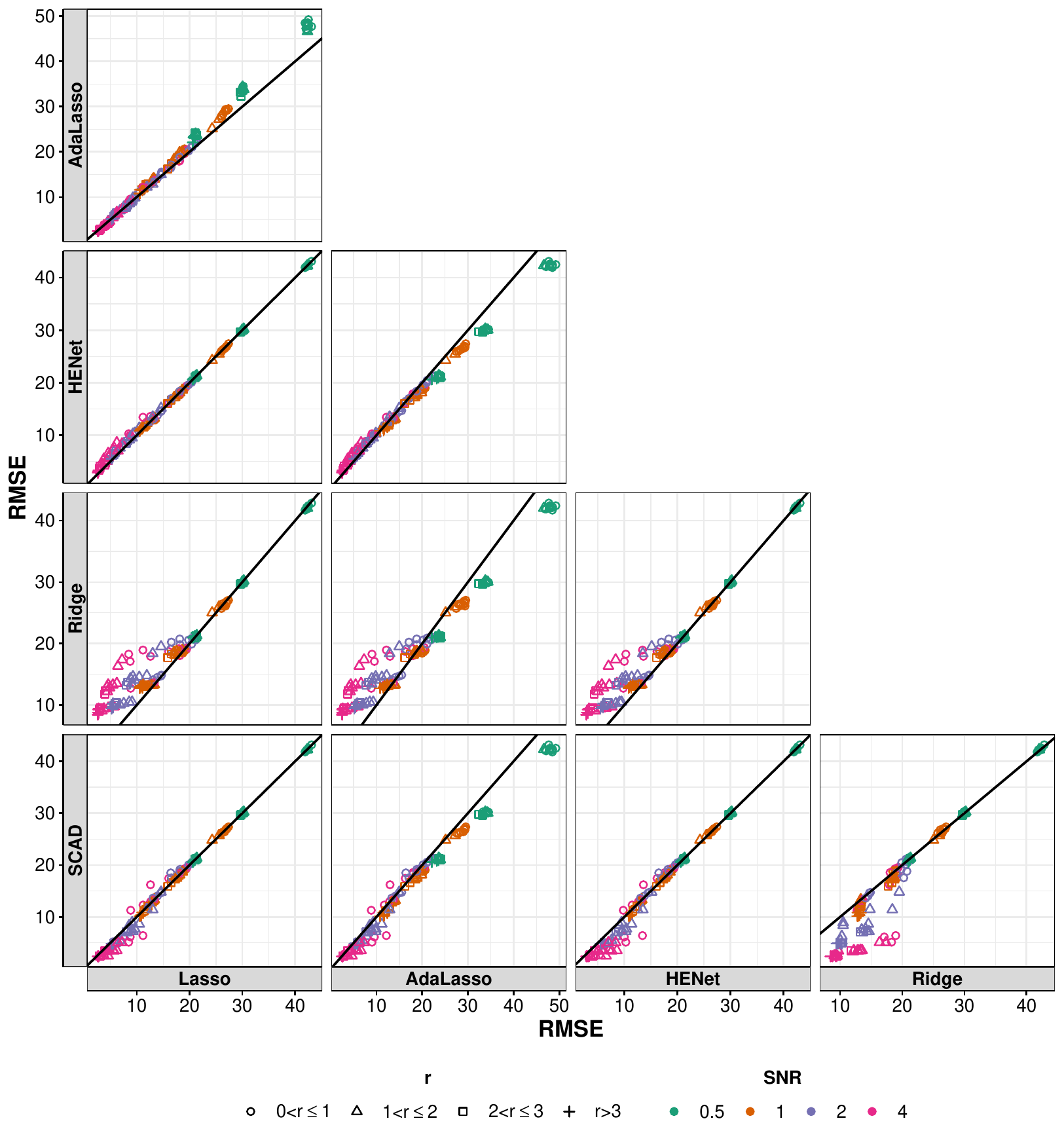}
\caption{A comparison of method performance in synthetic independence design scenarios: prediction.
Each panel plots the prediction performance of one method versus the prediction performance of another method. Each data point within a panel corresponds to an independence design scenario with color indicating SNR and symbol representing the value of the rescaled sample size $r$ (categorized).
}
\end{figure*}

\begin{figure*}
\centering
\includegraphics[height=7in]{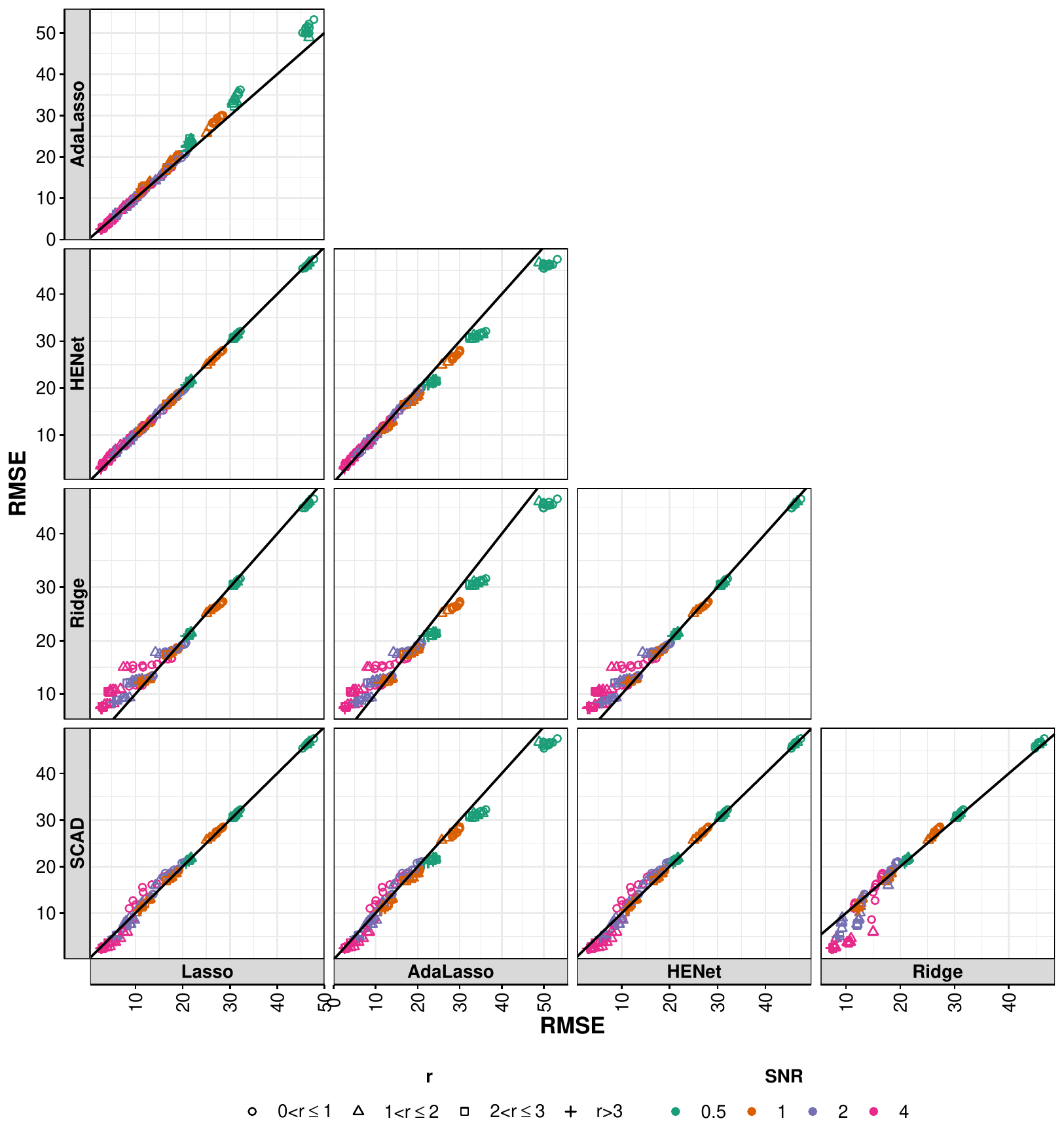}
\caption{A comparison of method performance in semisynthetic ``low''-correlation design scenarios: prediction.
Each panel plots the prediction performance of one method versus the prediction performance of another method. Each data point within a panel corresponds to a scenario with color indicating SNR and symbol representing the value of the rescaled sample size $r$ (categorized).
}
\end{figure*}

\begin{figure*}
\centering
\includegraphics[height=8in]{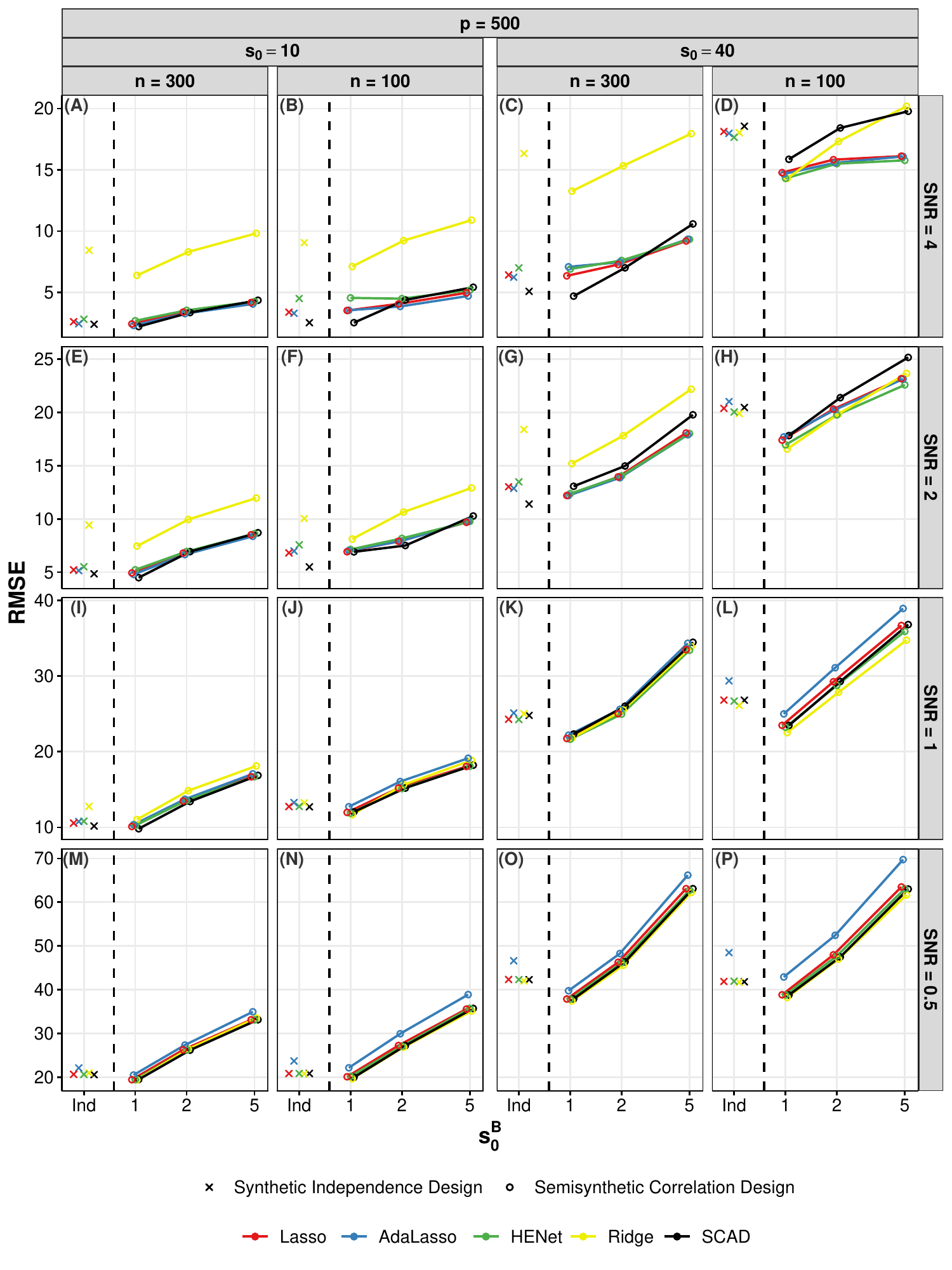}
\caption{Prediction performance (RMSE) versus $s_0^B$ (number of signals per block) for a subset of semisynthetic ``high''-correlation designs.
As Figure~6 in Main Text, but with $p{=}500$ (instead of $p{=}2000$) and all values of SNR are shown.
}
\end{figure*}

\begin{figure*}
\centering
\includegraphics[height=8in]{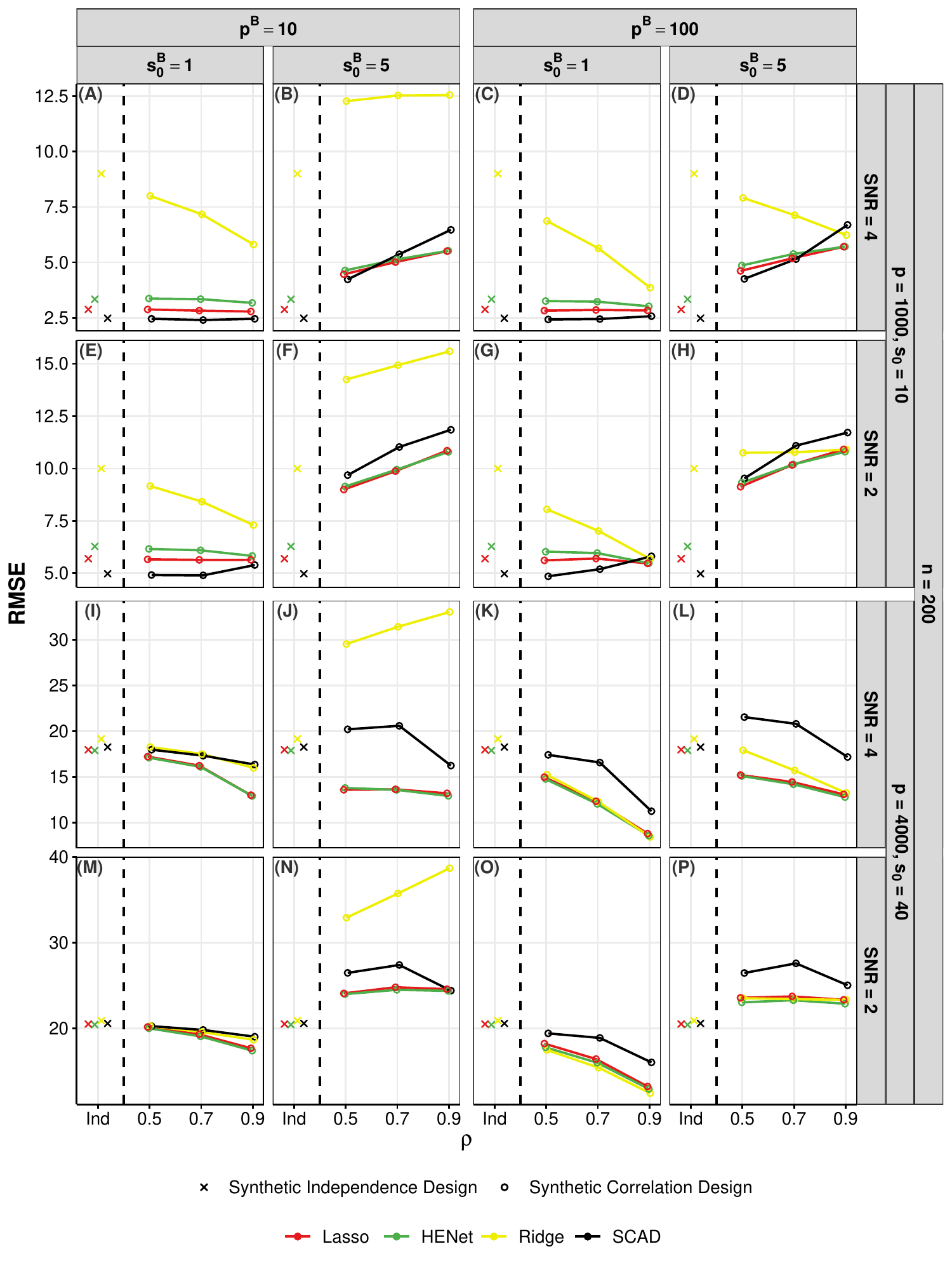}
\caption{Prediction performance (RMSE) versus $\rho$ (correlation strength) for a subset of synthetic pairwise correlation designs.
As Figure~7 in Main Text, but with SNR=2 and 4 (instead of SNR=1).
}
\end{figure*}

\begin{figure*}
\centering
\includegraphics[height=7in]{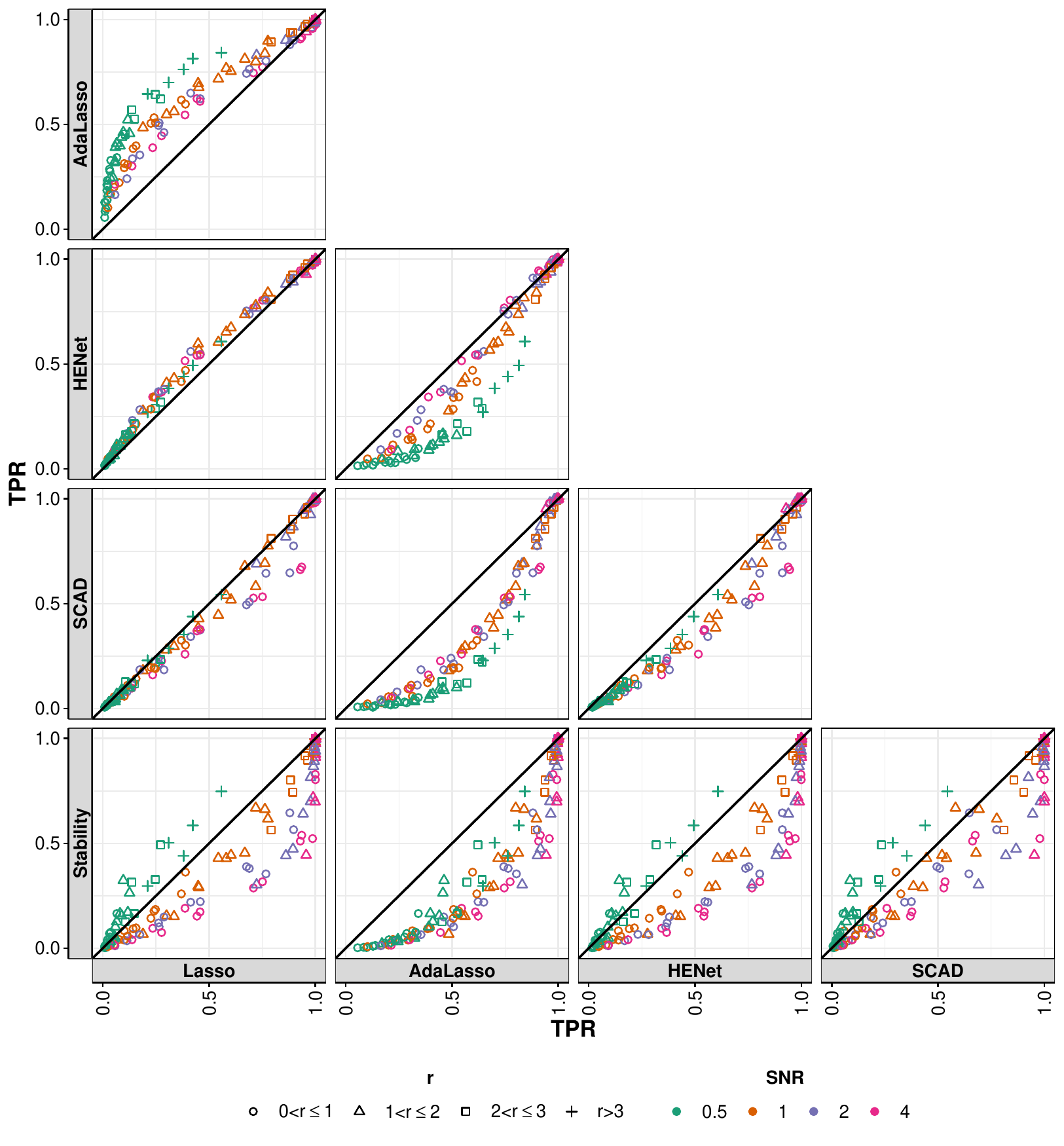}
\caption{A comparison of method performance in synthetic independence design scenarios: selection - TPR.
Each panel plots TPR of one method versus TPR of another method. Each data point within a panel corresponds to an independence design scenario with color indicating SNR and symbol representing the value of the rescaled sample size $r$ (categorized).
}
\end{figure*}

\begin{figure*}
\centering
\includegraphics[height=7in]{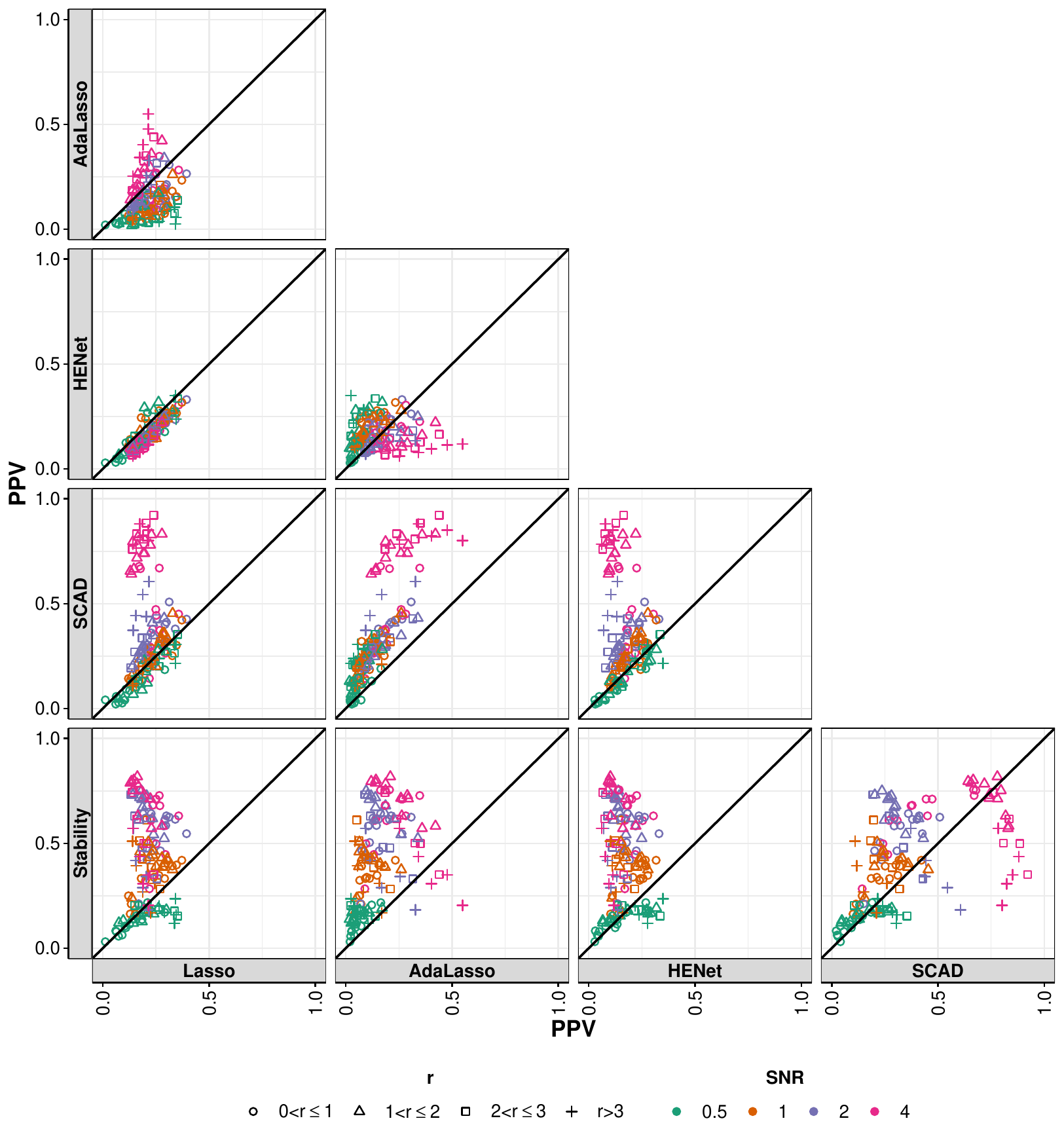}
\caption{A comparison of method performance in synthetic independence design scenarios: selection - PPV.
Each panel plots PPV of one method versus PPV of another method. Each data point within a panel corresponds to an independence design scenario with color indicating SNR and symbol representing the value of the rescaled sample size $r$ (categorized).
}
\end{figure*}

\begin{figure*}
\centering
\includegraphics[height=7in]{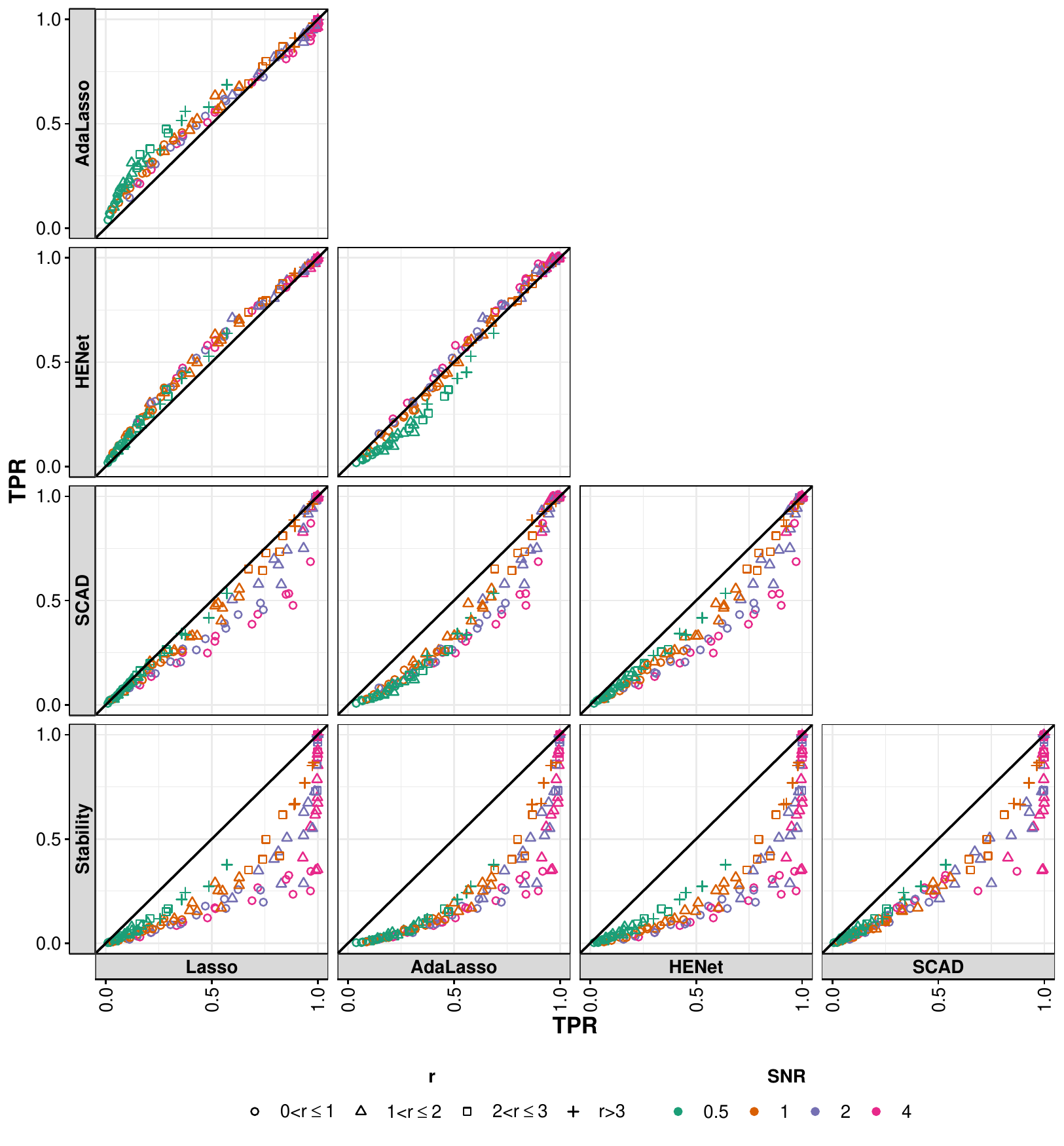}
\caption{A comparison of method performance in semisynthetic ``low''-correlation design scenarios: selection - TPR.
Each panel plots TPR of one method versus TPR of another method. Each data point within a panel corresponds to a scenario with color indicating SNR and symbol representing the value of the rescaled sample size $r$ (categorized).
}
\end{figure*}

\begin{figure*}
\centering
\includegraphics[height=7in]{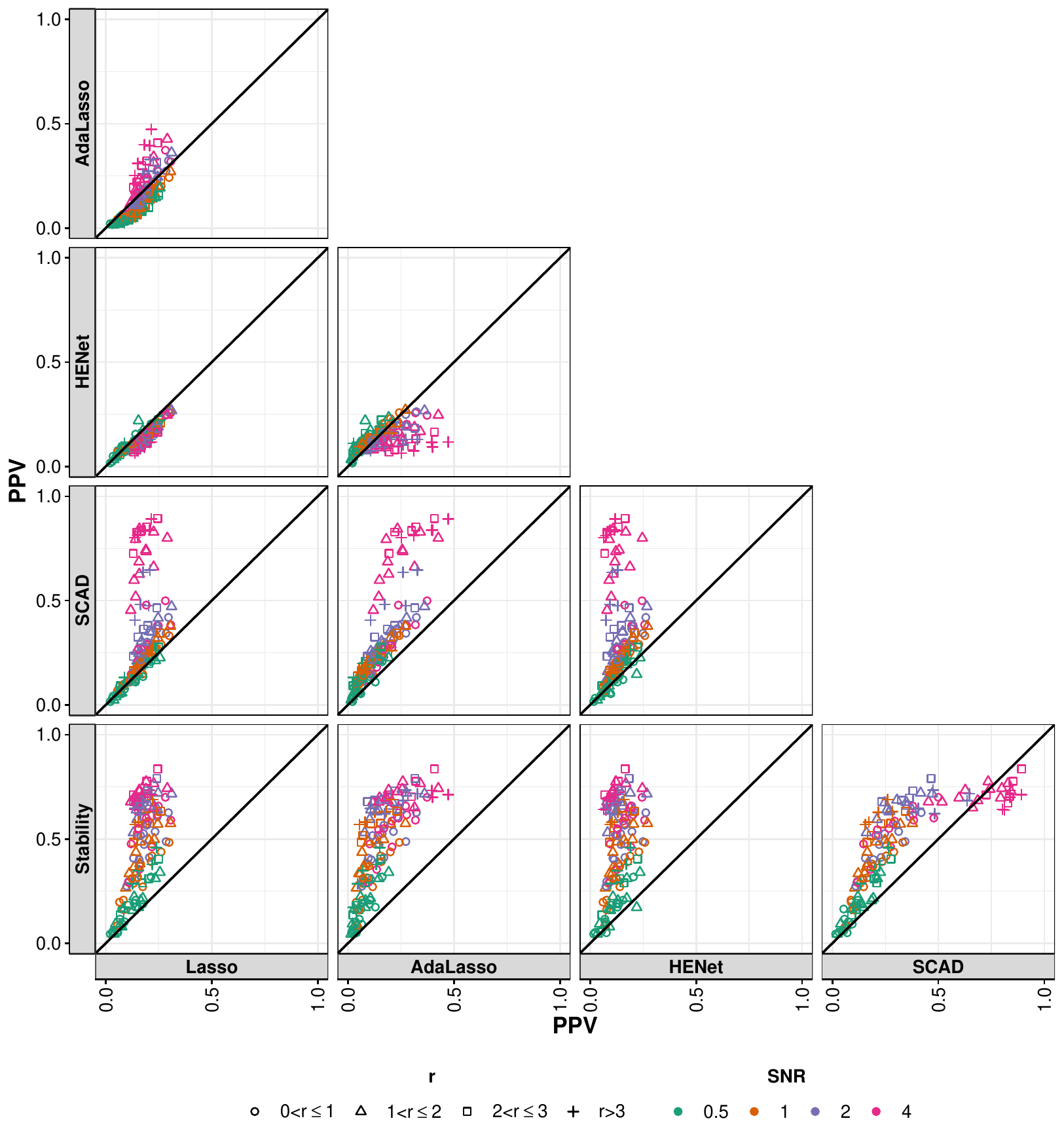}
\caption{A comparison of method performance in semisynthetic ``low''-correlation design scenarios: selection - PPV.
Each panel plots PPV of one method versus PPV of another method. Each data point within a panel corresponds to a scenario with color indicating SNR and symbol representing the value of the rescaled sample size $r$ (categorized).
}
\end{figure*}

\begin{figure*}
\centering
\includegraphics[height=8.5in]{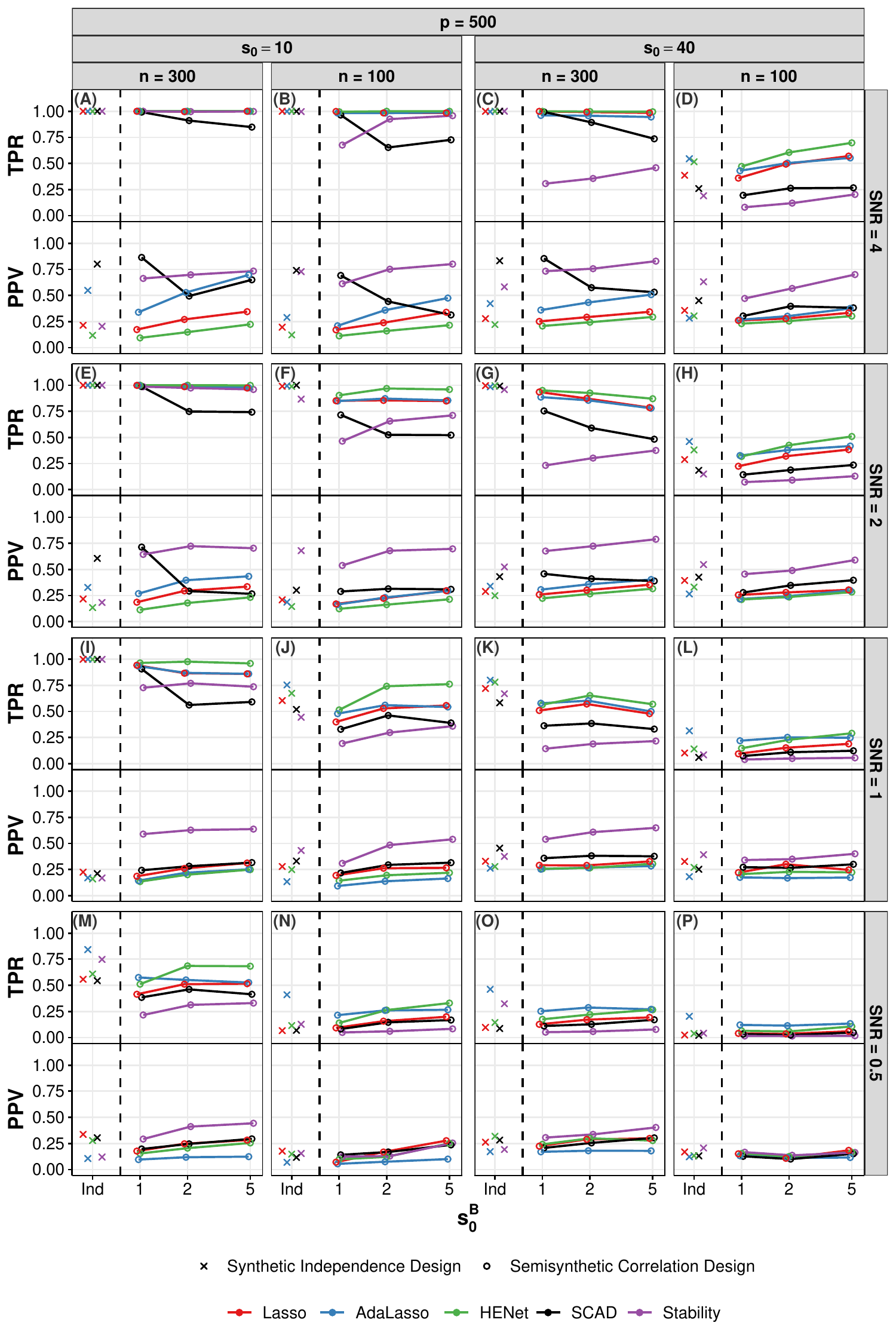}
\caption{Selection performance (TPR and PPV) versus $s_0^B$ (number of signals per block) for a subset of semisynthetic ``high''-correlation designs.
As Figure~9 in Main Text, but with $p{=}500$ (instead of $p{=}2000$) and all values of SNR are shown.
}
\end{figure*}

\begin{figure*}
\centering
\includegraphics[height=8.5in]{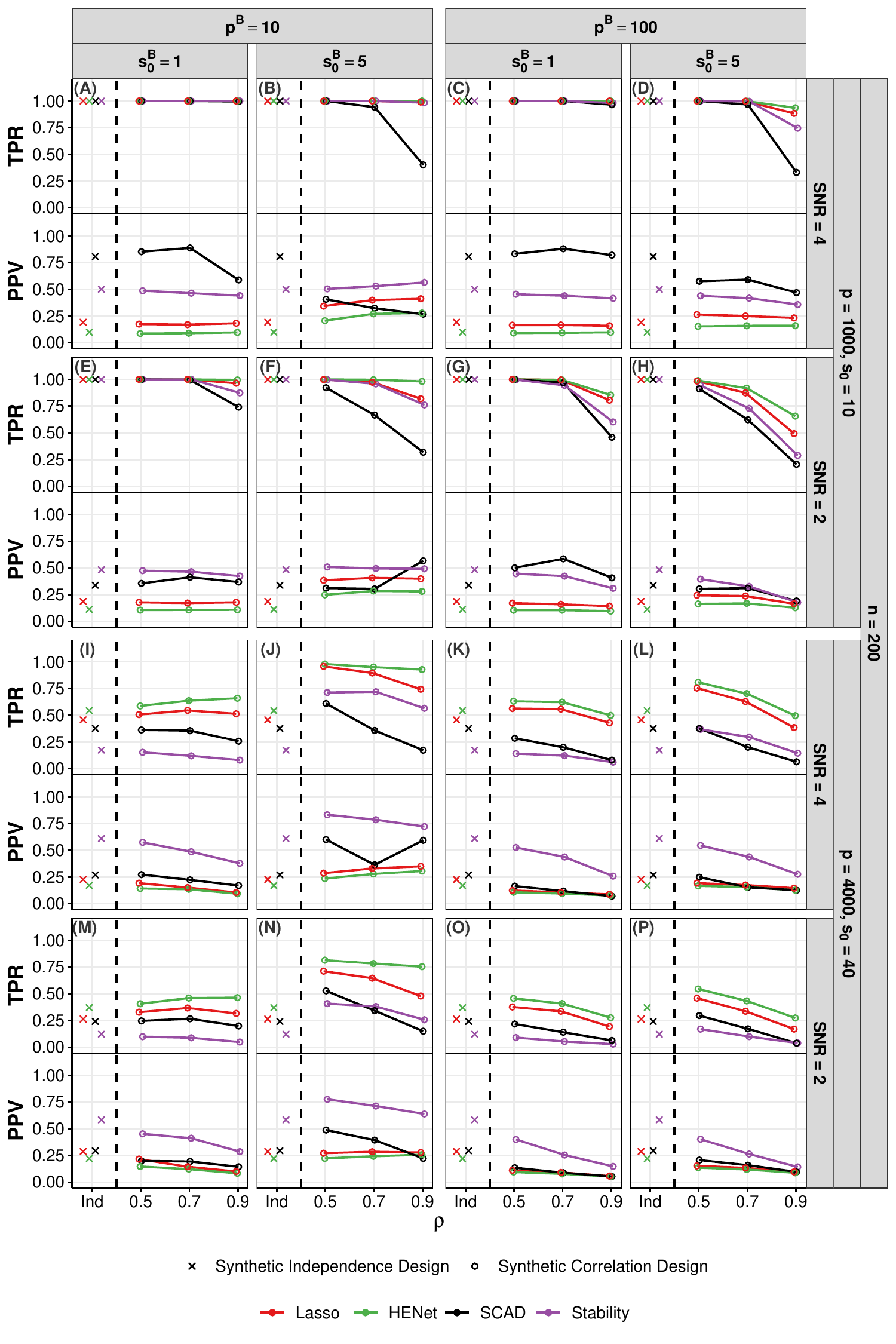}
\caption{Selection performance (TPR and PPV) versus $\rho$ (correlation strength) for a subset of synthetic pairwise correlation designs.
As Figure~10 in Main Text, but with SNR=2 and 4 (instead of SNR=1).
}
\end{figure*}

%%%%%%%%%%%%%%%%%%%%%%%%%%%%%%%%%%%%%%%%%%%%%%%%%%%

\begin{figure*}
\centering
\includegraphics[width=6in,height=6in]{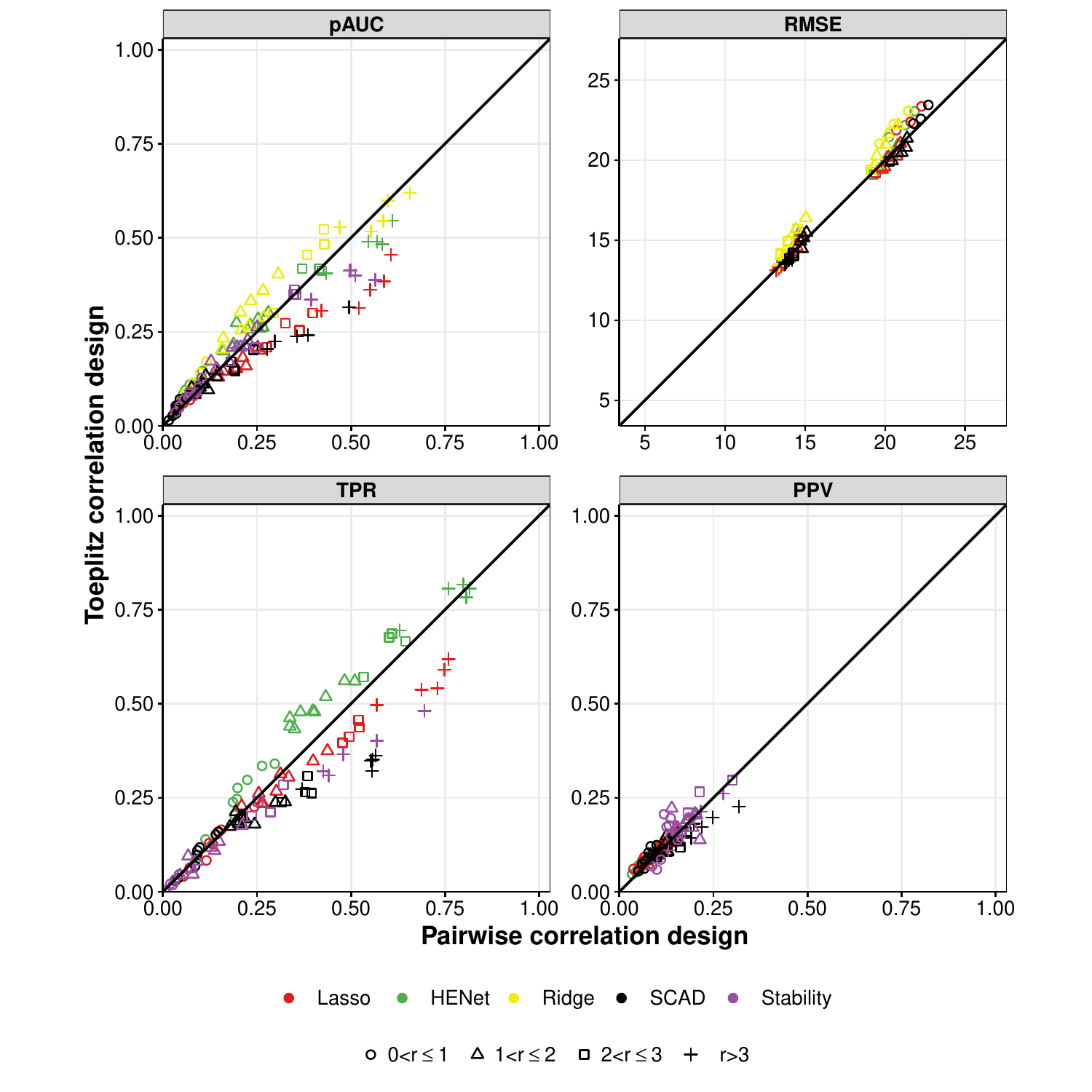}
\caption{Comparison between Toeplitz correlation and pairwise correlation designs for ranking, prediction and selection performance.
As Figure~11 in Main Text, but with SNR=1 (instead of SNR=2).}
\label{fig:pairwise_vs_toeplitz_snr1}
\end{figure*}

\begin{figure*}
\centering
\includegraphics[width=6in,height=6in]{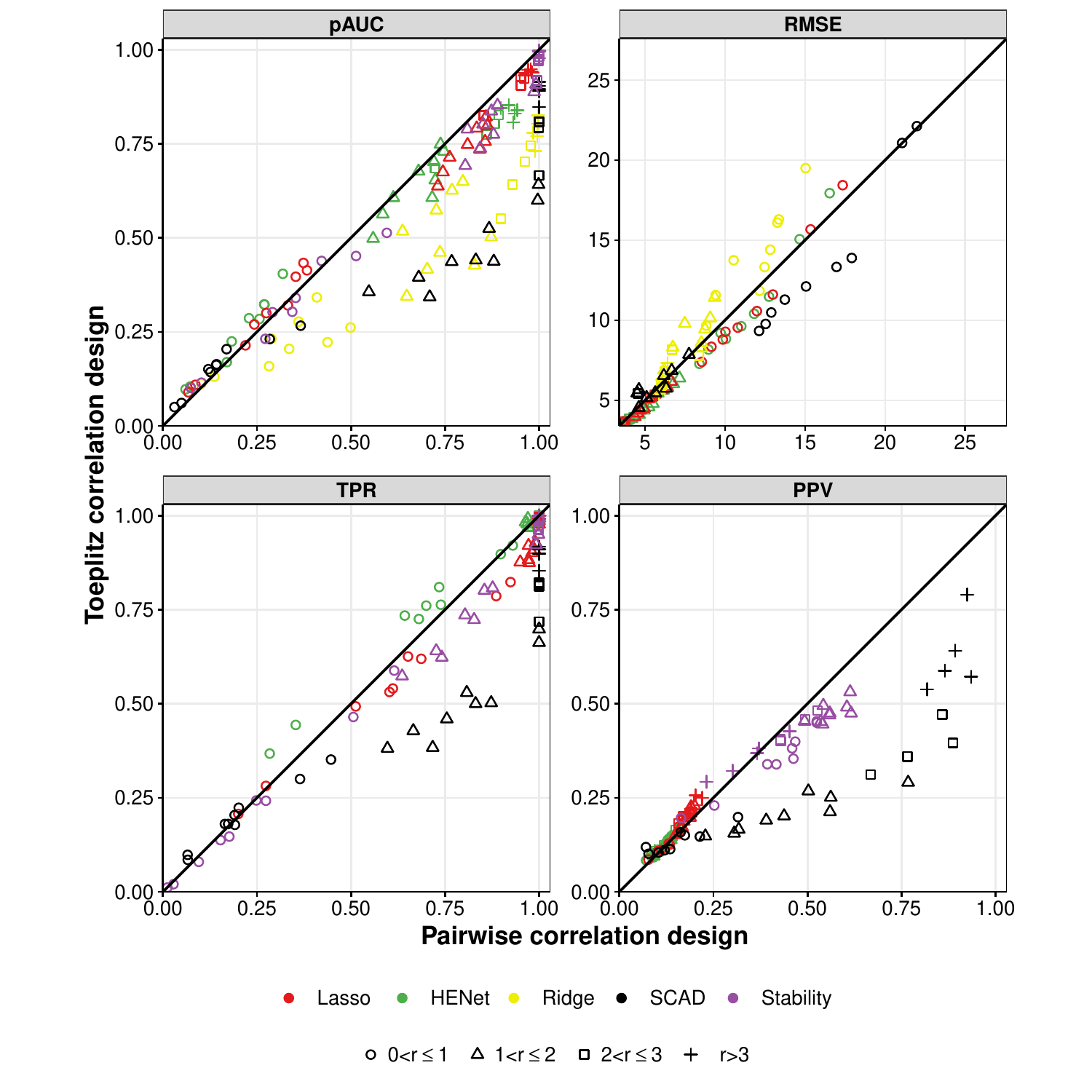}
\caption{Comparison between Toeplitz correlation and pairwise correlation designs for ranking, prediction and selection performance.
As Figure~11 in Main Text, but with SNR=4 (instead of SNR=2).}
\label{fig:pairwise_vs_toeplitz_snr4}
\end{figure*}

\begin{figure*}
\centering
\includegraphics[width=6in,height=6in]{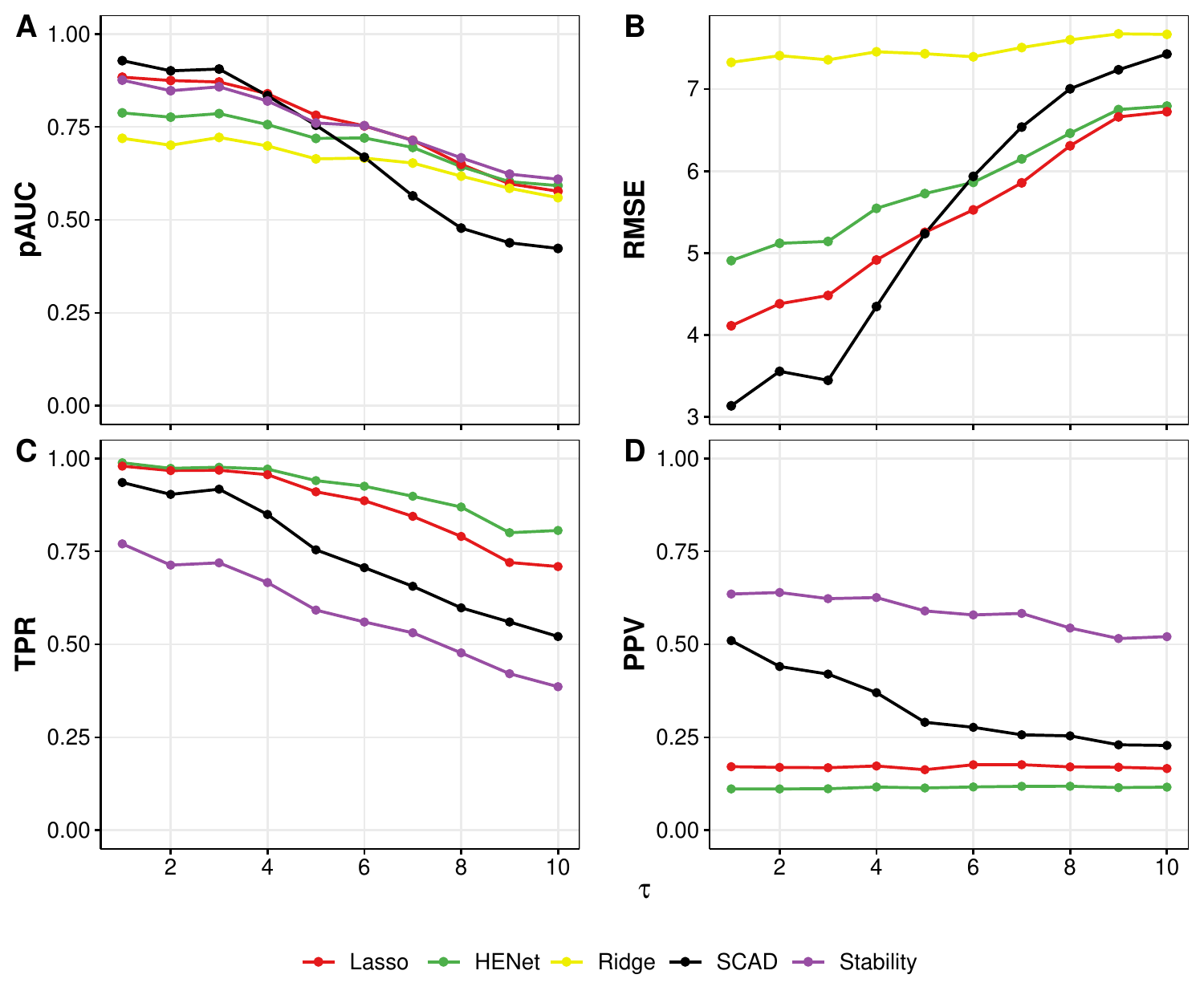}
\caption{Semisynthetic (TCGA ovarian cancer expression) data analysis: ``low''-correlation scenario with non-Gaussian error distribution.
Semisynthetic training and test datasets were generated as described in the Main Text for the ``low''-correlation scenario with $n=100$, $p=1000$ and $s_0=10$, but with 95$\%$ of error terms drawn from $N(0,\sigma^2)$ and the other 5$\%$ drawn from $N(0, {(\tau \sigma)}^2)$, with $\sigma$ set such that SNR=4 and $\tau\in\left\{1,\dots,10\right\}$. $\tau=1$ represents the standard set-up with noise drawn from a single Gaussian distribution.
Ranking (A), prediction (B) and selection (C,D) performance are plotted against $\tau$.
Line color indicates method and results are averages over 100 replicates. }
\label{fig:heavy_tailed_error}
\end{figure*}

\end{document}